\documentclass[%
 reprint,
 amsmath,amssymb,
 aps,
]{revtex4-1}

\usepackage{graphicx}
\usepackage{dcolumn}
\usepackage{bm}
\usepackage{caption}
\usepackage{soul}
\usepackage{color}
\usepackage{adjustbox}
\usepackage{subcaption}
\usepackage{cleveref}

\begin{document}

\preprint{APS/123-QED}

\title{Machine-learned Interatomic Potentials for Alloys and Alloy Phase Diagrams}

\author{Conrad W. Rosenbrock}
 \email{rosenbrockc@gmail.com}
 \affiliation{Department of Physics and Astronomy, Brigham Young University, Provo UT USA 84602}
 \author{Konstantin Gubaev}%
 \author{Alexander V. Shapeev}%
\affiliation{Skolkovo Institute of Science and Technology, Skolkovo Innovation Center, Nobel str. 3, Moscow,
143026 Russia}
\author{Livia B. P\'artay}%
\affiliation{%
Department of Chemistry, University of Reading, Whiteknights, Reading, RG6 6AD, UK}%
\author{Noam Bernstein}%
\affiliation{Center for Computational Materials Science,  U.~S. Naval Research Laboratory, Washington DC 20375, USA}
 \author{G\'abor Cs\'anyi}%
\affiliation{%
 Department of Engineering, University of Cambridge, Trumpington Street, Cambridge, CB2 1PZ, UK}
\author{Gus L. W. Hart}
 \affiliation{Department of Physics and Astronomy, Brigham Young University, Provo Utah USA 84602}

\date{\today}

\begin{abstract}
We introduce machine-learned potentials for Ag-Pd to describe the energy of alloy configurations over a wide range of compositions.  We compare two different approaches. Moment tensor potentials (MTP) are polynomial-like functions of interatomic distances and angles. The Gaussian Approximation Potential (GAP) framework uses kernel regression, and we use the Smooth Overlap of Atomic Positions (SOAP) representation of atomic neighbourhoods that consists of a complete set of rotational and permutational invariants provided by the power spectrum of the spherical Fourier transform of the neighbour density. Both types of potentials give excellent accuracy for a wide range of compositions and rival the accuracy of cluster expansion, a benchmark for this system. While both models are able to describe small deformations away from the lattice positions, SOAP-GAP excels at transferability as shown by sensible transformation paths between configurations, and MTP allows, due to its lower computational cost, the calculation of compositional phase diagrams. Given the fact that both methods perform as well as cluster expansion would but yield off-lattice models, we expect them to open new avenues in computational materials modeling for alloys.
\end{abstract}

\maketitle


\section{\label{sec:Intro}Introduction}

The technology frontier relies on the exceptional performance of next-generation materials. First-principles calculations of material properties provide one way to discover new materials or optimize existing ones. However, calculating properties from a first principles approach is resource intensive, restricting its applicability. For example, molecular dynamics simulations using Density Functional Theory (DFT) are currently capable of handling fewer than a thousand atoms at picosecond time scales. However, many interesting materials science problems and technologically important processes can only by described with millions of atoms on micro- to millisecond time scales.

Conventional interatomic potentials (IPs) such as Lennard-Jones, embedded atom method (EAM), modified EAM, Tersoff, Stillinger-Weber, etc., typically provide six to eight orders of magnitude speed-up compared to DFT calculations, and due to their simple, physically motivated forms, they are somewhat robust in the sense that their predictions for low energy structures are plausible.
However, their quantitative accuracy is typically quite poor compared to DFT, especially in reproducing macroscopic properties. Machine-learned IPs tend to be much more accurate, but the  are typically three to four orders of magnitude slower than conventional IPs. Importantly, the range of their applicability may be quite restricted. In typical parlance, their \emph{transferability} can be limited. This transferability problem requires researchers to take care in constructing, applying, and validating IPs, and in particular makes it a rather tenuous proposition to use them to discover and predict new structures and novel properties.

In 2010, the Gaussian Approximation Potential (GAP)~\cite{bartok2010gaussian} was introduced as an approach to create IPs with \emph{ab initio} accuracy, using kernel regression and invariant many-body representations of the atomic neighbourhood. Since their introduction, they have been effective at modeling potential energy surfaces \cite{Szlachta:2014jh,doi:10.1021/acs.jpcb.8b06476,doi:10.1021/acs.jpcb.7b09636} and reactivity \cite{doi:10.1021/acs.chemmater.8b03353} of molecules \cite{doi:10.1021/acs.chemrev.5b00644} and solids \cite{C6CP00415F,doi:10.1080/08927022.2018.1447107}, defects \cite{PhysRevMaterials.2.013808}, dislocations~\cite{Maresca:2018tu}, and grain boundary systems \cite{Rosenbrock:2017vda}. Recently Bart\'ok et al. showed that a GAP model using a Smooth Overlap of Atomic Positions (SOAP) kernel \cite{PhysRevB.87.184115} can be systematically improved to reproduce even complex quantum mechanical effects \cite{Bartoke1701816}. SOAP-GAP has thus become a standard by which to judge the effectiveness of numerical approximations to \emph{ab initio} data. There are a number of other machine-learned potentials that also perform well and have overlapping applications with SOAP-GAP, though applications of several of these methods to alloys are still nascent~\cite{pilania2013accelerating,handley2014next,lorenz2004representing,ishida1999local,mills2012polarisable,crespos2003multi,brown2003classical,hansen2013assessment,rupp2012fast,von2004optimization,montavon2012learning,gubaev2019accelerating,faber2015crystal}. Although the GAP framework can be used with arbitrary kernels, for simplicity we will use the GAP abbreviation to mean SOAP-GAP exclusively in the rest of this paper.

The Moment Tensor Potential (MTP) \cite{shapeev2016moment} is an another approach to learning quantum-mechanical potential energy surfaces. Due to the efficiency of its polynomial basis of interatomic distances and angles, MTP is significantly faster than GAP and has already been shown to be capable of reaching equivalent accuracy for modeling chemical reactions \cite{novikov2018-MTP-RPMD}, single-element systems \cite{podryabinkin2019-MTP,novoselov2019-MTP}, single-phase binary systems \cite{novikov2019-MTP-SiO2}, or ground states of multicomponent systems \cite{gubaev2019accelerating}.
In this work, we demonstrate that both GAP and MTP are capable of fitting the potential energy function of a binary metallic system, the Ag-Pd alloy system, with DFT accuracy across the full space of configuration and composition for solid and liquid systems. In addition to reproducing energies, forces and stress tensor components with near-DFT accuracy, we show that these potentials can also approximate phononic band structure quite well and can be used to model compositional phase diagrams.
These new capabilities of quantum-accurate IPs for alloys would pave the way to accelerated materials discovery and optimization.

The Ag-Pd system provides a stringent test for a machine-learned interatomic potential that shows whether it can compete with the cluster expansion method despite the much simpler ``lattice gas'' formalism of the latter. The chemical similarity of silver and palladium and their similar atomic sizes (leading to small atomic mismatch)\cite{nguyen2017} make it an ideal system for cluster expansion and a challenging test for competing methods.

Phonon band structures directly describe phase stability at moderate temperatures via the quasi-harmonic approximation. We first show that both SOAP-GAP and MTP potentials can accurately reproduce DFT-calculated phonon band structures for alloy configurations that are not in the training set.
As a demonstration of speed and transferability, we use the MTP potential to calculate melting lines and transition temperatures for the Ag-Pd phase diagram using the nested sampling (NS) method~\cite{our_NS_paper,PhysRevB.93.174108,pymatnest_paper}. We then compare the performance of GAP and MTP across a low energy transition pathway between two stable configurations to demonstrate the importance of regularization and active learning.

\section{Datasets and Fitting}

\subsection{Datasets}
\label{sec:Datasets}

In this section, we describe the datasets used to fit and validate the potentials. Both the MTP and GAP potentials were fitted to the \emph{same} active-learned dataset, while a liquid dataset provided validation for energies, forces and virials. Although only the active-learned dataset was used for building the models, there was some overlap between the \emph{seed} configurations in the active-learned dataset and the configurations for which phonons are predicted (discussed later), both having their origin in enumerated supercells.

\subsubsection{Active-Learned Dataset}
\label{sec:ActiveDB}

We use the MTP potential and its associated tools to create a database via active learning~\cite{podryabinkin2017active}.  We start with a catalog of small fcc- and bcc-based derivative superstructures. The energies, forces, and virials of these structures are computed by DFT and are then used to fit an MTP potential. This potential is then used to perform structural relaxation for all structures in the database. If, during the relaxation of a particular structure, the estimated extrapolation error of the potential is too large, that (partially relaxed) structure is computed with DFT and added to the training set. When the potential can reliably relax all structures in the enumerated database, the database is expanded to include larger unit cells, and the process is repeated.

For this work,
an initial catalog of 58 enumerated structures \cite{PhysRevB.77.224115} with bcc and fcc derivative superstructures containing 4 atoms or less were calculated.  We iterated the active learning process until the MTP was able to successfully relax all enumerated structures with cell sizes up to 12 (a total of 10,850 structures). This final active learning dataset has 774 configurations.

All the DFT data for these potentials was calculated with VASP~\cite{kresse1993ab,kresse1996efficiency,kresse1994ab,kresse1999ultrasoft,blochl1994projector} using the PBE functional \cite{PhysRevLett.77.3865}. The $k$-points were selected using either Monkhorst-Pack \cite{PhysRevB.13.5188} or WMM \cite{PhysRevB.93.155109} schemes as described below. {\tt PREC=Accurate} and {\tt EDIFF=1e-4} were used for all calculations unless otherwise specified.

During active learning, we used a $k$-point density setting of {\tt MINDISTANCE=55} and an energy convergence target of {\tt EDIFF=1e-4} for the self-consistent loop. However, for the final fit, we found it necessary to recompute the DFT for this dataset with higher $k$-point density and a tighter {\tt EDIFF} setting in order to get good convergence of phonon dispersions. In our experience, a linear $k$-point density of 0.015 $k$-points per \AA$^{-1}$ is a reliable density for alloy fits.

The final dataset for training the GAP and MTP potentials used the original 774 configurations discovered through active learning but computed with \verb|MINDISTANCE=65| in Mueller's scheme and \verb|EDIFF=1e-8|.

\subsubsection{Liquid Dataset}
\label{sec:LiquidDB}
We built a dataset of liquid-like configurations by performing MD simulations using VASP at a high temperature. These calculations were performed at compositions of 25, 50 and 75 at-\% Ag in cells with 32 atoms. The temperature for each simulation was set around the theoretical melting point (linearly interpolated from atomic melting points). Thus, 2766 K, 3063 K, and 3360 K were set as target temperatures for the MD runs and the thermostat parameters were \verb|SMASS=3| and \verb|POTIM=1.0|. The simulation ran for 100,000 fs with snapshots taken every 50 fs. \verb|NELMIN=4| ensured sufficient electronic steps were taken at each MD step. For this MD data, only the $k$-point at $\Gamma$ was used. After the MD runs, each independent snapshot was evaluated again with VASP, but using a $4\times 4 \times 4$ MP $k$-point grid.

\subsection{Potential Fitting}
\label{sec:Fitting}

The  GAP model was fitted to the active-learned dataset using the {\tt QUIP}~\footnote{https://github.com/libAtoms/QUIP} package, using a sum of a 2-body term with Gaussian kernels of pairwise distances and a many-body term with a SOAP kernel, a combination that has produced successful fits of materials in the past~\cite{GAP_aC,GAP_Si,GAP_B,GAP_P}. Parameters for the two-body and many-body parts of the GAP model are summarized in Table \ref{tab:GAP} and are broadly in line with what were used in the previous works. The $\sigma$ values control regularisation in the GAP model, and can be broadly thought of as target accuracies; they were set to $10^{-3}$~eV for energies (per atom), $10^{-3}$~eV/\AA\ for forces and $0.02$~eV for virial stresses (per atom). Their relative magnitudes also control the tradeoff between the fit accuracies in energies, forces and virials.

\begin{table*}
\begin{subtable}[H]{3.8in}
GAP: 2-body Term
\begin{tabular}{|l|l|l|}
\hline
Parameter                                       & Value & comment\\ \hline
kernel                                          & Gaussian & functional form of kernel\\
$\theta$                                        & 1.0 & kernel length scale\\
$r_{\textrm{cut}}$                              & 6.0 \AA& cutoff distance \\
cutoff\_transition\_width                       & 1.0 \AA& cutoff smoothing length scale \\
$\delta$                                        & 2.0 eV & typical contribution to energy \\
$n_{\textrm{sparse}}$                           & 25 & number of basis functions \\
\hline
\end{tabular}
\caption{Parameters for the 2-body GAP term.}
\label{tab:GAP}
\end{subtable}\hfill
\vspace{0.2in}
\begin{subtable}[H]{2.in}
MTP Model
\begin{tabular}{|l|c|}
\hline
Parameter                                       & \multicolumn{1}{c|}{Value} \\ \hline
radial functions   & 4                        \\
radial basis size  & 6                        \\
fitting parameters & 188                        \\
cutoff             & 5.0 \AA                       \\
stress weight      & $5\cdot10^{-4}$                         \\
force weight       & $5\cdot10^{-3}$ \AA$^{2}$                        \\
BFGS iterations    & 500                         \\\hline
\end{tabular}
\caption{Parameters for the MTP model. }
\label{tab:MTP_params}
\end{subtable}\hfill
\vspace{0.01in}
\begin{subtable}[H]{0.45\textwidth}
GAP: SOAP (many-body) term

\begin{tabular}{|l|l|l|}
\hline
Parameter                               & Value & comment\\ \hline
$r_{\textrm{cut}}$                      & 4.5 \AA & cutoff distance                      \\
cutoff\_transition\_width               & 0.5 \AA & cutoff smoothing length scale                        \\
$\delta$                                & 0.2 eV &  typical contribution to energy                       \\
$n_{\textrm{sparse}}$                   & 500   & number of basis functions                      \\
$n_{\textrm{max}}$                      & 8     & radial basis truncation                    \\
$l_{\textrm{max}}$                      & 8     & angular basis truncation                    \\
$\zeta$                           & 2      & power SOAP kernel is raised to                   \\
$\sigma_{\textrm{atom}}$             & 0.5 \AA & smoothing of atoms in neighbour density                         \\
 \hline
\end{tabular}
\caption{Parameters for the GAP SOAP (many-body) term.}
\label{tab:GAP-soap}
\end{subtable}
\hspace{3.7in}
\caption{Fitting parameters for GAP (Tables \subref{tab:GAP} and \subref{tab:GAP-soap}) and MTP models (Table \subref{tab:MTP_params}).}
\end{table*}

The MTP model with polynomial degree up to 16 \cite{gubaev2019accelerating} with 188 adjustable fitting parameters was trained on the same dataset as GAP. Table \ref{tab:MTP_params} summarizes the parameters needed to recreate the MTP model.
The fitting weights (roughly corresponding to $\sigma$ parameters of GAP) were $10^{-3}$~eV, $1.4\times 10^{-2}$ eV/\AA\ and $0.04$~eV for energy, force, and stress, respectively.
This is somewhat different from the parameters used for GAP; however, as we verified, this does not significantly affect the results.

\section{Nested Sampling}
\label{sec:nested_sampling}

\subsection{Dataset Augmentation}
\label{sec:NSDataset}

As described below, nested sampling simulates atoms at extremely high temperatures that are well outside of the typical active-learned dataset described above. The MTP potential used for nested sampling had to be trained using a slightly augmented dataset, to avoid the formation of dimers in the gaseous phase.

As the first step to constructing the augmented dataset, we identified 67 structures that are within 5 meV/atom from the convex hull of stable Ag-Pd structures. These structures were periodically repeated to form supercells with 32 -- 64 atoms. These structures were used as initial configurations for molecular dynamics, running for 0.1 ns, while the MTP potential was trained on-the-fly \cite{podryabinkin2017active,gubaev2019accelerating} at the range of temperatures from nearly zero to temperatures ensuring melting.

\subsection{Methodology}

The constant pressure nested sampling (NS) method~\cite{PhysRevB.93.174108,pymatnest_paper} was used to calculate phase diagrams by sampling the entire potential energy surface with corresponding configuration space volumes to calculate the canonical partition function.  The specific heat, which is the second derivative with respect to temperature of the partition function, shows peaks at phase transitions, and we use temperatures of specific heat maxima as estimates of the corresponding transition temperatures. While the nested sampling method has previously been applied to multicomponent systems, those simulations assumed constant composition.  However, it is possible for phase separation to occur in temperature-composition space, which would be neglected by this constraint.  Here we have extended the constant pressure NS method to a semi-grand-canonical (sGC) version~\cite{Kofke1987}, where the total number of atoms is constant, but the numbers of the individual species is allowed to vary. This is implemented by carrying out the nested sampling procedure on a free energy $F$ defined as
\begin{equation}
    F = E + \sum_i N_i \mu_i,
\end{equation}
where $N_i$ and $\mu_i$ are the number of atoms and chemical potential of species $i$, respectively, and the sum is carried out over all species.  To explore these degrees of freedom we
also added Monte-Carlo steps that propose the changing of the species of a randomly selected atom. Note that since the procedure is invariant to shifts in the total energy, the total number of particles is conserved, and only two species are present, the simulation is entirely characterized by the difference in chemical potentials $\Delta \mu$.

To calculate the phase diagram in temperature-composition space, as it is usually plotted, we carry out sGC NS runs at a range of values of $\Delta \mu$.  In the sGC framework the composition is an \emph{output} of the simulation, and its value can be calculated as a function of temperature using the same ensemble average (with NS phase space volumes and Boltzmann weights) as any other quantity in the NS approach. For phase transitions that cross phase separated regions we would in principle expect discontinuous changes in composition (analogous to discontinuous changes in structure, internal energy, etc.) across the transition, but these will be broadened by finite size effects. We also compare these results to constant composition NS runs, where a single transition temperature is identified with the peak of the $C_p(T)$ curve. For one composition, 50\%, we continue one of those NS runs to sufficiently low temperature to search for solid state phase transitions.  The parameters used for both types of NS runs are listed in Table~\ref{tab:NS_params} in the notation of Ref.~\cite{pymatnest_paper}, and in all NS simulations one configuration per NS iteration was removed ($K_r = 1$).

\begin{table}
\begin{tabular}{|l|cc|} \hline
          & constant & \\
Parameter & compos.  & sGC \\ \hline
total number of particles $N$ & 96 & 64 \\
number of configurations $K$ & 1080 & 1152 \\
number of evaluations per walk $L$ & 640 & 1166 \\
positions steps (number $\times$ length) & $ 3 \times 8$ & $1 \times 8$ \\
cell steps (volume:shear:stretch) & 3:3:3 & 16:8:8 \\
swap steps & 6 & 8 \\
composition steps & 0 & 8 \\ \hline
\end{tabular}

    \caption{Nested sampling parameters in the notation of Ref.~\cite{pymatnest_paper} for constant composition and semi-grand canonical (sGC) NS runs. Swap steps exchange the species of a pair of atoms, while composition steps change the species of a single atom (sGC only).}
    \label{tab:NS_params}
\end{table}

\section{Results and Discussion}
\label{sec:Results}
\subsection{Energy, Force and Virial Predictions}
\label{sec:EFS-Results}
We now compare the performance of GAP and MTP models for the Ag-Pd system. Both the MTP and GAP models were validated against the liquid dataset for energy, force and virial predictions. Table \ref{tab:EFS_predictions} summarizes the Root Mean Square Error in each of these properties for both GAP and MTP. \emph{No liquid data was included in the original active-learned dataset}. Thus, these predictions represent severe extrapolation. The fact that both machine-learned IPs perform so well in this dataset is strong evidence of their significant transferability. The relatively simple approach to building the training set (iterative fitting and relaxing of enumerated superstructures) resulted in IPs that would be reliable in most solid or liquid simulations.

\begin{figure*}
\centering
\begin{subfigure}[b]{2.3in}
     \label{subfig:potatoE}
     \includegraphics[width=2.3in]{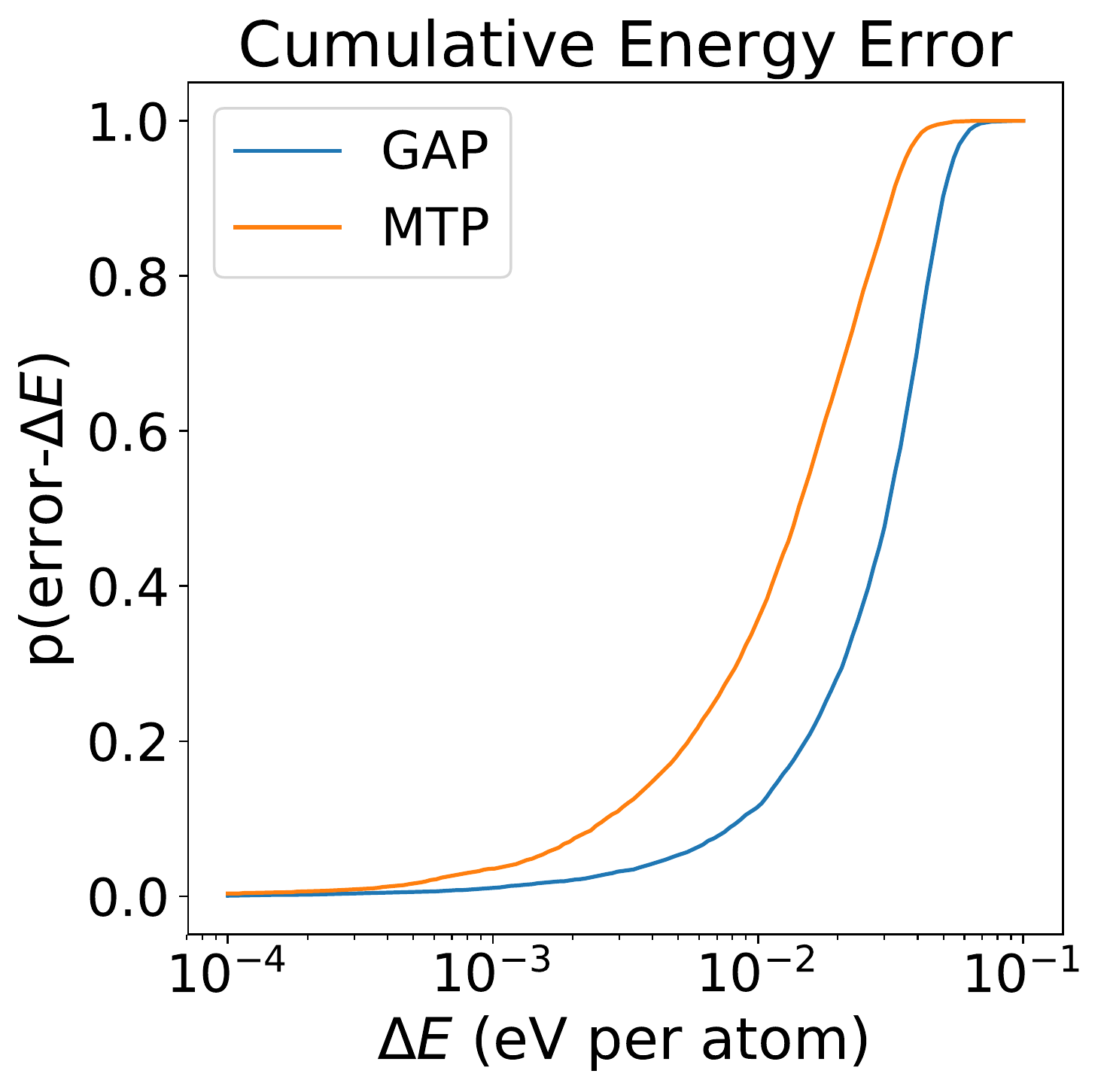}
\end{subfigure}
\begin{subfigure}[b]{2.3in}
     \label{subfig:potatoF}
     \includegraphics[width=2.3in]{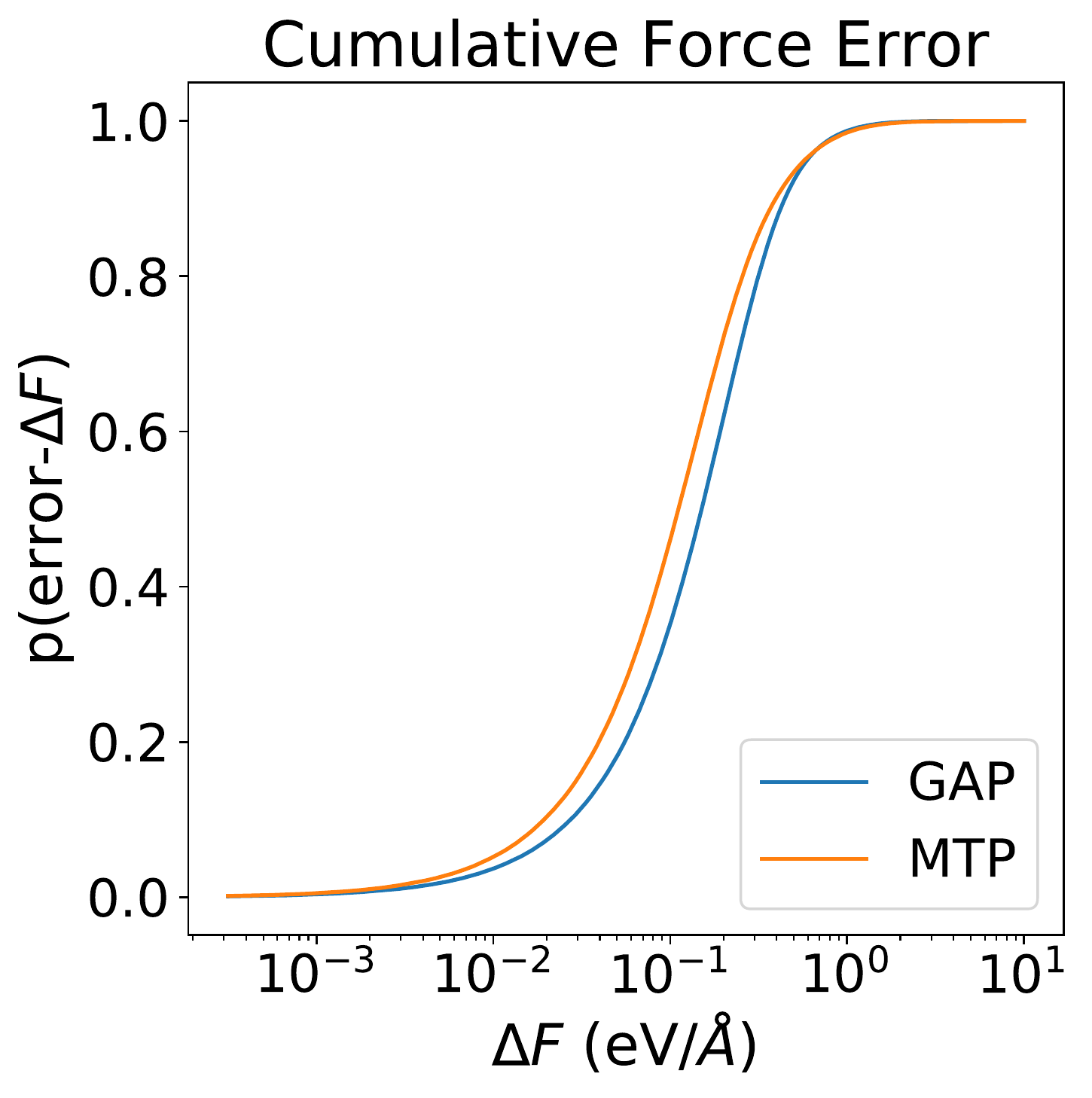}
\end{subfigure}
\begin{subfigure}[b]{2.3in}
     \label{subfig:potatoV}
     \includegraphics[width=2.3in]{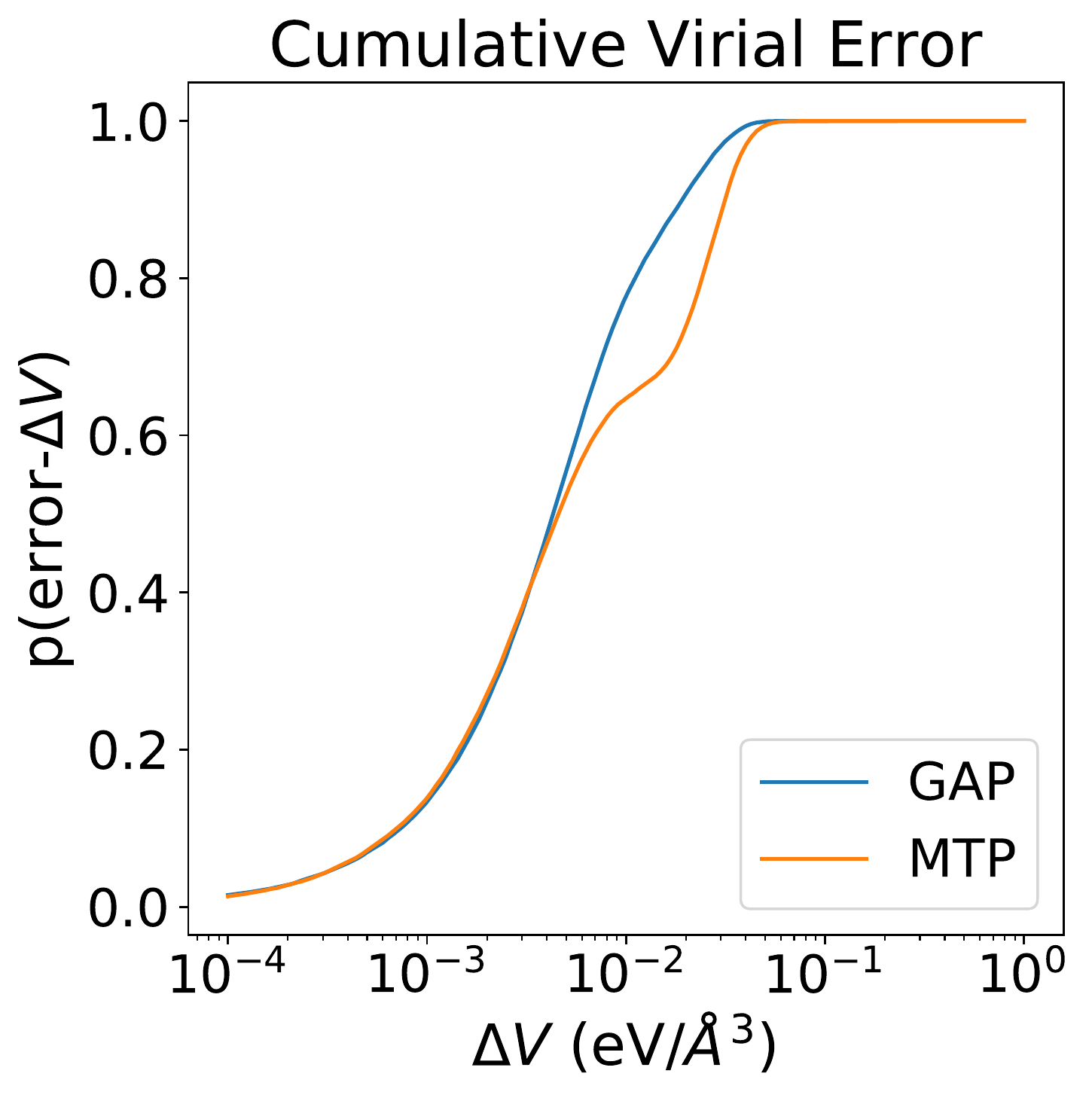}
\end{subfigure}
     \caption{Cumulative probability of error plots for the liquid dataset consisting of $\sim$$6\,000$ configurations from a sub-sampled MD run. From left to right, energy, force and virial error probabilities are plotted, where error is calculated relative to DFT. Mean errors are given in Table \ref{tab:EFS_predictions}. }
     \label{fig:potato_plots}
\end{figure*}

\begin{table}[htbp]
\vspace{0.2in}
\begin{subtable}{0.45\textwidth}
\begin{tabular}{|l|c|c|}
\hline
RMS Error & GAP & MTP \\ \hline
Energy (meV)                                         & 15.4 & 10.9                        \\
Force (meV/\AA)                               & 224 & 241   \\
Virial (meV/\AA$^3$)                           & 8.3 & 12.7                        \\\hline
\end{tabular}
\caption{Root Mean Square Error of energy, force and virial predictions for GAP and MTP interatomic potentials validated against the Ag-Pd dataset of $\sim$ $6\,000$ liquid configurations subsampled from \emph{ab initio} molecular dynamics using VASP. The models were both trained using the same active learning dataset.
}
\label{tab:EFS_predictions}
\end{subtable}
\\
\vspace{0.2in}
\begin{subtable}{0.45\textwidth}
\begin{tabular}{|l|c|c|}
\hline
    & Training Error (THz) & Prediction Error (THz) \\ \hline
GAP & $0.13 \pm 0.05$ & $ 0.13 \pm 0.04$   \\ \hline
MTP & $0.12 \pm 0.03$ & $0.11 \pm 0.01$   \\ \hline
\end{tabular}
\caption{Mean RMSE for phonon predictions when integrating errors over the entire Brillouin Zone. Eigenvalues were sampled at $13 \times 13 \times 13$ $k$-points for 65 structures. Bands for each of the 65 structures are plotted in the Supplementary Information. Both GAP and MTP have good agreement with DFT across the full phonon spectrum.}
\label{tab:PhononStats}
\end{subtable}

\caption{Model validation for GAP and MTP models using the Liquid dataset, and RMSEs for phonon dispersion curve predictions.}
\end{table}

In Figure \ref{fig:potato_plots}, a cumulative probability distribution of errors for energy, force and virial predictions are plotted for GAP and MTP, where errors are calculated relative to DFT. For energy, MTP has lower cumulative probability of error overall. This difference is less pronounced for the force errors where MTP is only marginally better.
Interestingly, the probability of error for virial predictions in MTP deviates significantly from GAP at larger errors. These error statistics are consistent with the ratio of energy, force and virial weights ($\sigma$ parameters) for the two models: on the one hand MTP has more than 10x higher weights (lower $\sigma$ values) for energies relative to forces, whereas GAP uses the same; meanwhile GAP uses twice the weights (half $\sigma$) for the virial stress.

\subsection{Phonon Predictions}
\label{sec:Phonon-Results}
The phonon eigenvalues computed from a force-constants matrix for a crystal structure describe the energy required to excite a specific vibrational mode within the crystal. For a potential to closely reproduce a phonon spectrum, it must accurately approximate the \emph{curvature} of the potential energy surface for small deformations from its relaxed form. Thus, while energy and force validation provides useful insights into the accuracy of a potential for specific \emph{points} and their \emph{slopes} on the potential energy surface, validating against phonons gives insight into the second derivatives of the energy surface. Historically, potentials have successfully reproduced phonon spectra for certain compositions and configurations, but not generally for an entire alloy system. To demonstrate the ability of our potentials to produce accurate phonon band structures, we compare DFT- and IP-calculated bandstructures for fcc-type derivative superstructures~\cite{PhysRevB.77.224115} of cell sizes from 2 through 6; there are 65 of these cells.

\subsubsection{DFT Phonon Dispersion Curves}
First, with DFT, we relaxed each configuration twice using \verb|IBRION=2| and \verb|ISIF=3|, which allows both cell shape and volume to change during relaxation.

We then used \verb|phonopy| to generate frozen phonon displacements. For selecting the supercell, we enumerated the list of all possible Hermite Normal Form matrices (HNF)\footnote{See the discussion in the appendix of Ref.~\cite{hart2018robust} for the utility of using HNF matrices in this context.} for each structure and selected the HNF in each case that maximized the distance between periodic images with a supercell size of no more than 32 atoms. When two HNFs were equivalent for both size and distance metric, we selected the one with the larger point-group. This procedure allowed us to choose the smallest possible supercell with highest symmetry subject to the constraint of maximal distance between periodic images.

Each of the displaced structures from \verb|phonopy| were computed using \verb|EDIFF=1e-8|, \verb|ADDGRID=TRUE|, \verb|ENCUT=400| and \verb|MINDISTANCE=55| in Mueller's $k$-point scheme.
\vfill
\subsubsection{Machine-learn Phonon Dispersion Curves}

 Using both GAP and MTP, we demonstrate here that a single, machine-learned potential can simultaneously approximate phonons across the full compositional space for many configurations, and with good accuracy. In the Supplementary Information, we include additional phonon plots that cover a broad structure-composition range. In this section, we have chosen two that are interesting for discussion purposes.

\begin{figure}
    \centering
    \includegraphics[width=0.55\textwidth]{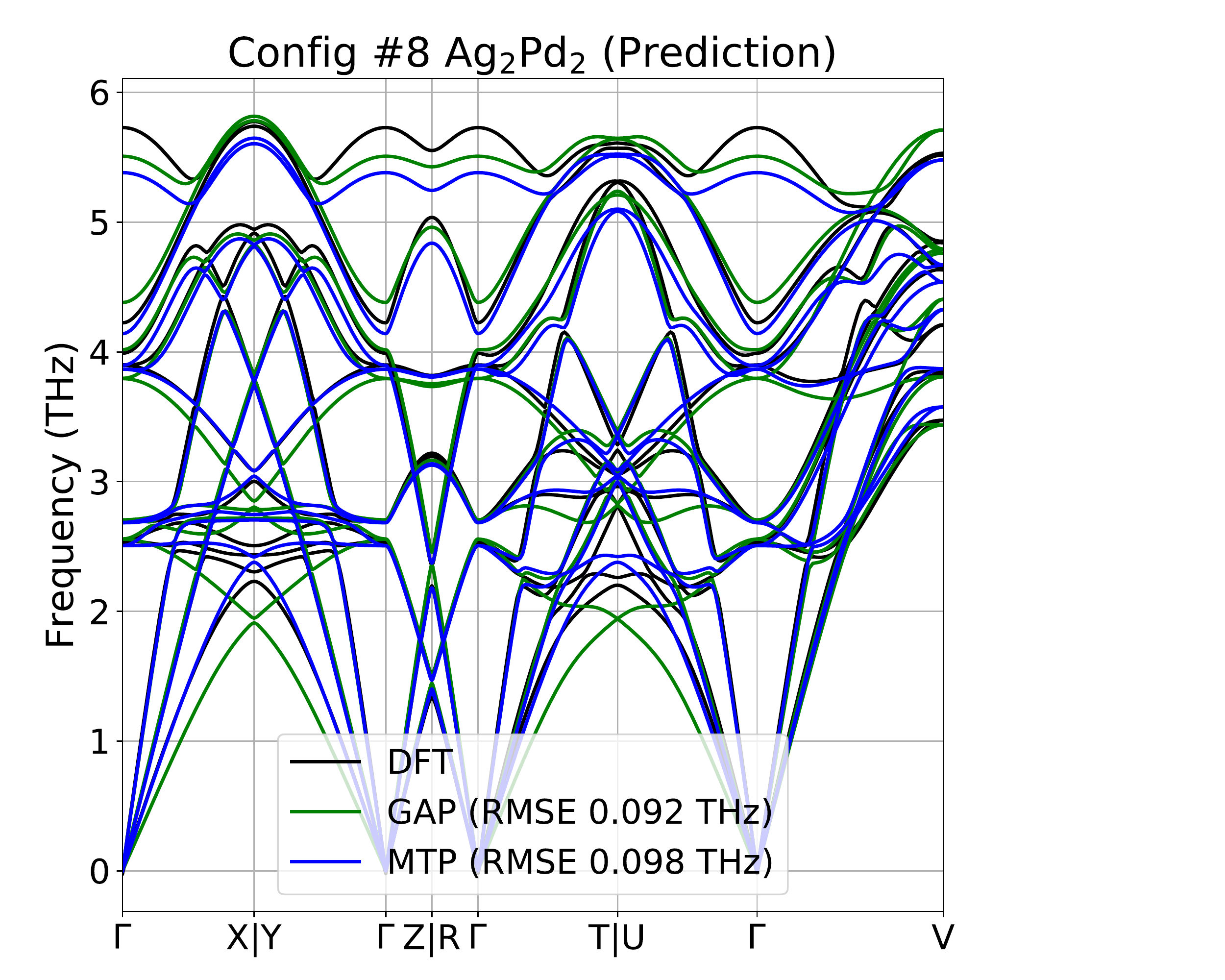}
    \caption{Prediction of the phonon dispersion curves for a 4-atom structure with 50 at-\% Ag. Both the RMS errors in parenthesis represent the integrated error across all eigenvalues sampled on a $13 \times 13 \times 13$ grid in the Brillouin Zone. GAP and MTP both provide good approximations to the curves along the special path, though each makes different errors in over- and underestimating the eigenvalues at different $k$-points. Note that this configuration was not included in the fitting dataset.}
    \label{fig:Phonon8}
\end{figure}

\begin{figure}
    \centering
    \includegraphics[width=0.55\textwidth]{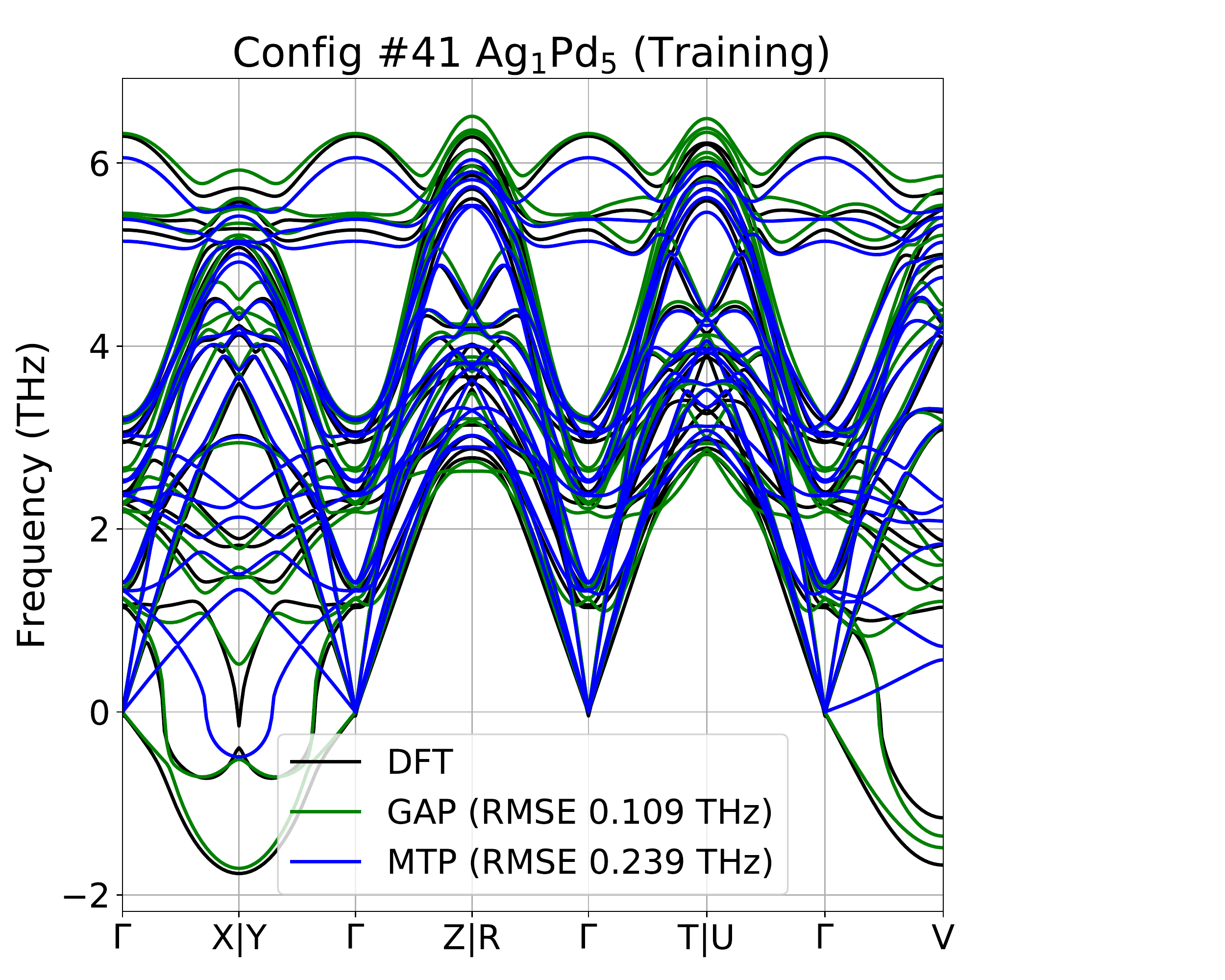}
        \caption{Phonon dispersion curves for a 6-atom structure with ~16 at-\% Ag. Both the RMS errors in parenthesis represents the integrated error across all eigenvalues sampled on a $13 \times 13 \times 13$ grid in the Brillouin Zone. According to DFT, this structure is not dynamically stable. Both GAP and MTP recover this dynamic instability, although GAP is more accurate. Note that this seed configuration was included in the active-learned dataset that the potentials were fitted to.}
    \label{fig:Phonon41}
\end{figure}

In Figure \ref{fig:Phonon8}, we plot a typical phonon spectrum for a 50 at-\% Ag configuration with 4 atoms. For this structure, both GAP and MTP approximate the eigenvalues along the special path well. The RMSE, reported in parentheses, is the integrated error across all eigenvalues in the Brillouin Zone (BZ), sampled on a $13 \times 13 \times 13$ grid. Figure \ref{fig:Phonon41} shows a structure with a dynamic instability (i.e., negative phonons), as reported by DFT. For this structure, both GAP and MTP learn and reproduce this instability, albeit with different accuracies. Both figures demonstrate ability of the IPs predict dynamic changes due to small perturbations. In the Supplementary Information we similarly plot 65 phonon spectra.

Table \ref{tab:PhononStats} provides statistics for the integrated error across all 65 structures for GAP and MTP. Both GAP and MTP predictions for integrated error are close to 0.1 THz across the full validation set. Similar training and prediction errors indicate a good bias/variance trade-off (not over-fitting). Importantly, these aggregated results show that across a broad range of alloy compositions and structures for Ag-Pd, machine-learned IPs are in good agreement with DFT in the harmonic approximation for vibrational modes.

\subsection{Transition Pathway}
\label{sec:Pathway-Results}
\begin{figure}
    \centering
    \includegraphics[width=0.45\textwidth]{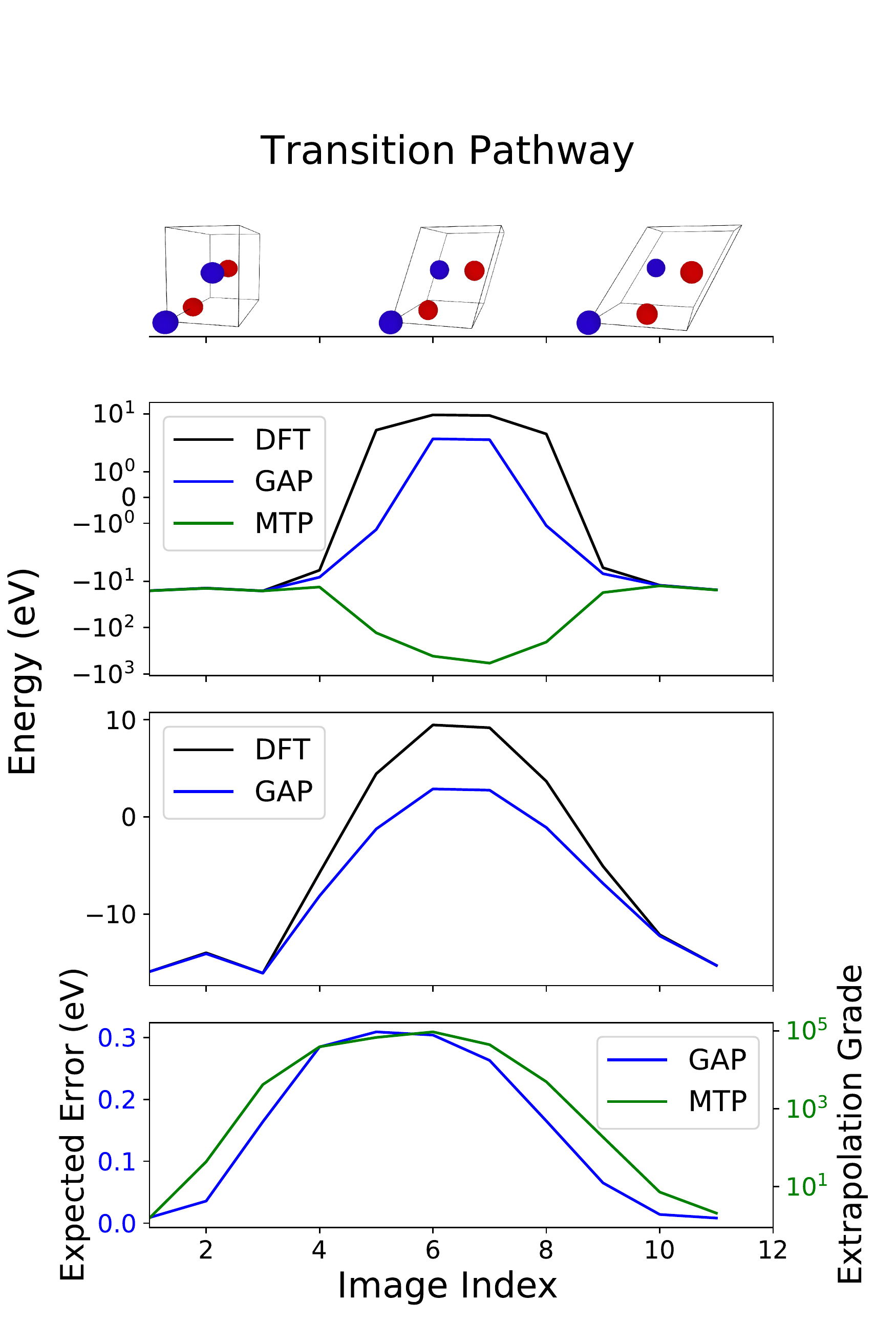}
    \caption{GAP and MTP predictions for the transition pathway. Since there were no configurations in the training set along the middle of the path, the MTP prediction is purely extrapolative and presents a false, deep, local minimum. Because of its tight regularization, the GAP prediction for this transition is reasonable. It underestimates the barrier significantly due to the lack of training data, but doesn't introduce a false minimum. Note the symmetric log scale (linear between $-1$ and $+1$) on the y-axis in the upper plot. Above the plots are configurations along the pathway. The upper Pd atom (blue) at index 1 moves further into the page as the unit cell distorts. The furthest Ag atom (red) simultaneously moves out of the page. The lowest panel plots the expected variance of the GAP model and the extrapolation grade of the MTP model. Predictive variance is essentially the expected error per atom (in eV) along the transition pathway and the extrapolation grade is a dimensionless quantity indicating that if it exceeds $\sim10$ then a configuration must be added to the training set because the prediction of its energy is not reliable.
    }
    \label{fig:pathway-combined}
\end{figure}

As discussed in the introduction, transferability is the price we pay for approximating quantum mechanics with high accuracy. In general, a machine-learned IP is only valid within the subspace in which it was trained. Although it is possible to apply an IP outside of that space, the results will not be trustworthy. We demonstrate this by computing the energy along a smooth transition between two structures. Figure \ref{fig:pathway-combined} shows two AgPd structures that are connected by a smooth transition (essentially these two structures are identical except that the upper two atoms switch places). Although the cell must enlarge slightly and distort, the atoms have a clear path to transition from the starting configuration (Index 1) to the final configuration (Index 11) without colliding. The total energy along the path is also shown in Figure \ref{fig:pathway-combined}. Note that in the figure, the y-axis is the total energy, not the energy difference between distorted and undistorted cases. Also note that y-axis scale is linear between $-1$ and $+1$ eV and logarithmic elsewhere in the upper plot. In the starting configuration, the total energy is approximately $-16$ eV. At its highest point on the transition path, the energy is about 9.5 eV, a total difference of about 25 eV. Such a high energy structure is not problematic for DFT, but it's a big ask to expect a machine-learned IP to accurately extrapolate to this kind of a structure if similar structures were not included in the training dataset. Nevertheless, the GAP does quite well. Although the absolute error of its prediction for the top of the barrier is several eVs, the qualitative behavior is correct.

Due to its more local basis functions and built-in regularization, the GAP model provides reasonable physical behavior for the transition between the starting and final configurations. MTP, with its global polynomial basis functions, relies on active learning to ensure that its predictions fall within the interpolation regime. As part of its framework, MTP (like GAP) provides the extrapolation grade \cite{podryabinkin2017active,gubaev2019accelerating} to distinguish between configurations that can be evaluated reliably and configurations that should be added to the training set to avoid large extrapolation errors. In our test MTP correctly detects extrapolation, but has a much poorer extrapolation behaviour compared to GAP, and this is the price one pays for using unregularised global basis functions. The active-learning approach is general and could be applied to GAP too (using the predicted variance of the underlying Gaussian process as a proxy for extrapolation~\cite{GAP_Si}), which would be expected to make its predictions better too. This demonstration
should be seen as a warning in the application of machine-learned IPs generally. Using such models safely requires understanding the properties of the basis functions, how the training set was built, which parts of the configuration space were included, etc. In the case of MTP, the extrapolation grade should be used to checked against representative samples from the configurations that are expected to be encountered  before embarking on large scale molecular dynamics simulations.
\section{Phase Diagram Results}
\label{sec:NS-Results}

The success of the models in learning basic properties and phonons motivates examining the temperature-concentration  phase diagram for the alloy system. We used nested sampling with the MTP model to find the liquidus-solidus line, calculate order-disorder transition temperatures, etc. Although GAP can also be used, the MTP model is significantly faster and makes the exploration of the phase space more practical. For example, investigation of a single temperature slice of the phase space (for fixed composition and pressure) requires more than 2 billion evaluations of the potential. This cost is presently prohibitive for GAP but reasonable with MTP.

\subsection{Liquid-Solid Transition}

\begin{figure}[h]
    \centering
    \includegraphics[width=0.5\textwidth]{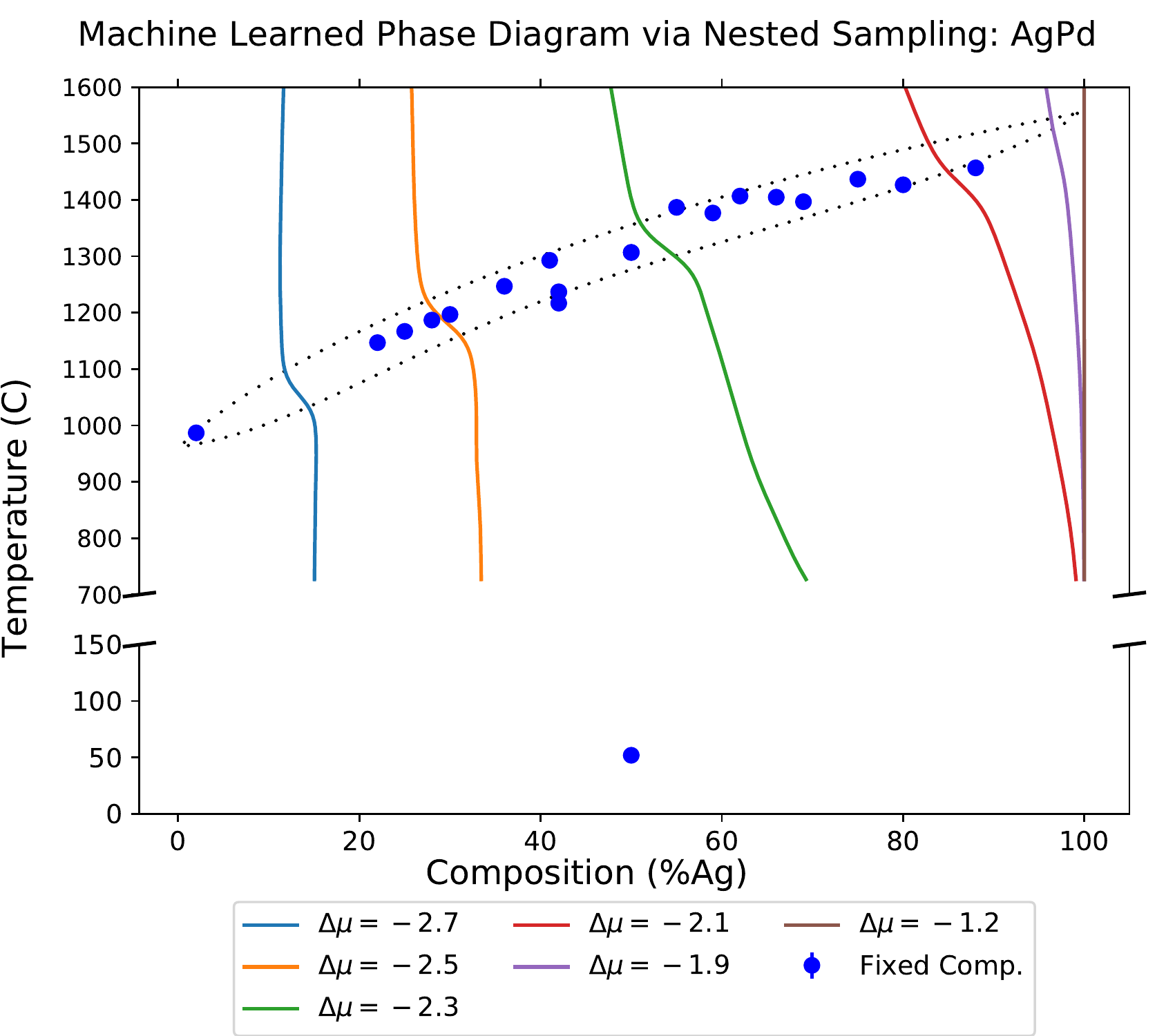}
    \caption{Phase diagram determined by a semi-grand canonical (sGC) and fixed composition ensemble  nested sampling (NS).  Solid lines show sGC NS composition as a function of temperature for each $\Delta \mu = \mu_\mathrm{Ag}-\mu_\mathrm{Pd}$, with approximately horizontal jogs indicating finite-size broadened discontinuities associated with the liquidus-solidus (L-S) gap. Dotted lines indicate experimentally observed L-S gap. Circles indicate constant composition NS specific heat maxima corresponding to liquid-solid and solid-solid phase transition temperatures. We have shifted all simulation results up by 200~K to make it easier to compare the width of the L-S gap with experiment.  The experimental liquidus and solidus  are adapted from several sources, Refs.~\cite{morioka1999thermodynamic,ghosh1999thermodynamic,karakaya1988ag}.}
    \label{fig:L-S-Gap}
\end{figure}

Inasmuch as nested sampling cools down from a high temperature gas phase, we first reproduce the liquid transition behavior as a function of temperature. Each solid line
in Figure~\ref{fig:L-S-Gap} shows a trace of the ensemble averaged composition as a
function of temperature that results from a NS run at fixed $\Delta \mu = \mu_\mathrm{Ag}-\mu_\mathrm{Pd}$.  In the
high temperature liquid and low temperature solid regions the composition varies
smoothly with temperature.  The solid-liquid transition is indicated by a sharper
horizontal (along the composition $x$-axis) jump, which we expect would become
discontinuous in the large system size limit. The width of the approximate
discontinuity indicates the width of the phase-separated region.

As is clear from Figure~\ref{fig:L-S-Gap}, the melting behavior of our MTP potential qualitatively matches the experimental line \cite{savitskii1961kurnakov,okamoto2000phase,naidu1971x,ellner2004partial,dos1999high}. However, although the entire line has roughly the same shape, it is shifted from the experimental results by $\sim$200~K.

The liquidus-solidus gaps are also in reasonable agreement with experimental data when the same shift of 200~K is included (added in Figure \ref{fig:L-S-Gap} to facilitate comparison).

This shift in temperature is expected for DFT with a PBE functional and has been discussed in the context of other \emph{ab initio} studies \cite{2015PhRvB..92b0104H,pozzo2013melting,PhysRevB.75.214103}. Since the shift does not appear to be composition-dependent, the trends are still reliable.

\subsection{Solid-Solid Transition}

\begin{figure*}
    \centering
\begin{subfigure}[b]{2.in}
     \includegraphics[width=2.in]{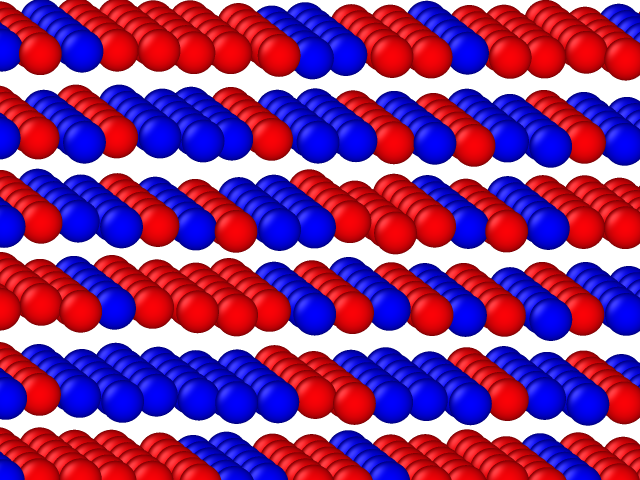}
     \caption{}
     \label{subfig:A150K}
\end{subfigure}
\begin{subfigure}[b]{2.in}
     \includegraphics[width=2.in]{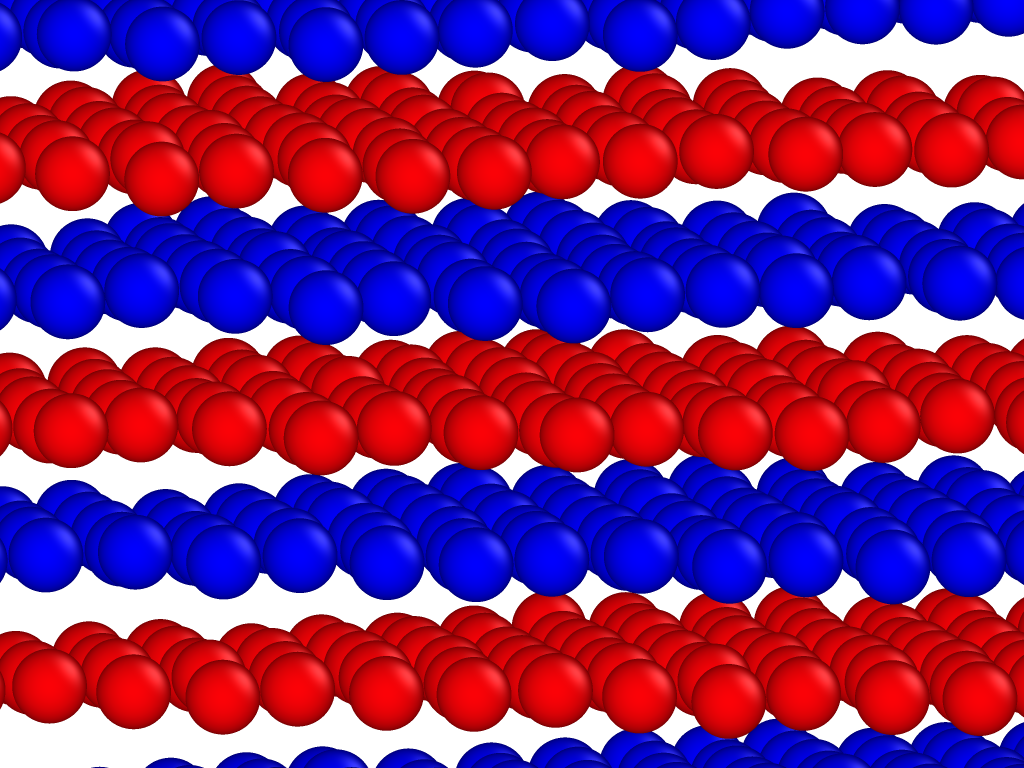}
     \caption{}
     \label{subfig:A100K}
\end{subfigure}
\begin{subfigure}[b]{2.8in}
     \includegraphics[width=2.8in]{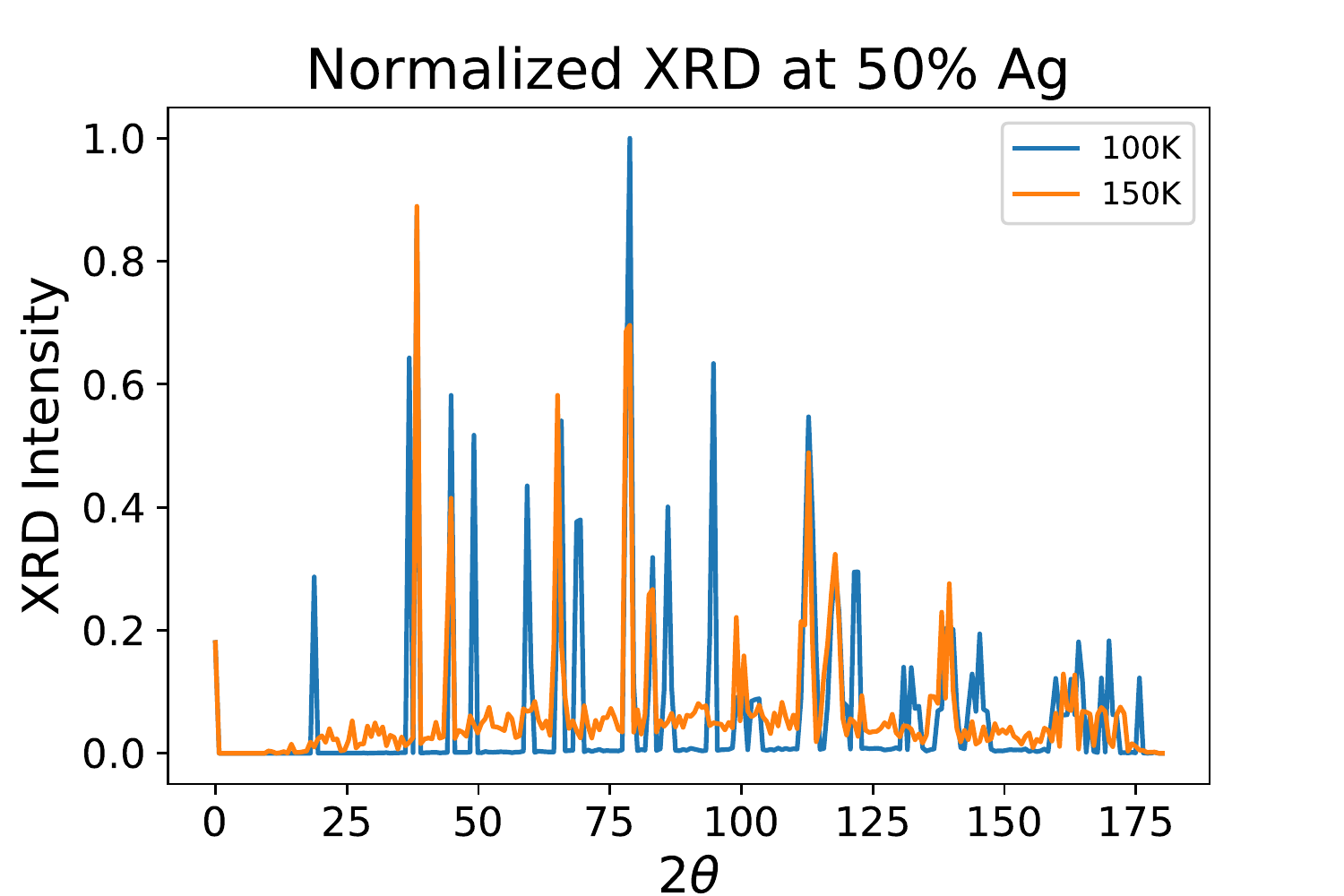}
     \caption{}
     \label{subfig:XRD}
\end{subfigure}
    \caption{Most-probable configurations from the nested sampling at 50\% Ag. Panel \protect\subref{subfig:A150K} is at 150K, while Panel \protect\subref{subfig:A100K} is at 100K. For both Panels \protect\subref{subfig:A150K} and \protect\subref{subfig:A100K}, the crystal is viewed edge-on at \big[111\big] planes.
    The lower temperature structure shows clear ordering as stacking planes. This ordering is confirmed by simulated XRD in Panel \protect\subref{subfig:XRD}. Ag atoms replaced H atoms to artificially boost the difference in structure factors for the simulation.}
    \label{fig:SS-transition}
\end{figure*}

Another stringent test of the potential is whether it can recover the transition from a disordered solid solution to an ordered phase. Experimentally, it appears that ordered phases are kinetically inhibited by low transition temperatures and would be unlikely to appear in experimental constitutional phase diagrams.
All reported phase diagrams~\cite{morioka1999thermodynamic,okamoto2000phase,ghosh1999thermodynamic,karakaya1988ag} typically show just solidus and liquidus lines and indicate a solid solution below the solidus (see Fig.~\ref{fig:L-S-Gap}) though one proposed phase diagram guesses two solid-solid transitions, based on some reports of ostensible ordered phases~\cite{savitskii1961kurnakov}, reliable evidence for first-order transformations in the solid state is lacking~\cite{darling1973some}\footnote{A careful review of relevant experimental literature\cite{allison1972structure,brouers1975temperature,feng2014thermodynamic,garino2000behavior,luef2005enthalpies,sopouvsek2010experimental,rao1968x} since 1961 (after Ref.~\cite{savitskii1961kurnakov}) suggests that, when Ag-Pd samples are annealed in air or otherwise exposed to oxygen, the formation of oxide phases can be misinterpreted as the effects of ordering. The experimental literature does not agree on the stoichiometry of these oxide phases, such phases have not been reported in samples not exposed to oxygen, and no structural information for these phases (from XRD studies, for example) have been reported.}. It seems reasonable that there is no formation of  intermetallic phases in the temperature ranges reported in the phase diagrams.

Since the melting transition was underestimated, we expected that any disorder-order transition would also take place at a reduced temperature, hence for the 1:1 composition system we continued the sampling well below the melting temperature. As shown in Figure \ref{fig:L-S-Gap}, which includes a region of the phase diagram outside of the experimental data, the order-disorder transition does exist at 125 K for this system. Other computational works find the transition temperatures to be similar to what we report here~\cite{ruban2007theoretical,muller2001first,gonis1983first,takizawa1989electronic}.

\subsection{Comparison to Cluster Expansion}

As a reference, we include a comparison of the GAP and MTP models to the a state-of-the-art Cluster Expansion (CE) of the same Ag-Pd system \cite{PhysRevB.85.054203}.

\begin{table}[htbp]
\begin{tabular}{|c|c|c|}
\hline
GAP & MTP & CE \\ \hline
5.7 & 4.2 & 2.1                        \\ \hline
\end{tabular}
\caption{Energy errors in meV on a reference dataset used to train a Cluster Expansion (CE) model for Ag-Pd. Details on building the dataset and model are given in \cite{PhysRevB.85.054203}. While the CE model is slightly better, both GAP and MTP perform well in that region of configuration space where the CE model is valid.}
\end{table}

\section{Conclusion}
\label{sec:Conclusion}

Cluster expansion has been a go-to tool for computing energies across configuration space for alloys. Because of its speed and applicability over the full range of compositions, it was useful for performing ground-state searches, and even for temperature-dependent phase mapping in certain systems. However, it cannot address dynamic processes that involve structural perturbations, which limits its usefulness.

This work demonstrates that machine-learned interatomic potentials are as good as cluster expansion for on-lattice computation of energies \cite{PhysRevB.85.054203}. But unlike cluster expansion, the machine-learned interatomic potentials can compute forces, virials and hessians across the compositional space as well. These additional derivatives of the potential energy surface are sufficiently accurate to approximate dynamic properties like phonon dispersion curves, as well as map out the temperature-composition phase diagram for an alloy. Software for creating datasets and fitting potentials is readily available and easy to use. These potentials therefore offer a viable alternative to cluster expansion models, and arguably represent the future direction of first principles computational alloy design.

\section*{Acknowledgements}
CWR and GLWH were supported under ONR (MURI N00014-13-1-0635). LBP acknowledges support from the Royal Society through a Dorothy Hodgkin Research Fellowship. NB acknowledges support from the U.~S. Office of Naval Research through the U.~S. Naval Research Laboratory's core research program, and computer time from the U.~S. DoD's High Performance Computing Modernization Program Office at the Air Force Research Laboratory Supercomputing Resource Center. AVS was supported by the Russian Science Foundation (grant number 18-13-00479).
This collaboration might not have been possible had the authors not met at the Institute of Pure and Applied Mathematics, UCLA. We thank Andrew Huy Nguyen for help with several NS calculations.

\section*{Data Availability}

Data and models are available at \href{https://github.com/msg-byu/agpd}{https://github.com/msg-byu/agpd}.

\section*{Author Contributions}

Datasets were created by CWR, GLWH, and AVS. GC, CWR and GLWH created GAP models and fine-tuned phonon calculations. KG and AVS created MTP models. LBP and CWR performed calculations for fixed-composition Nested Sampling calculations. sGC Nested Sampling calculations were done by NB. All authors contributed to the discussion, analysis and writing of the text.

\section*{Conflict of Interest}

The authors declare no conflicts of interest.


\end{document}


\preprint{APS/123-QED}

\title{Supplementary Information: Machine-Learned Interatomic Potentials for Alloys}

\author{Conrad W. Rosenbrock}
 \email{rosenbrockc@gmail.com}
\author{Gus L. W. Hart}
 \affiliation{Department of Physics and Astronomy, Brigham Young University, Provo UT USA 84602}
 \author{Konstantin Gubaev}%
 \author{Alexander V. Shapeev}%
\affiliation{Skolkovo Institute of Science and Technology, Skolkovo Innovation Center, Nobel str. 3, Moscow,
143026 Russia}
\author{Livia B. P\'artay}%
\affiliation{%
Department of Chemistry, University of Reading, Whiteknights, Reading, RG6 6AD, UK}%
\author{Noam Bernstein}%
\affiliation{Center for Computational Materials Science,  U.~S. Naval Research Laboratory, Washington DC 20375, USA}
 \author{G\'abor Cs\'anyi}%
\affiliation{%
 Department of Engineering, University of Cambridge, Trumpington Street, Cambridge, CB2 1PZ, UK}

\date{\today}

\begin{abstract}
Supplementary phonon dispersion curve plots for the machine-learned potentials discussed in the main paper.
\end{abstract}

\maketitle


\section{Appendix}

This section presents phonon dispersion curves for 65 enumerated FCC superstructures with cell sizes from 2 through 6. Special paths in $k$-space were calculated using the seekpath \cite{HINUMA2017140,2018arXiv180801590T} package. Integrated RMS error is displayed in parentheses next to each potential in the legend. It is calculated using all eigenvalues at each $k$-point on a $13 \times 13 \times 13$ grid in the Brillouin zone. Configurations that the potentials used during fitting have ``Training" in the title. Those configurations for which phonons are predicted have ``Prediction" in the title. Of the 65 configurations, 29 were used for fitting and 36 for validation.

\begin{figure*}
    \begin{adjustbox}{rotate=0}
    \begin{minipage}{\textwidth}  
    \begin{tabular}{ccc}
    \subfloat[]{\includegraphics[width = 3.2in]{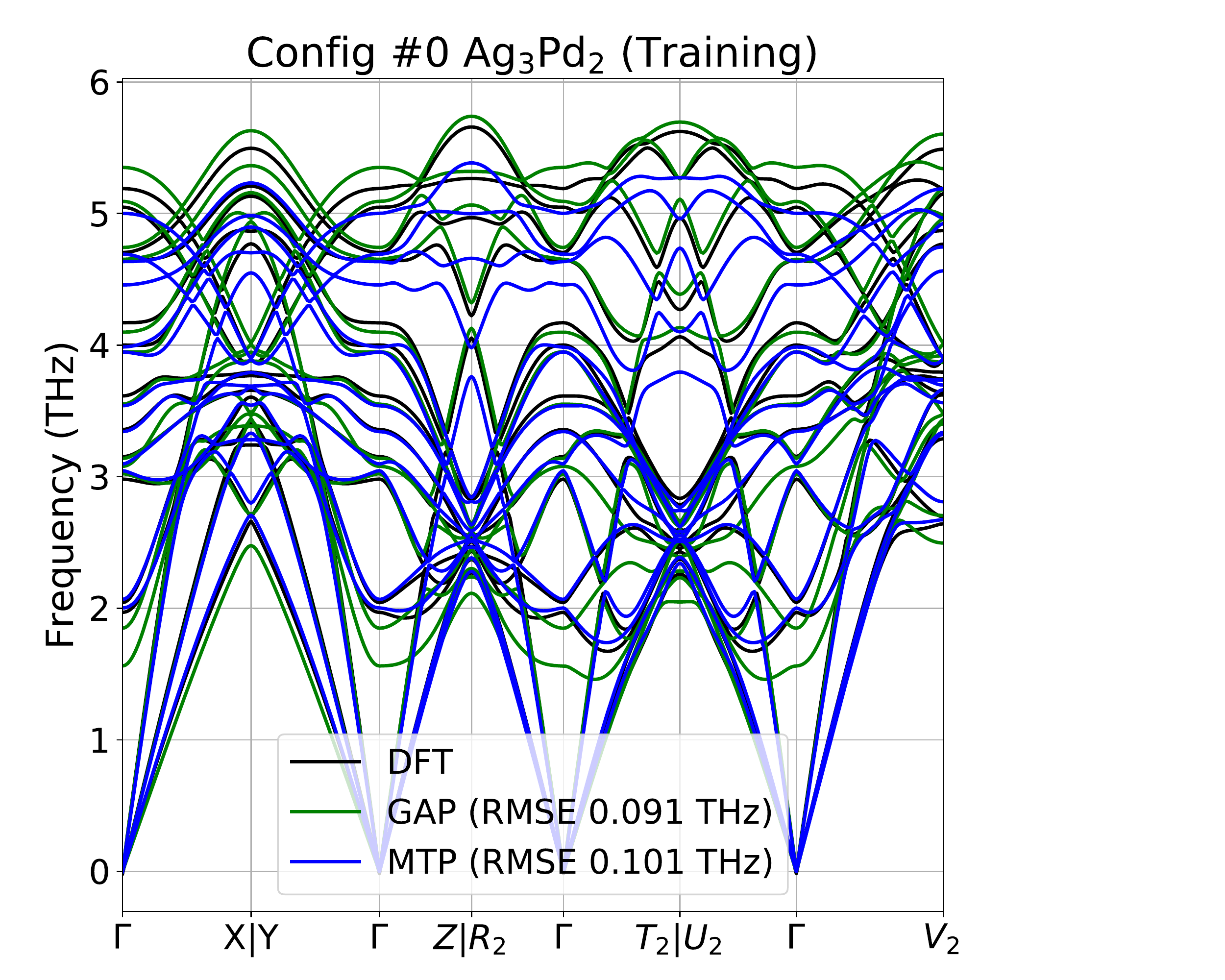}} &
    \subfloat[]{\includegraphics[width = 3.2in]{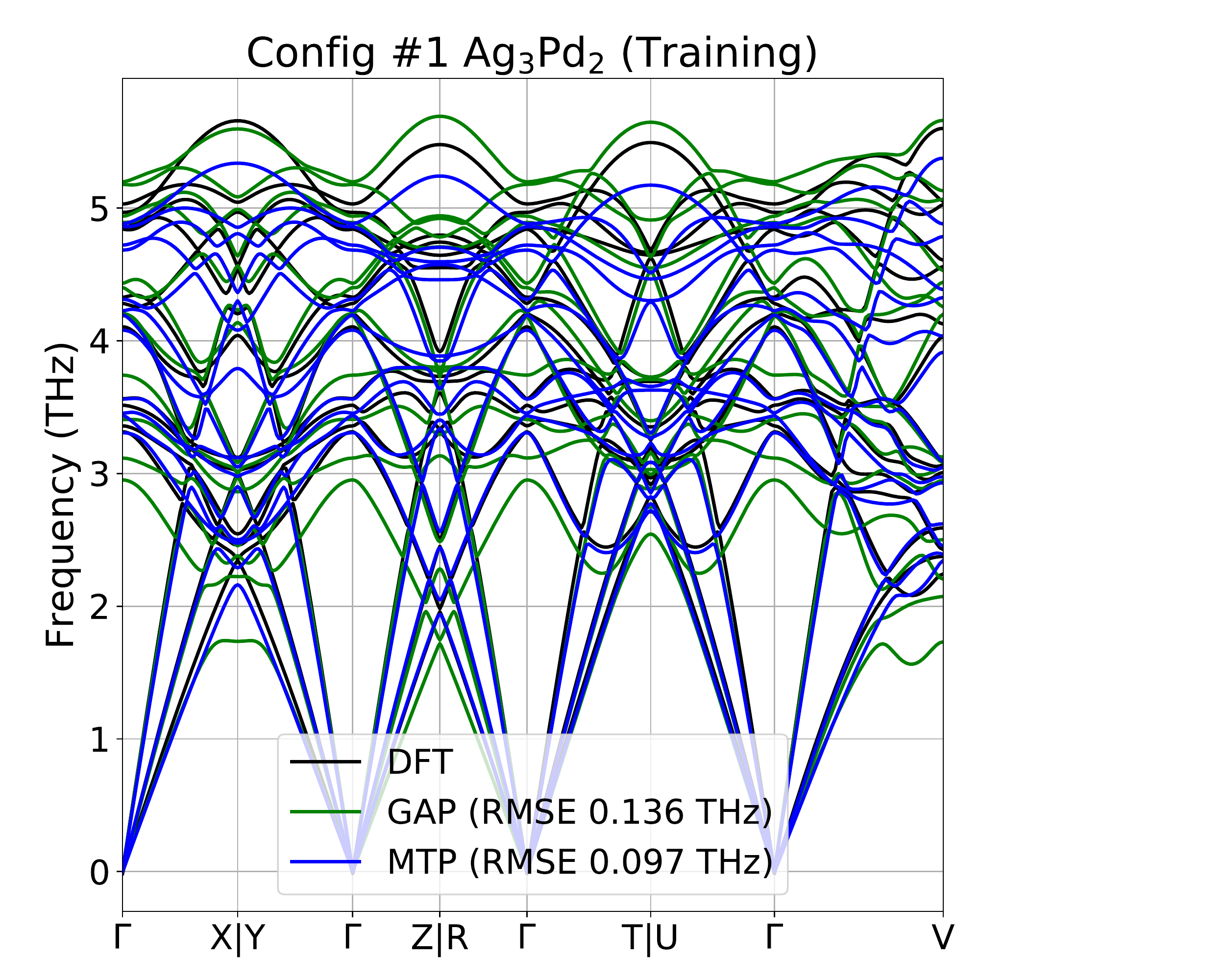}} \\
    \subfloat[]{\includegraphics[width = 3.2in]{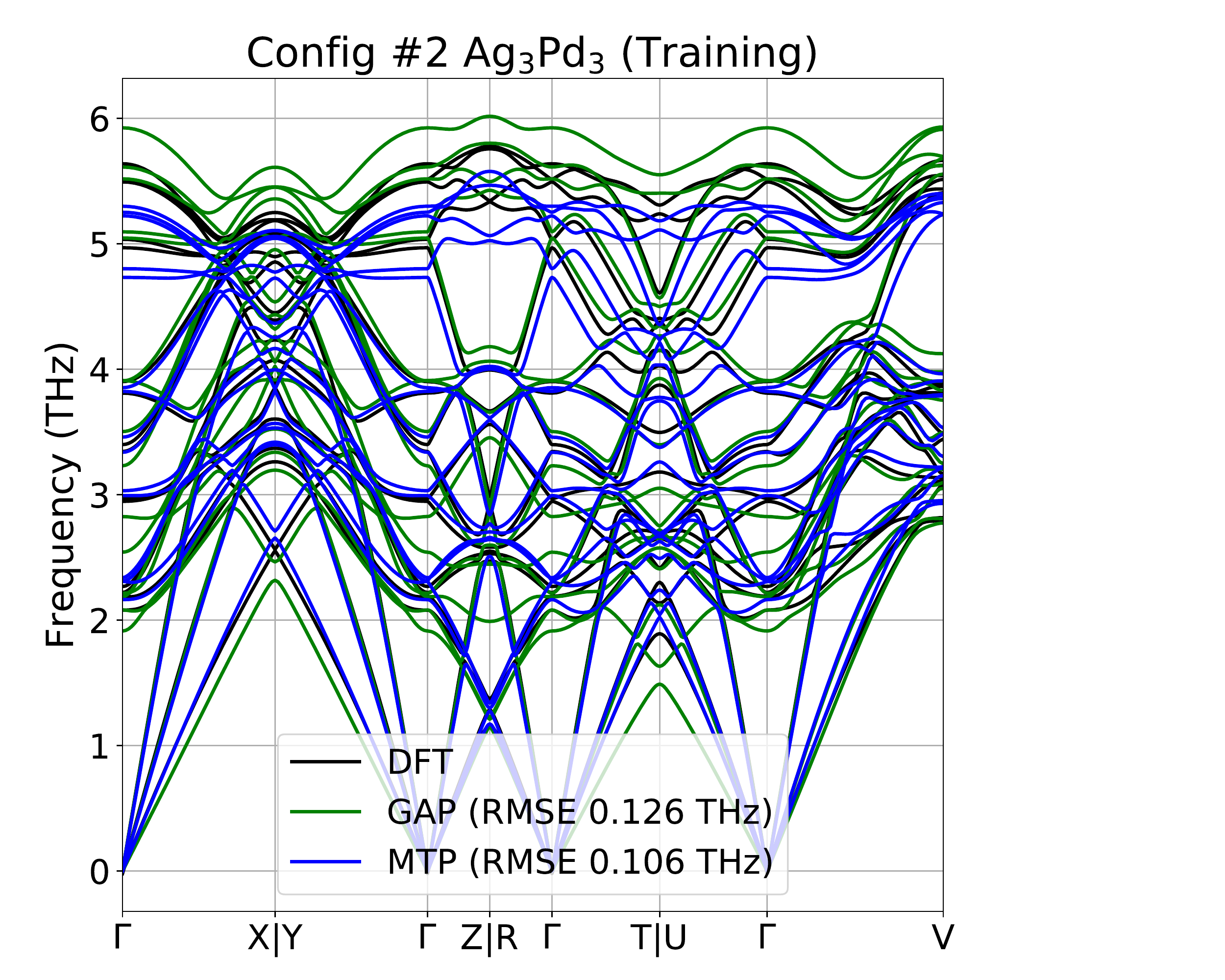}} &
    \subfloat[]{\includegraphics[width = 3.2in]{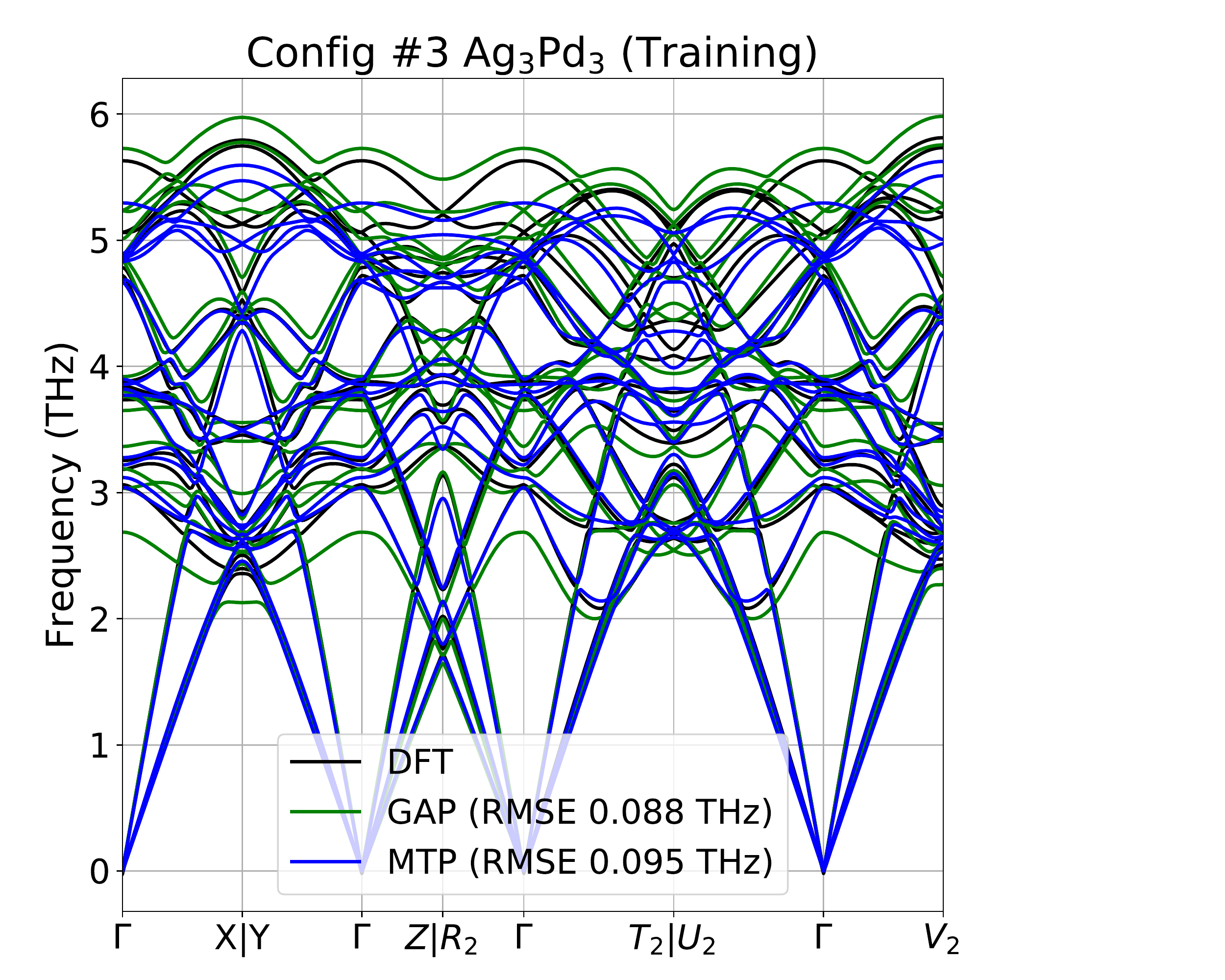}} \\
    \subfloat[]{\includegraphics[width = 3.2in]{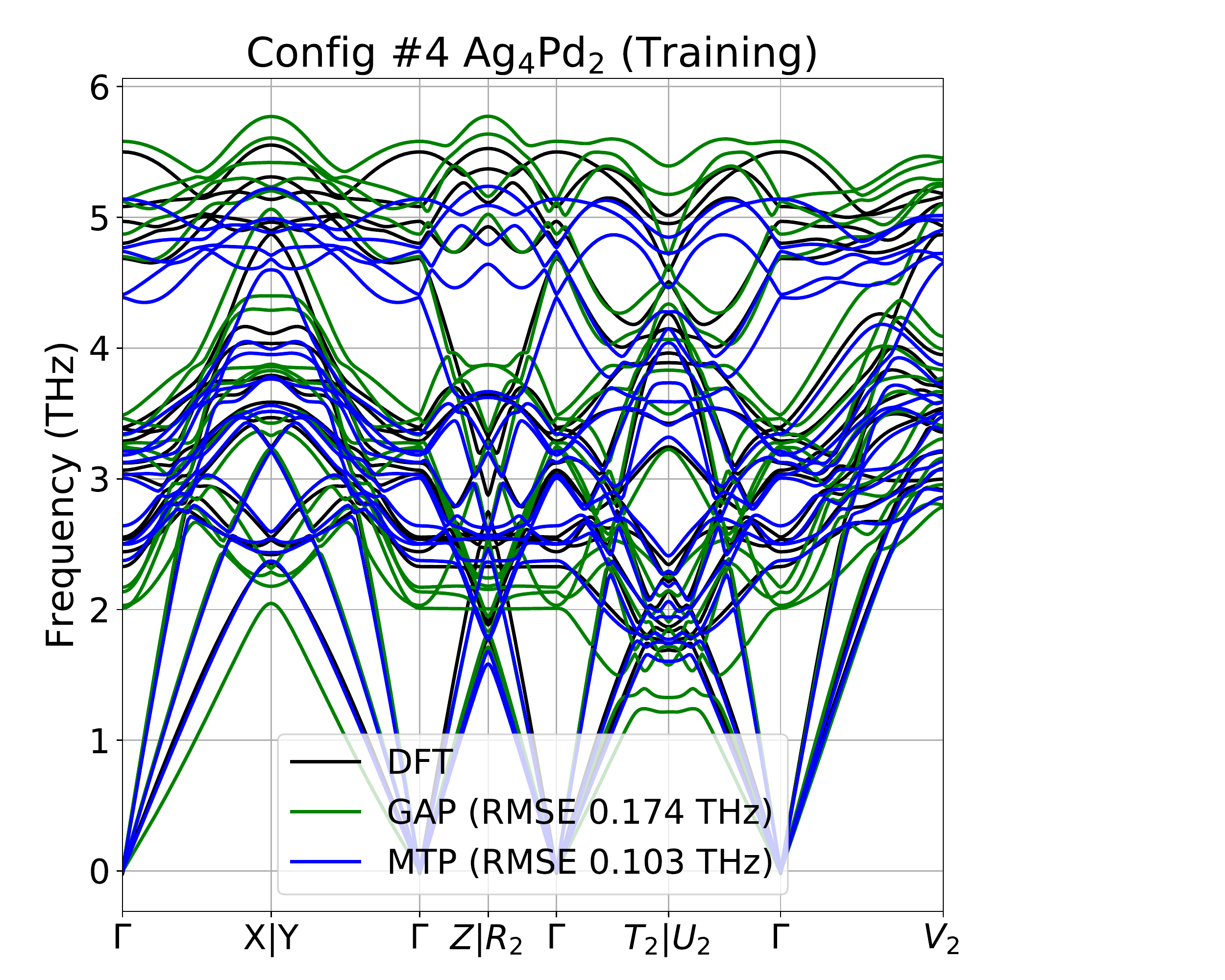}} &
    \subfloat[]{\includegraphics[width = 3.2in]{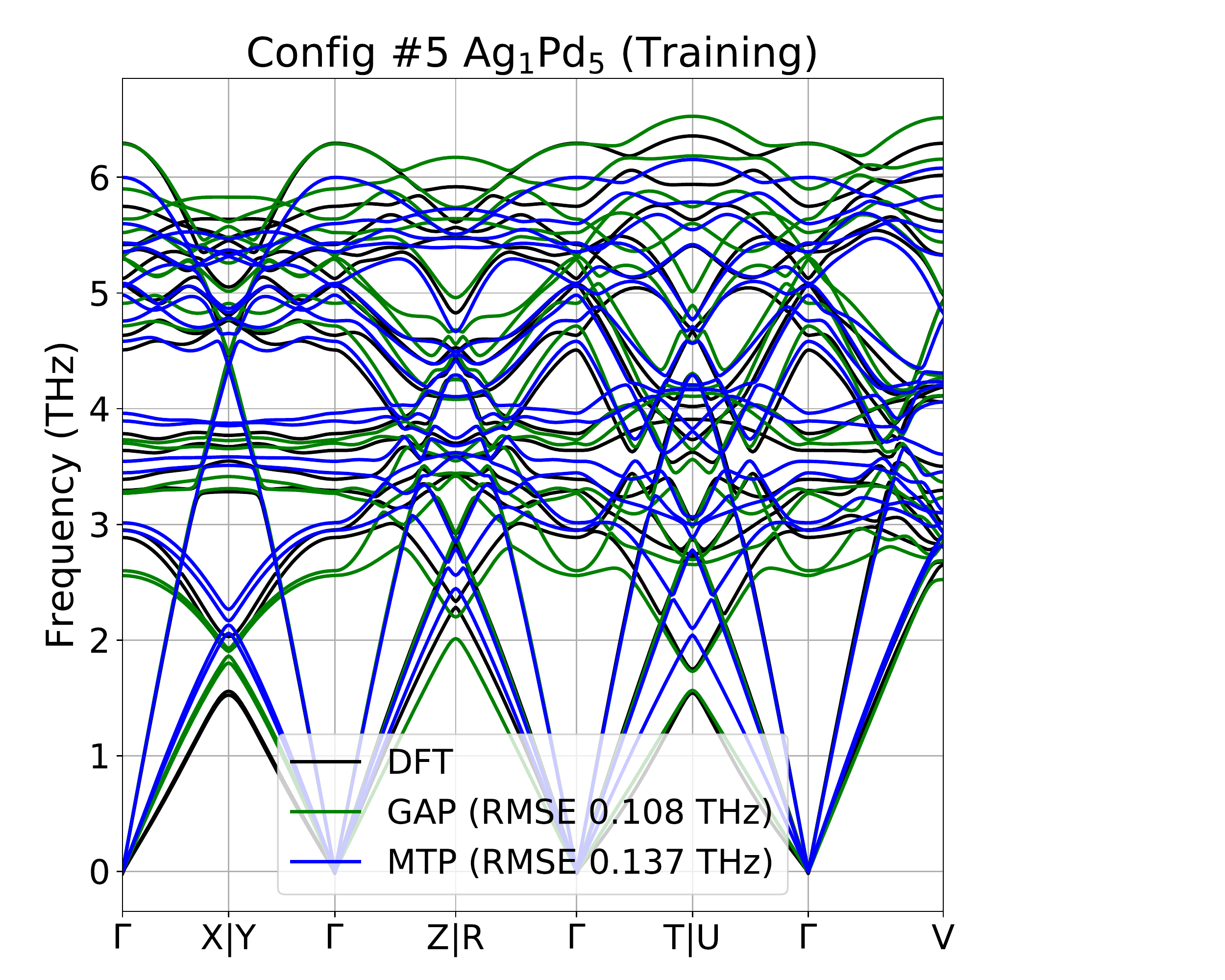}}
    \end{tabular}
    \end{minipage}  
    \end{adjustbox}
\end{figure*}
\begin{figure*}
    \begin{adjustbox}{rotate=0}
    \begin{minipage}{\textwidth}  
    \begin{tabular}{ccc}
    \subfloat[]{\includegraphics[width = 3.2in]{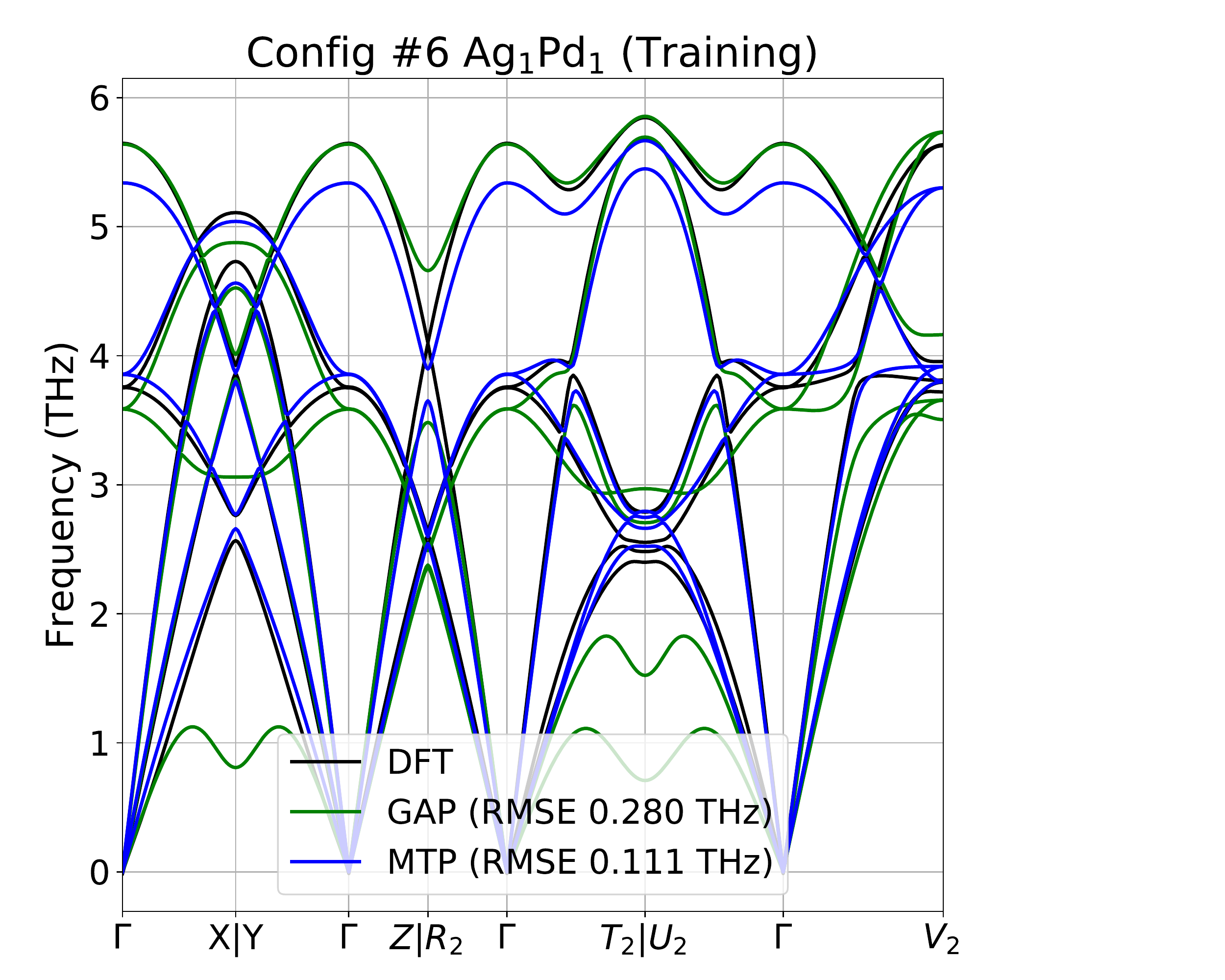}} &
    \subfloat[]{\includegraphics[width = 3.2in]{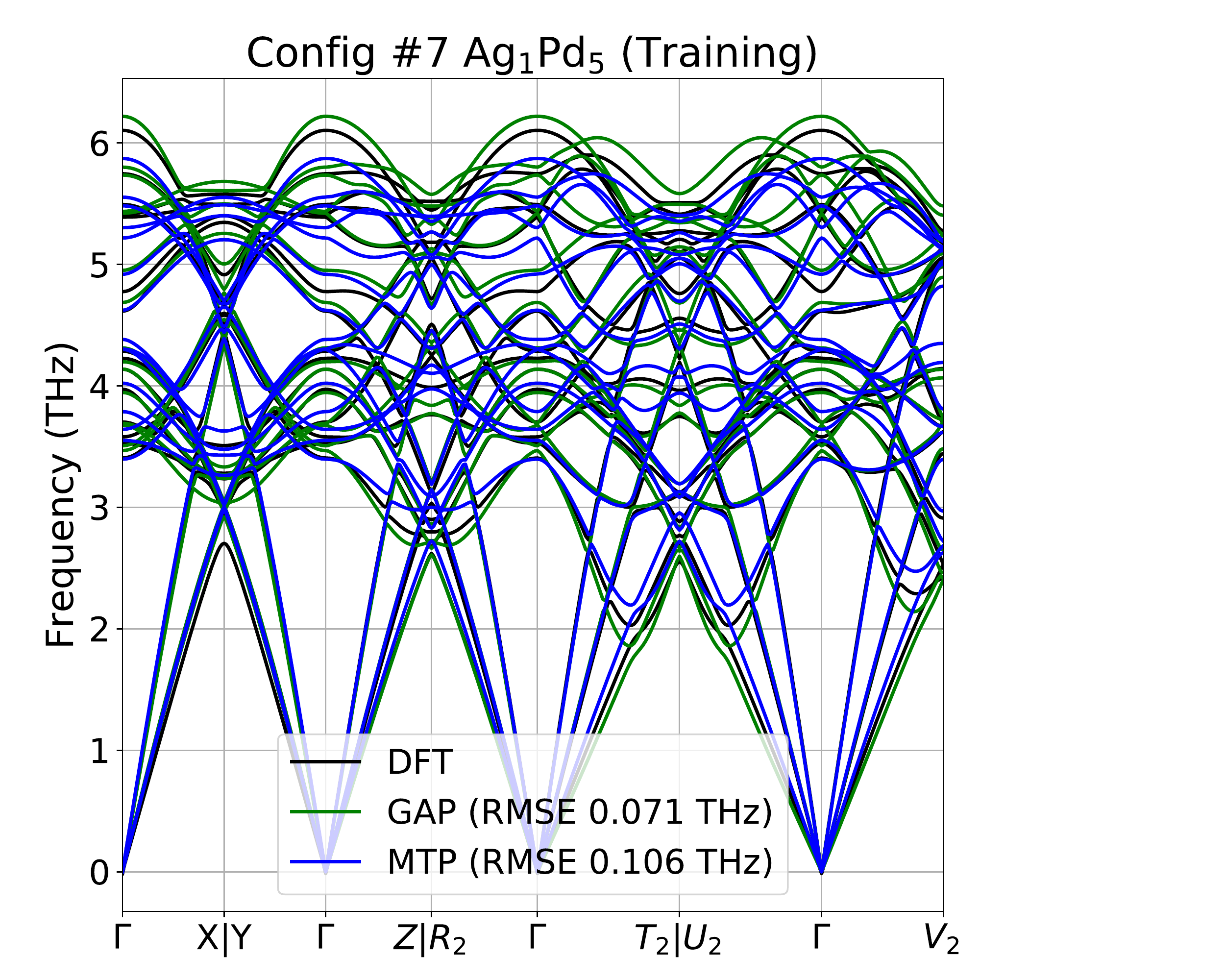}} \\
    \subfloat[]{\includegraphics[width = 3.2in]{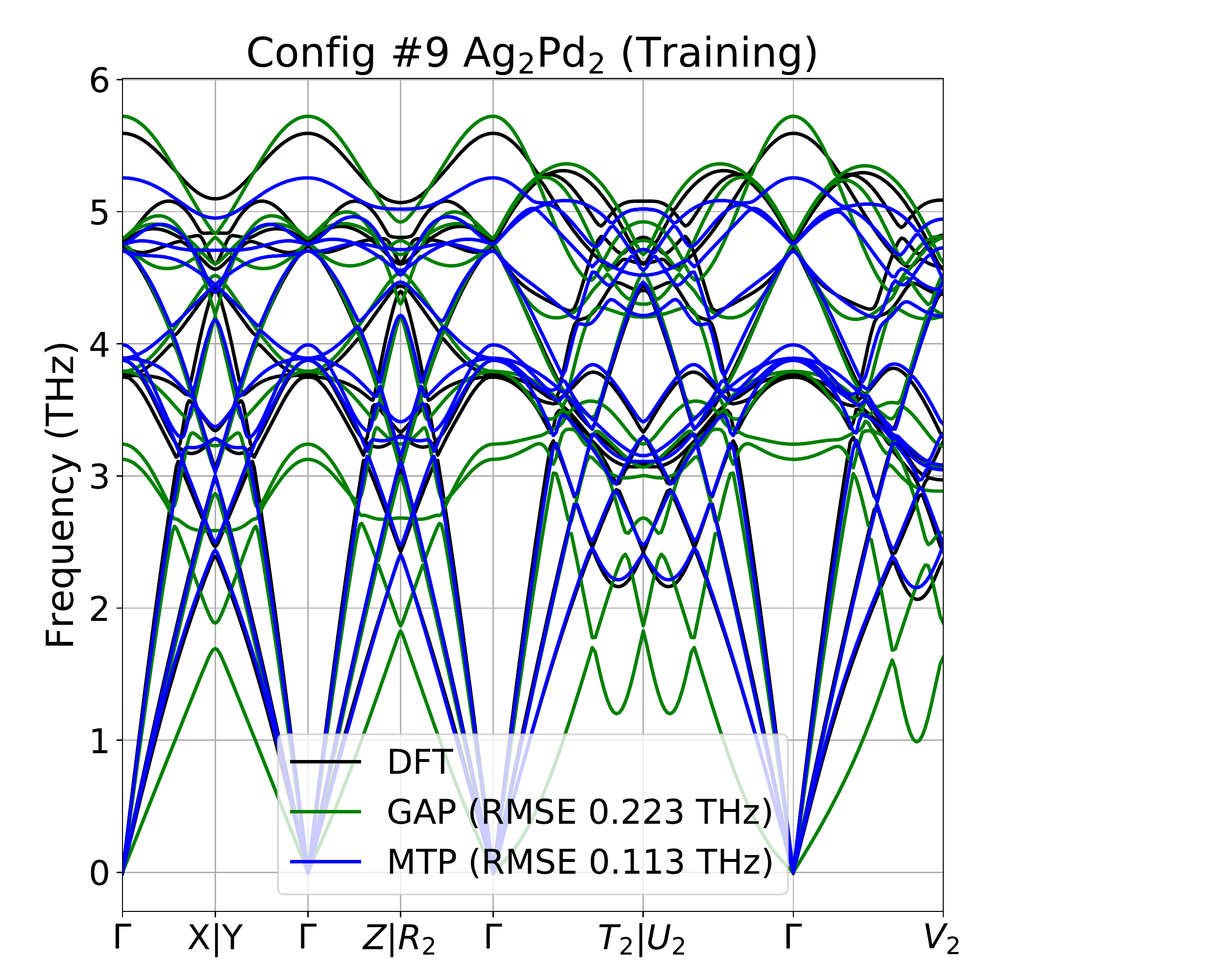}} &
    \subfloat[]{\includegraphics[width = 3.2in]{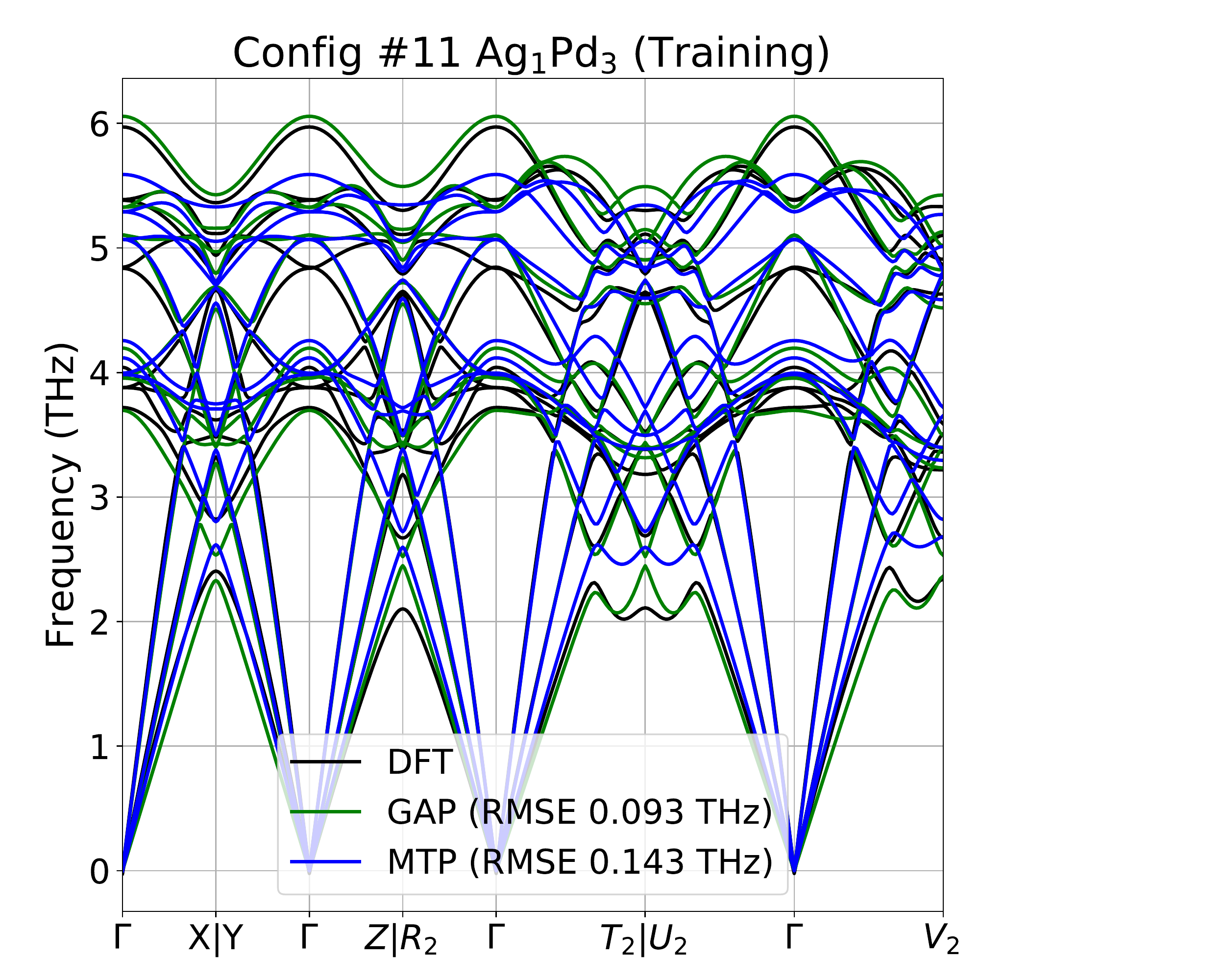}} \\
    \subfloat[]{\includegraphics[width = 3.2in]{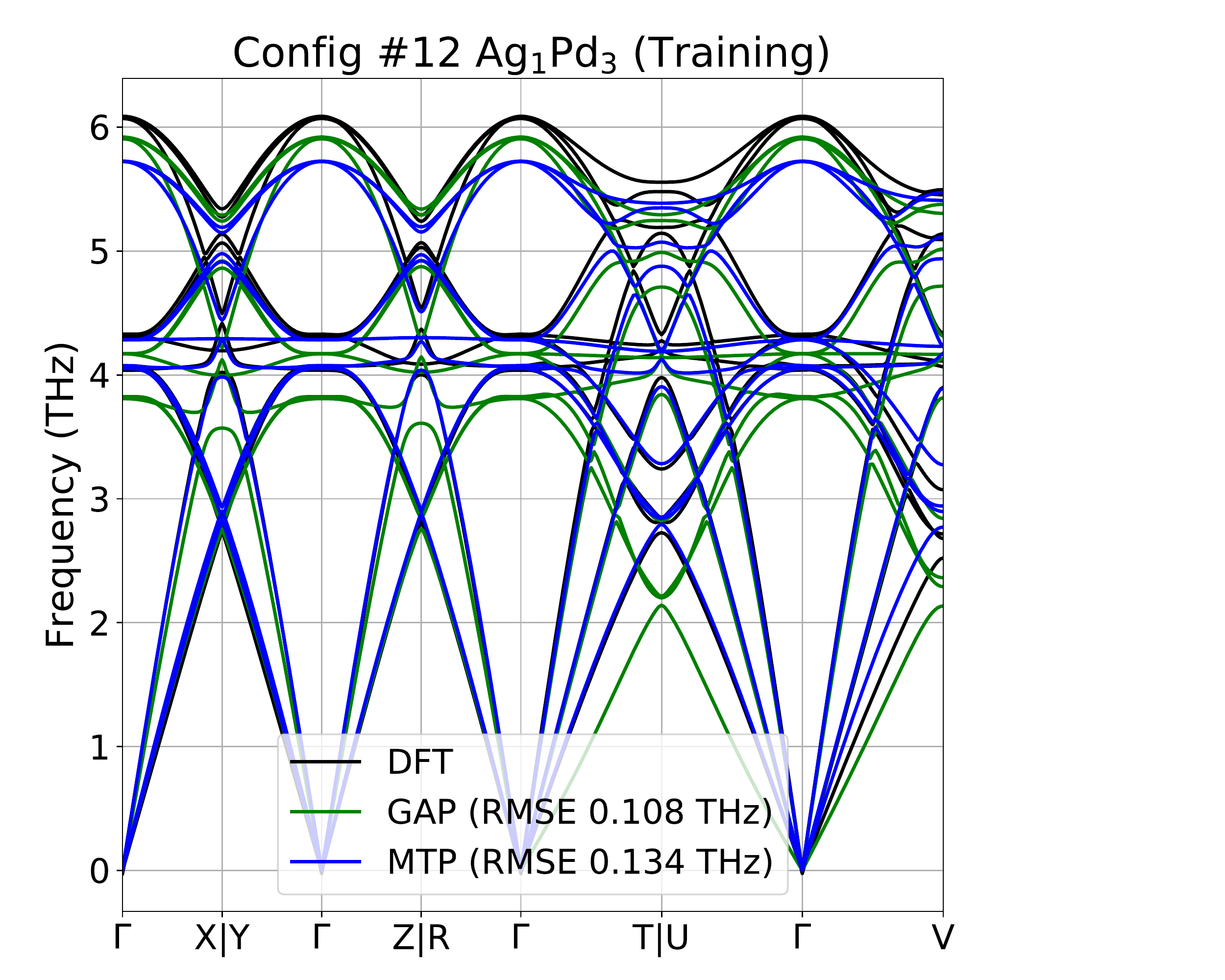}} &
    \subfloat[]{\includegraphics[width = 3.2in]{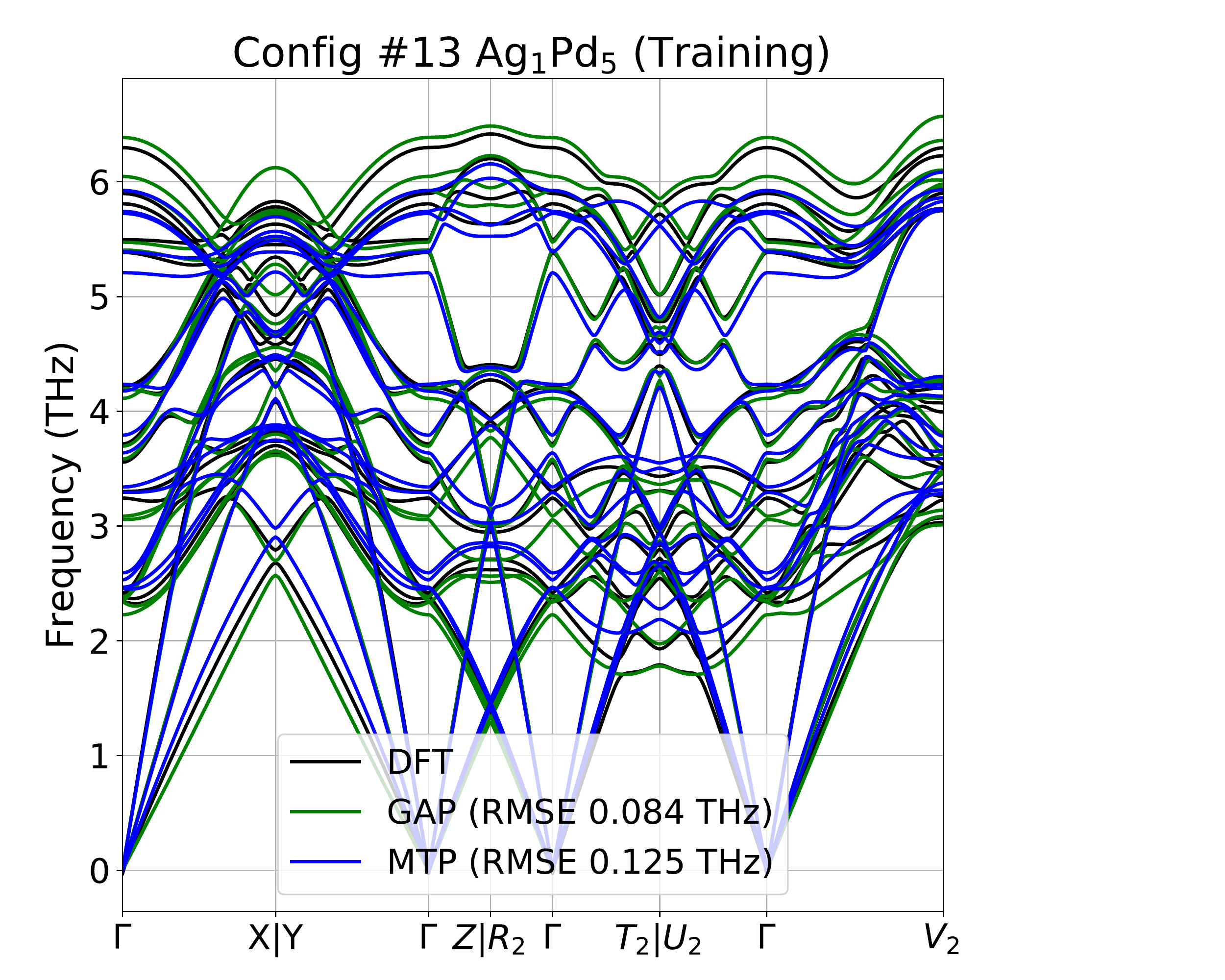}}
    \end{tabular}
    \end{minipage}  
    \end{adjustbox}
\end{figure*}
\begin{figure*}
    \begin{adjustbox}{rotate=0}
    \begin{minipage}{\textwidth}  
    \begin{tabular}{ccc}
    \subfloat[]{\includegraphics[width = 3.2in]{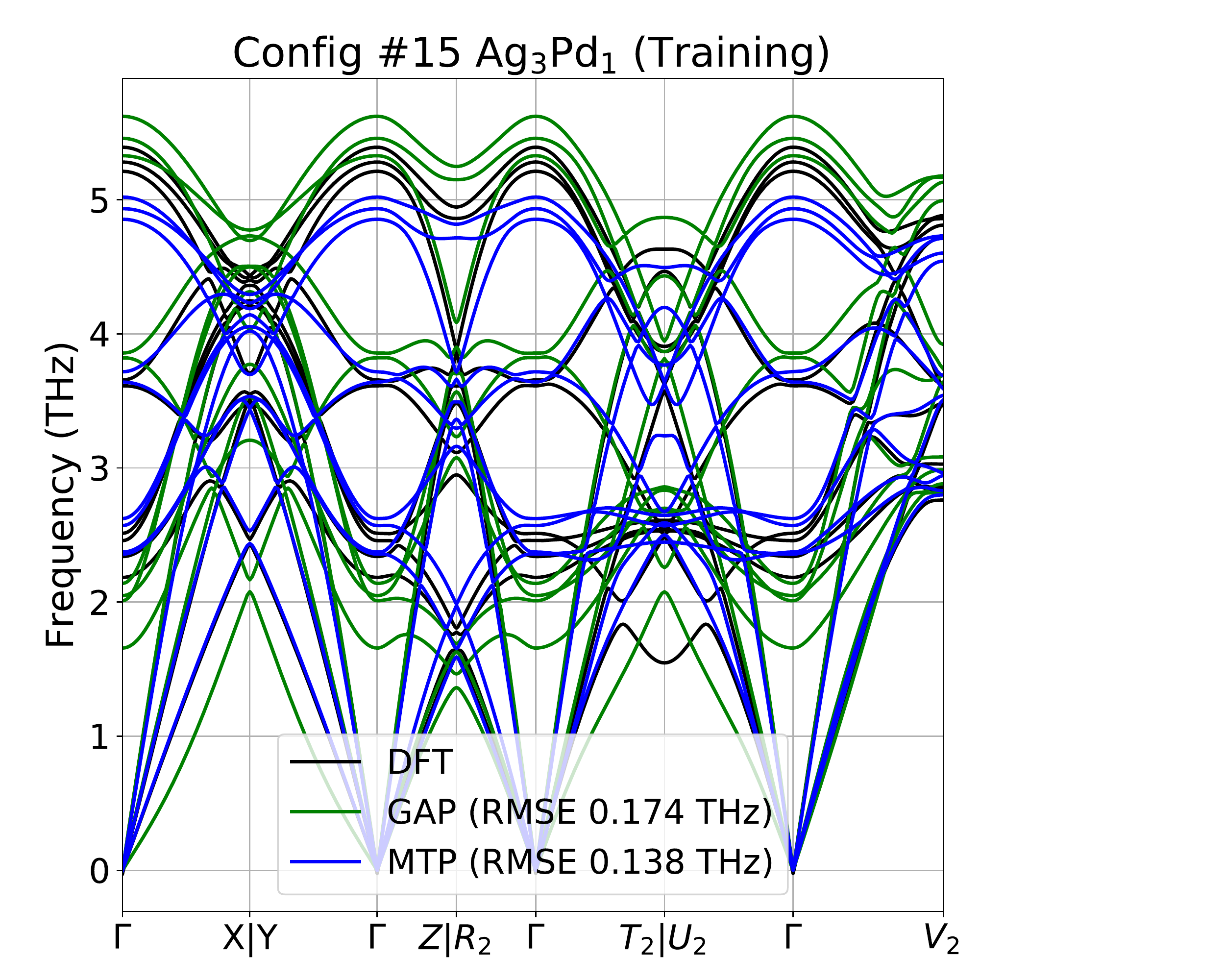}} &
    \subfloat[]{\includegraphics[width = 3.2in]{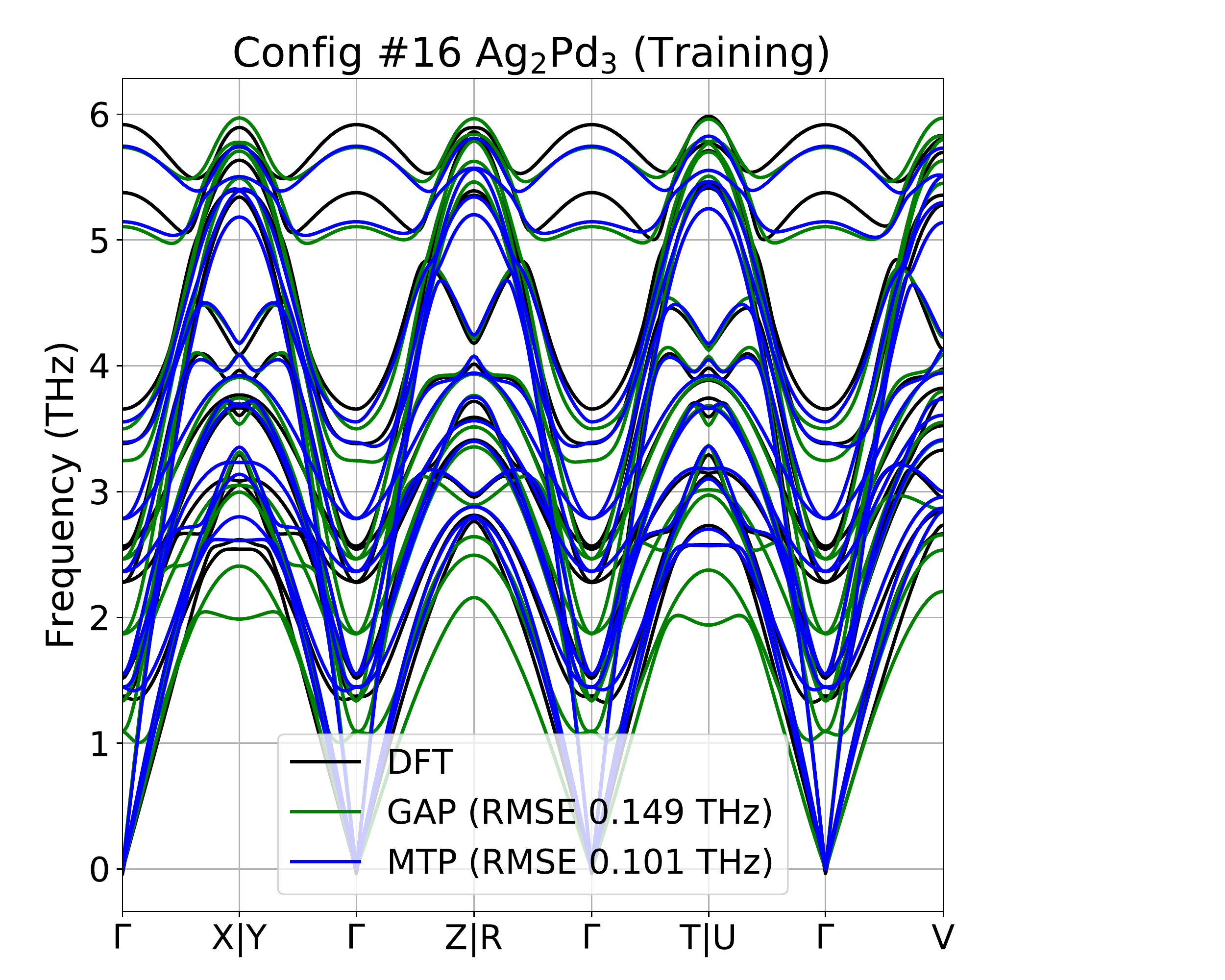}} \\
    \subfloat[]{\includegraphics[width = 3.2in]{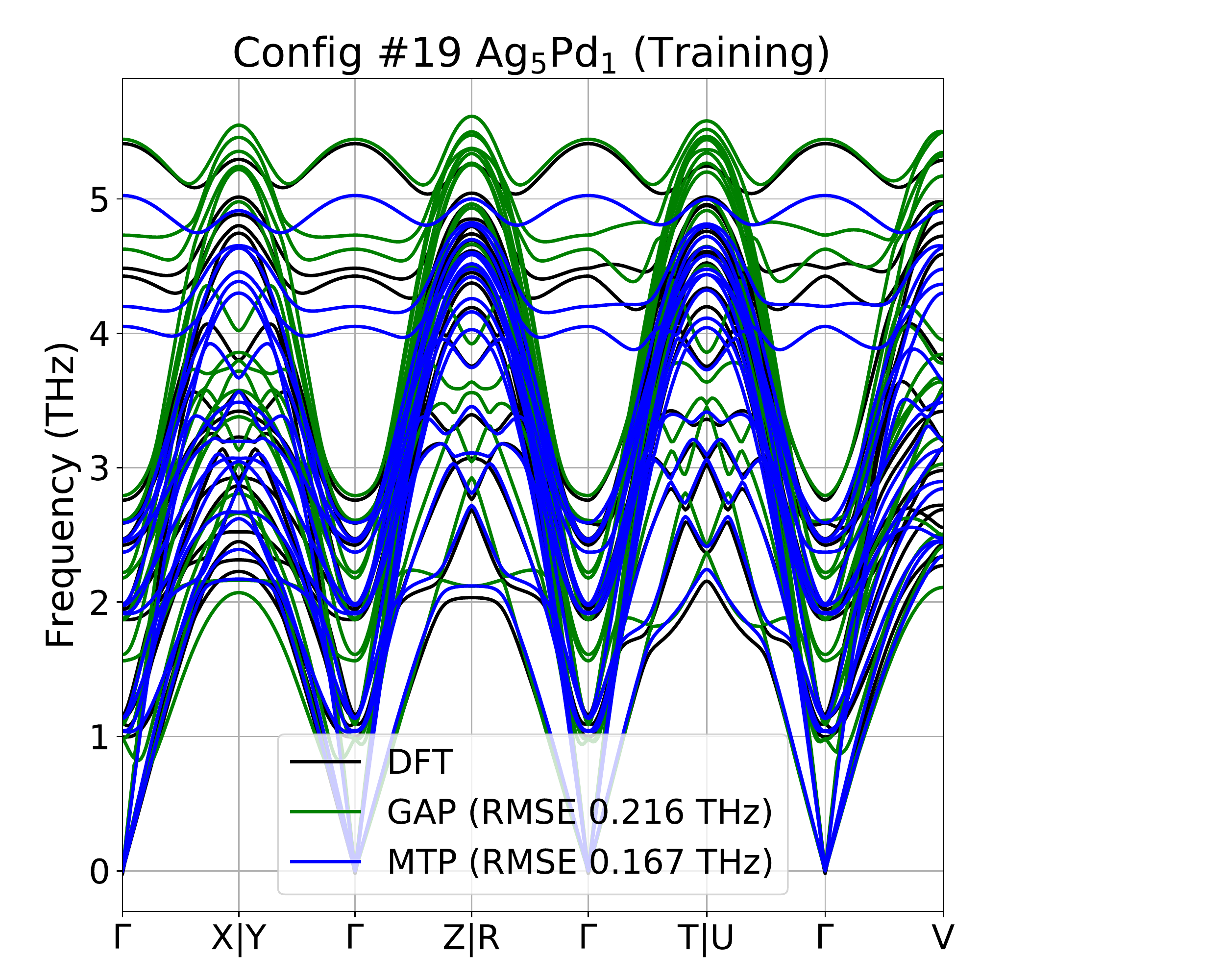}} &
    \subfloat[]{\includegraphics[width = 3.2in]{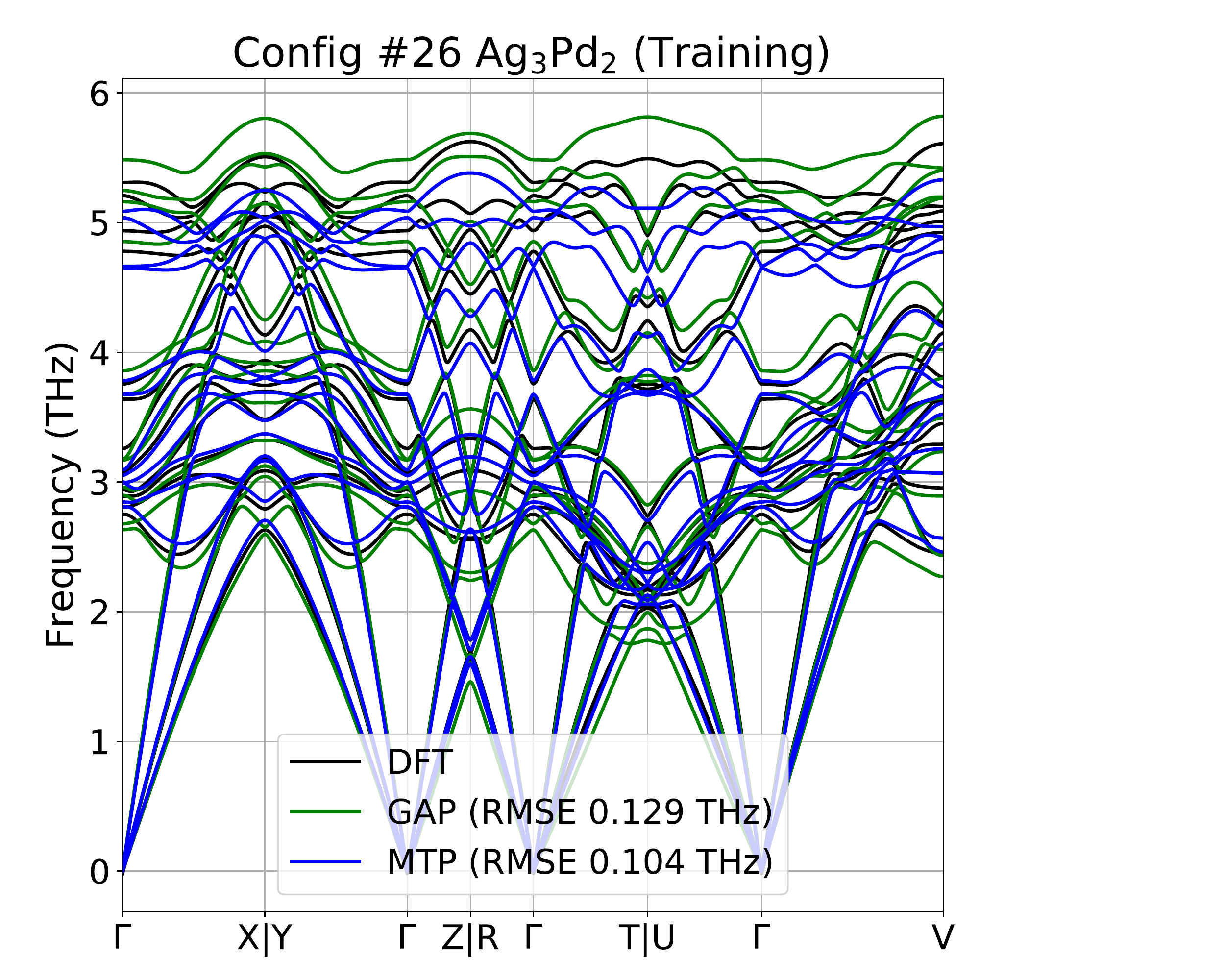}} \\
    \subfloat[]{\includegraphics[width = 3.2in]{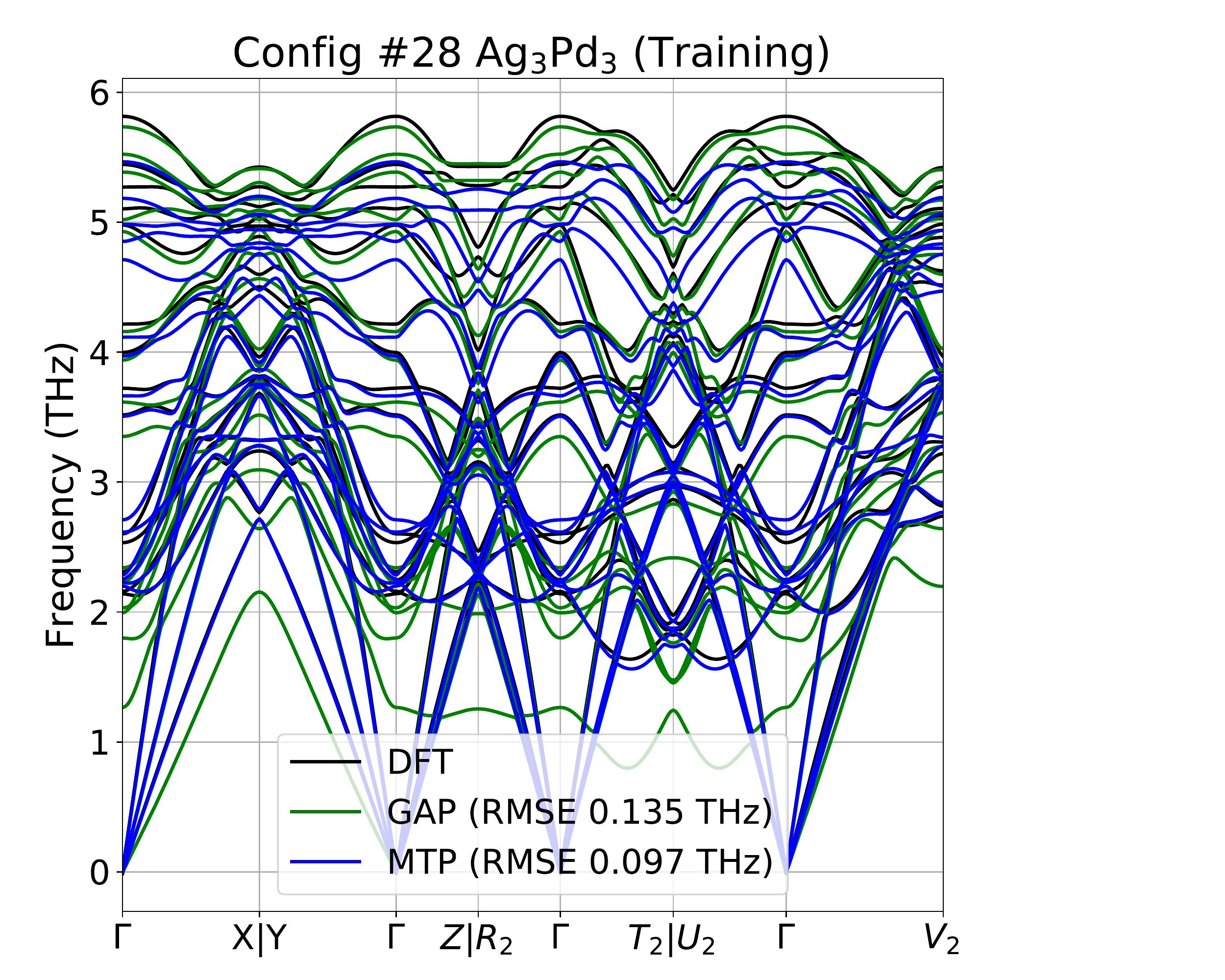}} &
    \subfloat[]{\includegraphics[width = 3.2in]{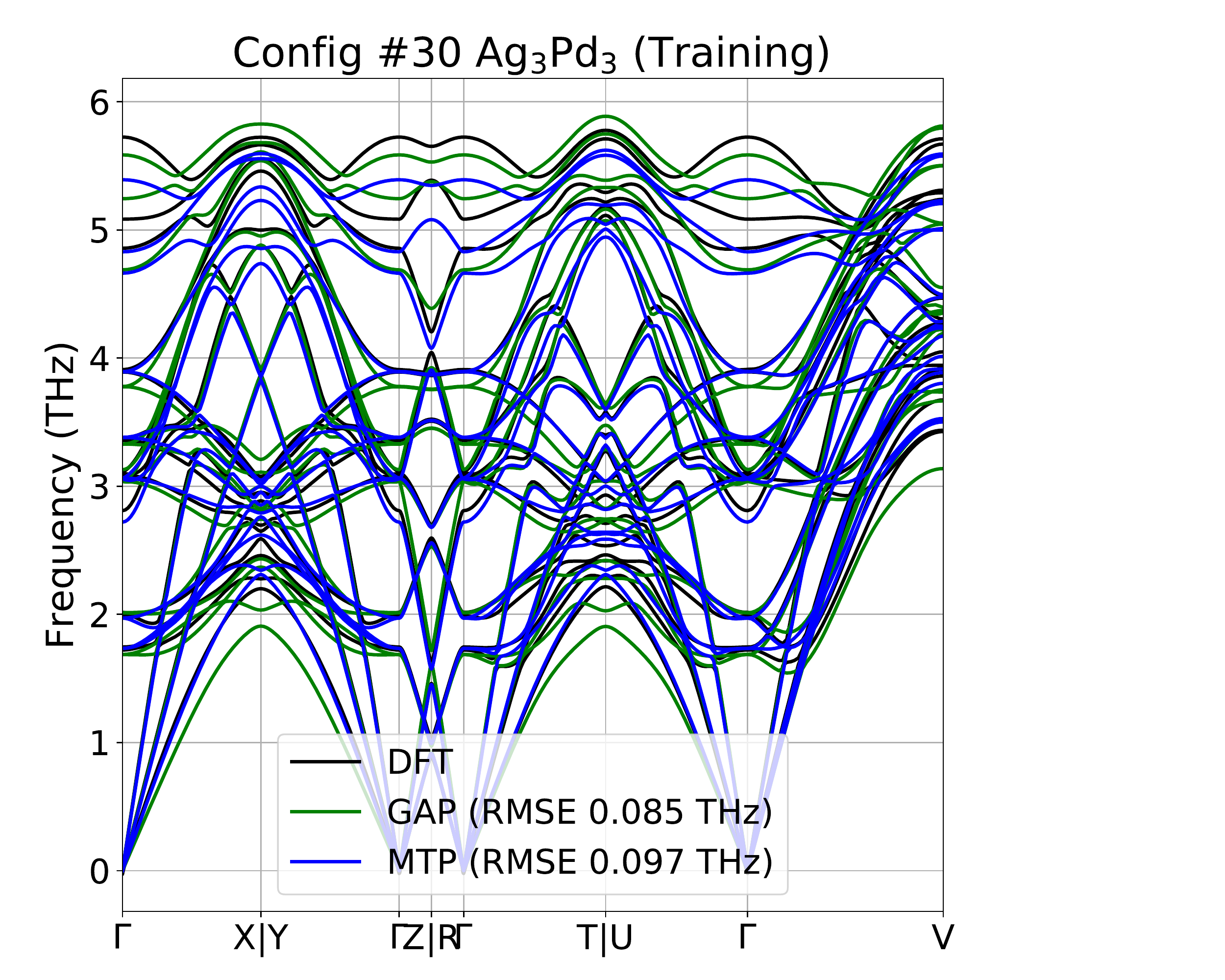}}
    \end{tabular}
    \end{minipage}  
    \end{adjustbox}
\end{figure*}
\begin{figure*}
    \begin{adjustbox}{rotate=0}
    \begin{minipage}{\textwidth}  
    \begin{tabular}{ccc}
    \subfloat[]{\includegraphics[width = 3.2in]{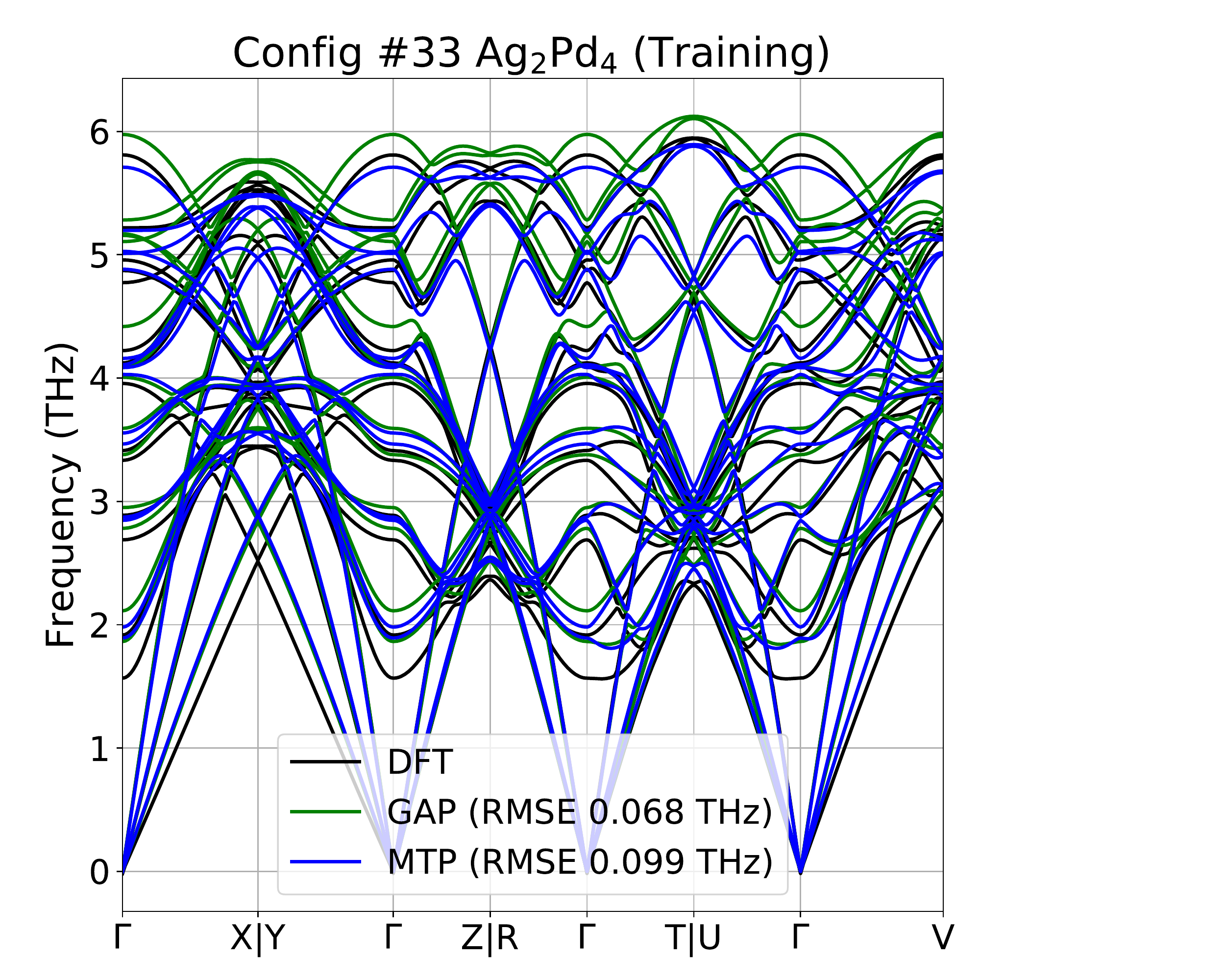}} &
    \subfloat[]{\includegraphics[width = 3.2in]{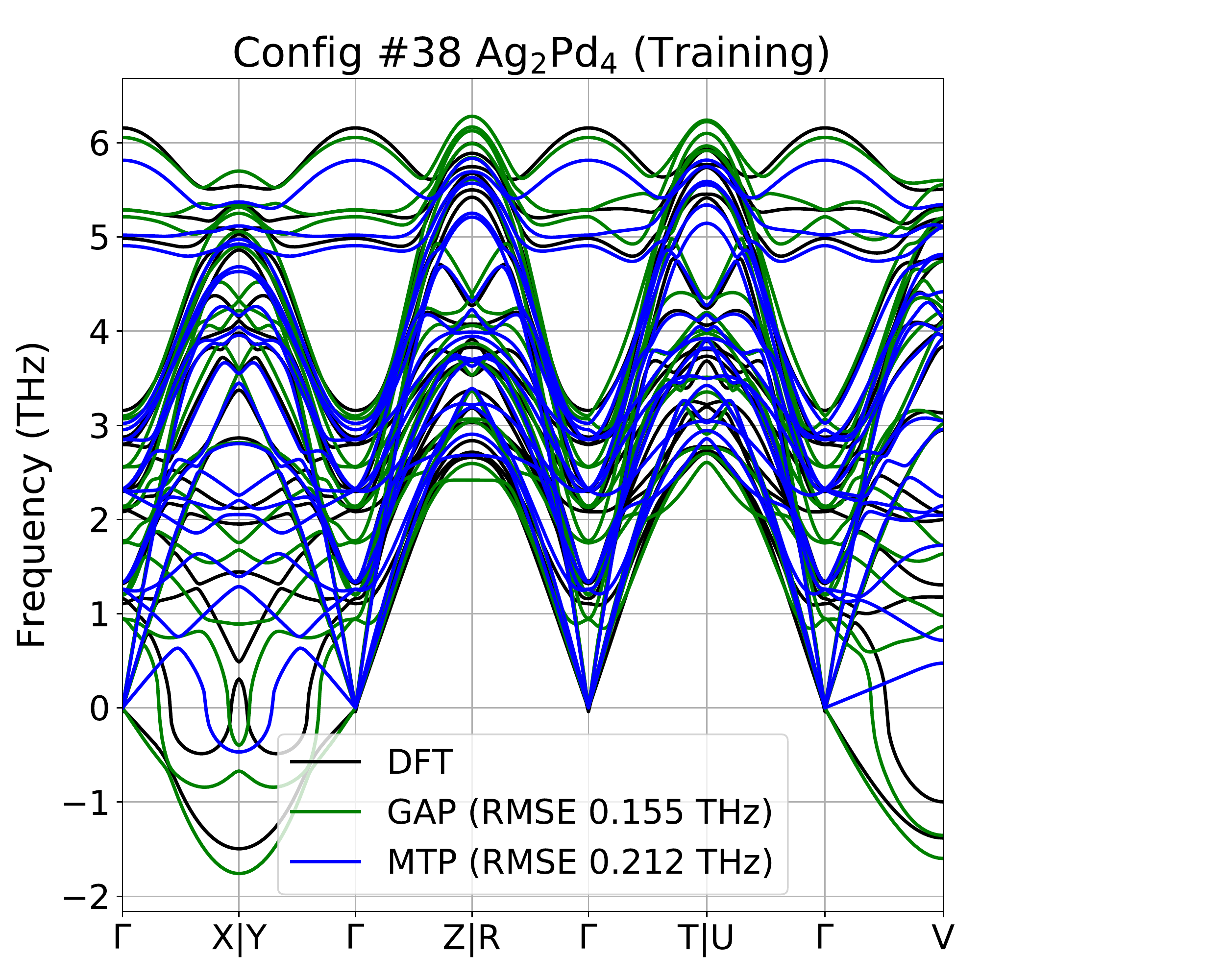}} \\
    \subfloat[]{\includegraphics[width = 3.2in]{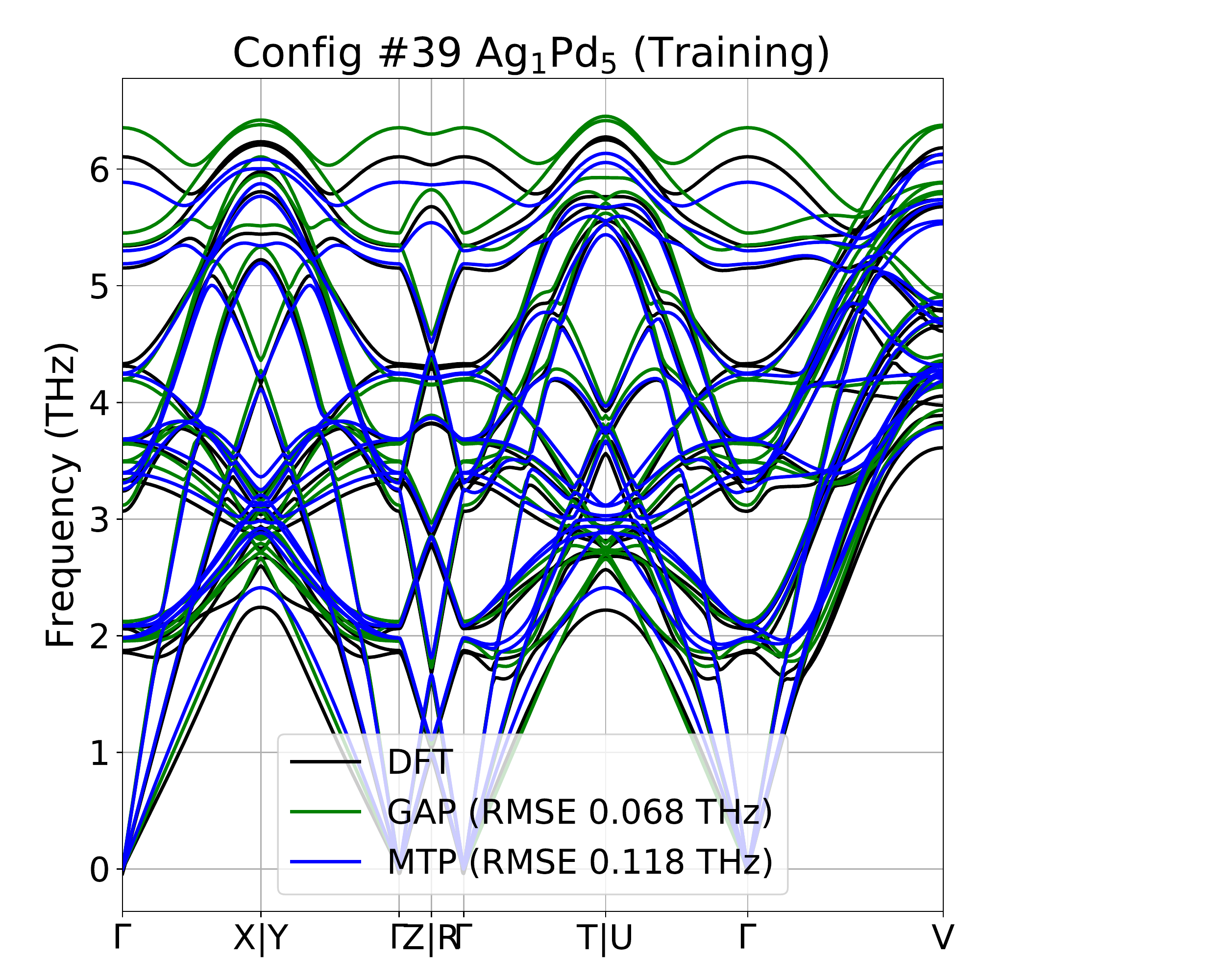}} &
    \subfloat[]{\includegraphics[width = 3.2in]{figures/phonons-paper-41.pdf}} \\
    \subfloat[]{\includegraphics[width = 3.2in]{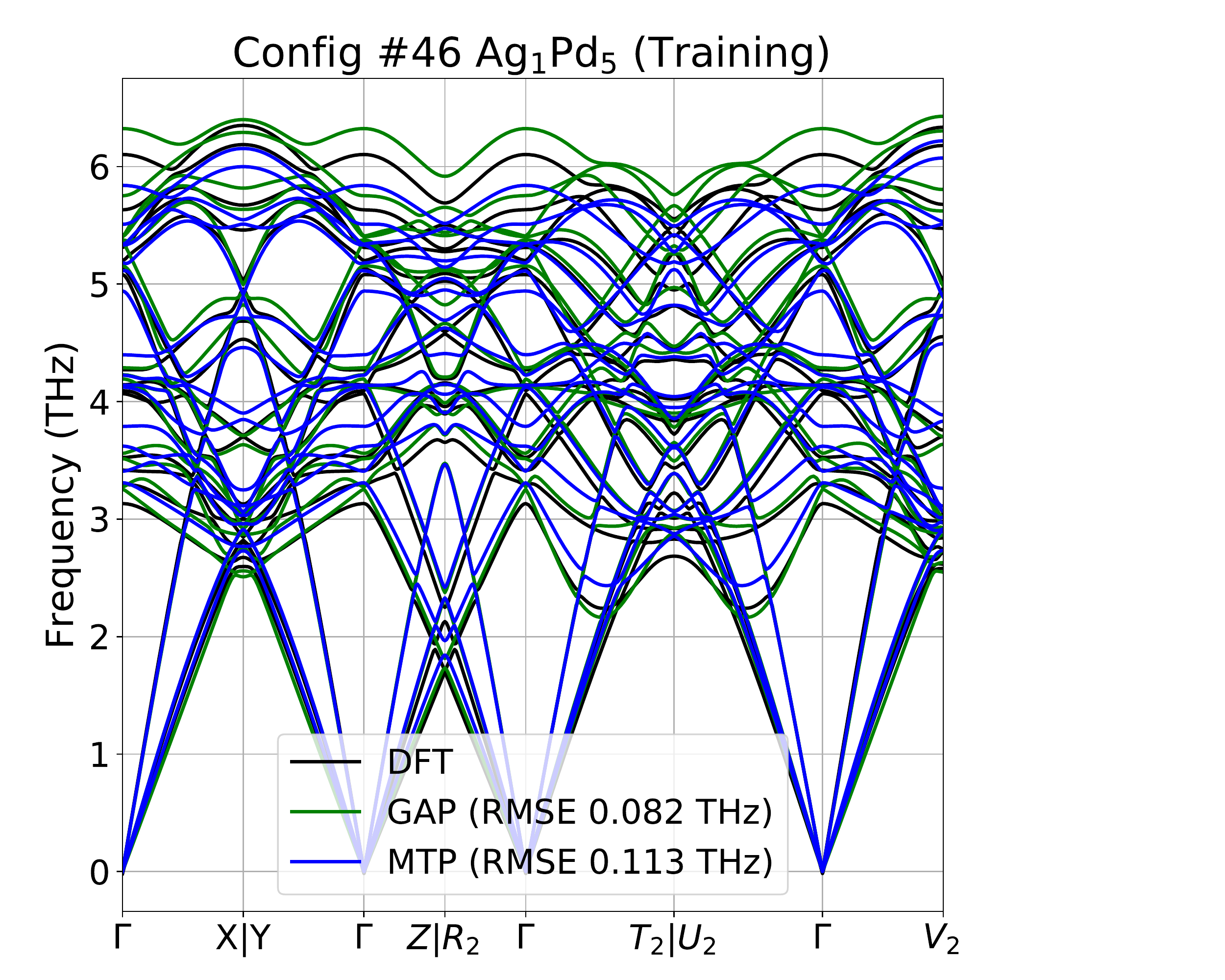}} &
    \subfloat[]{\includegraphics[width = 3.2in]{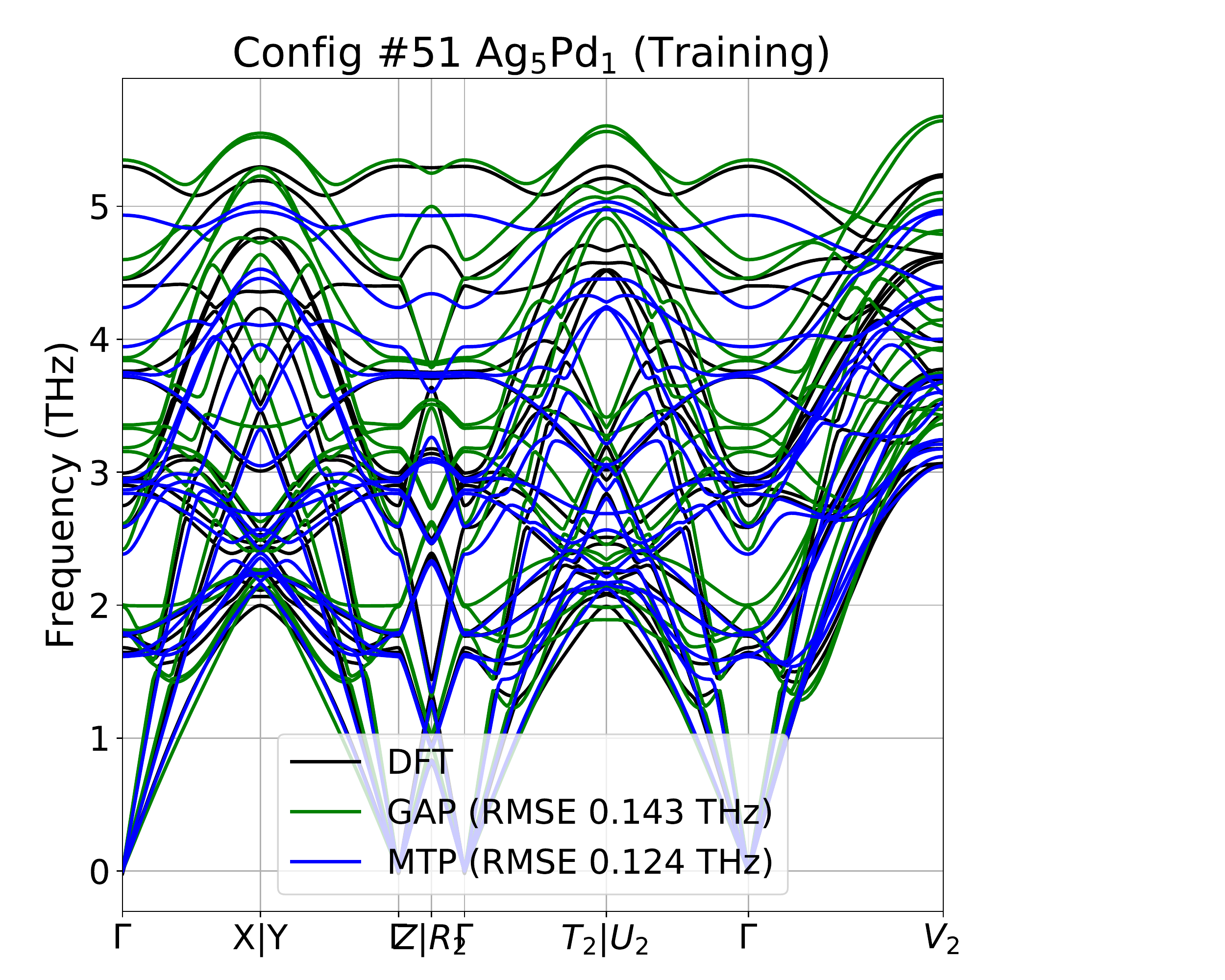}}
    \end{tabular}
    \end{minipage}  
    \end{adjustbox}
\end{figure*}
\begin{figure*}
    \begin{adjustbox}{rotate=0}
    \begin{minipage}{\textwidth}  
    \begin{tabular}{ccc}
    \subfloat[]{\includegraphics[width = 3.2in]{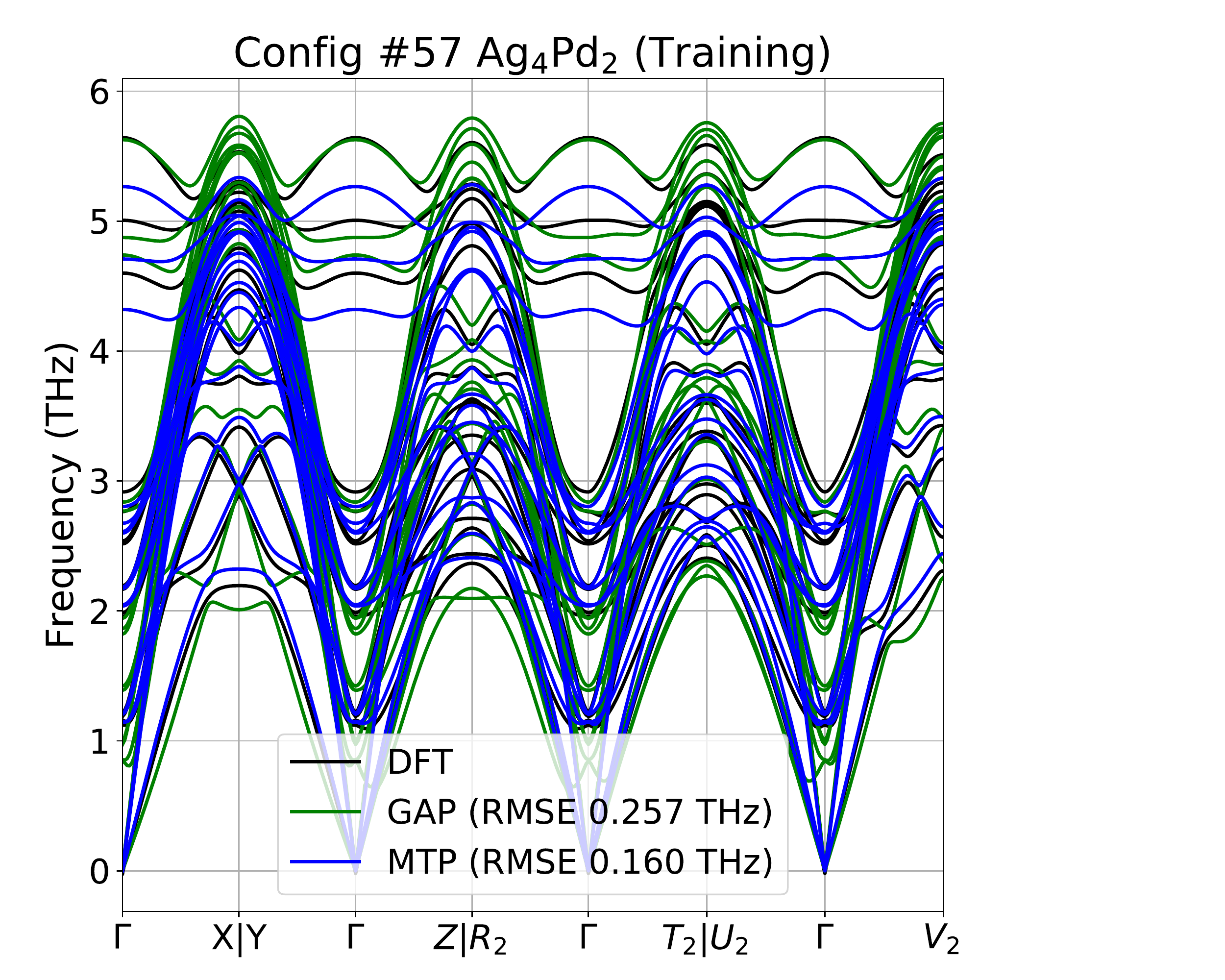}} &
    \subfloat[]{\includegraphics[width = 3.2in]{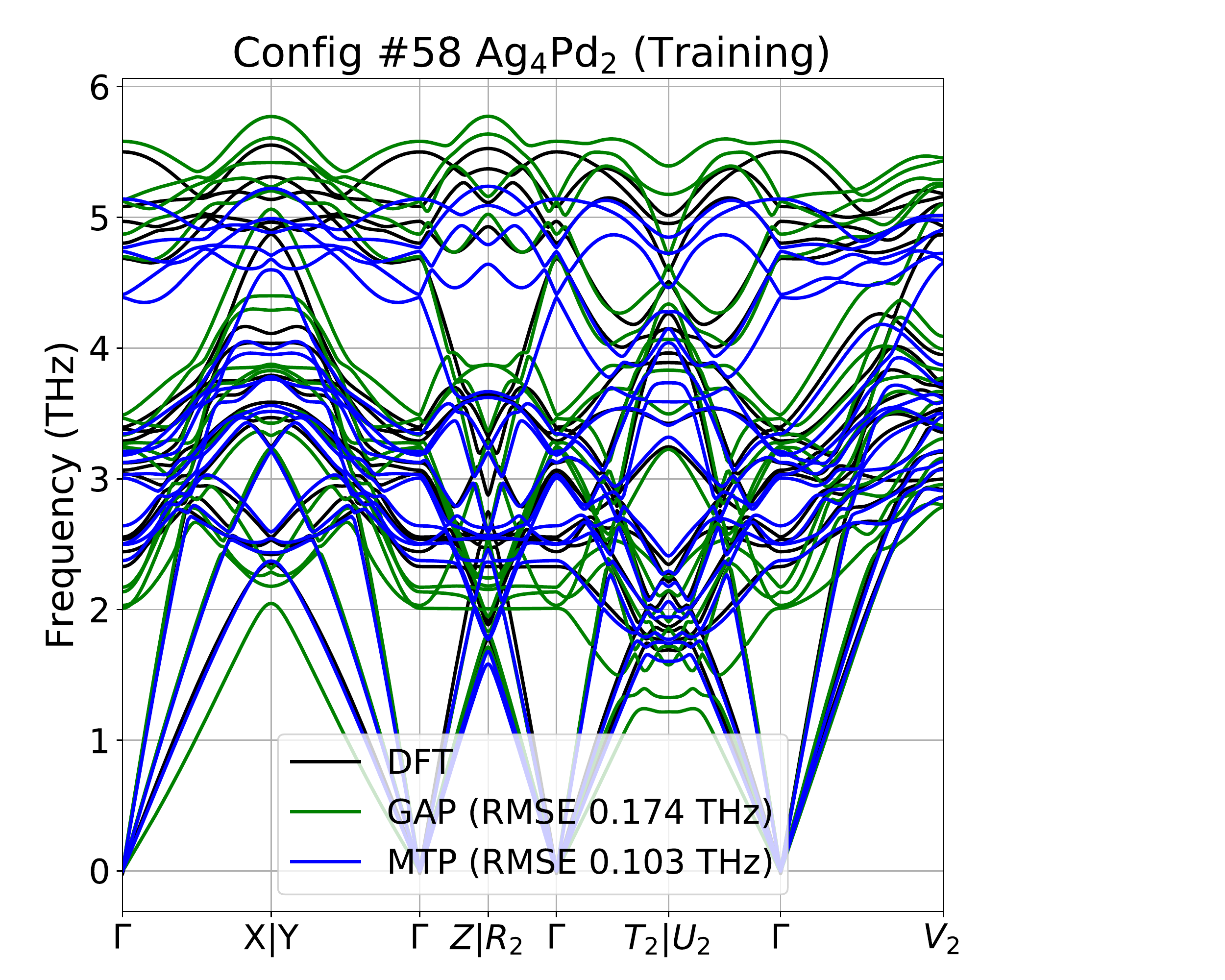}} \\
    \subfloat[]{\includegraphics[width = 3.2in]{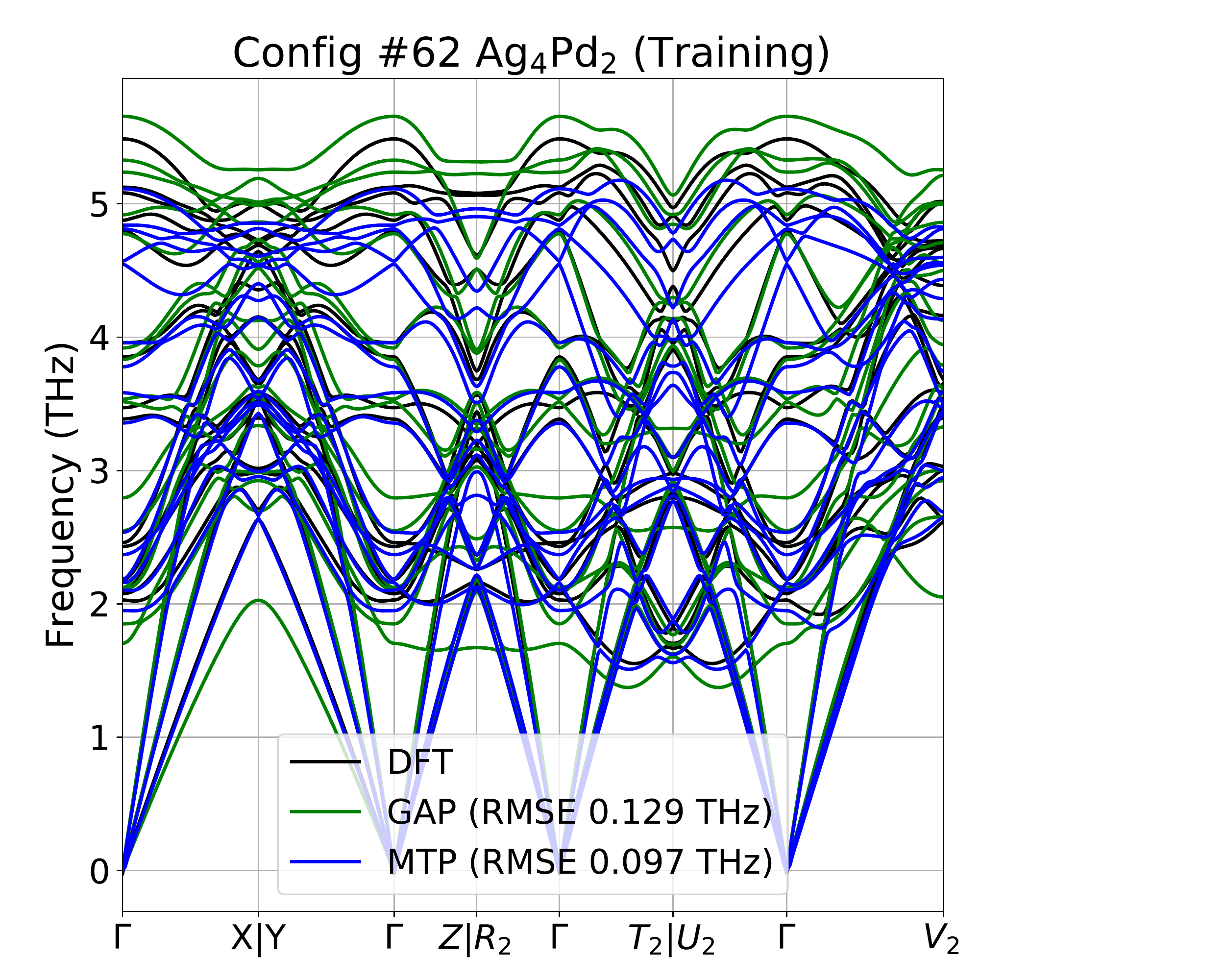}} &
    \subfloat[]{\includegraphics[width = 3.2in]{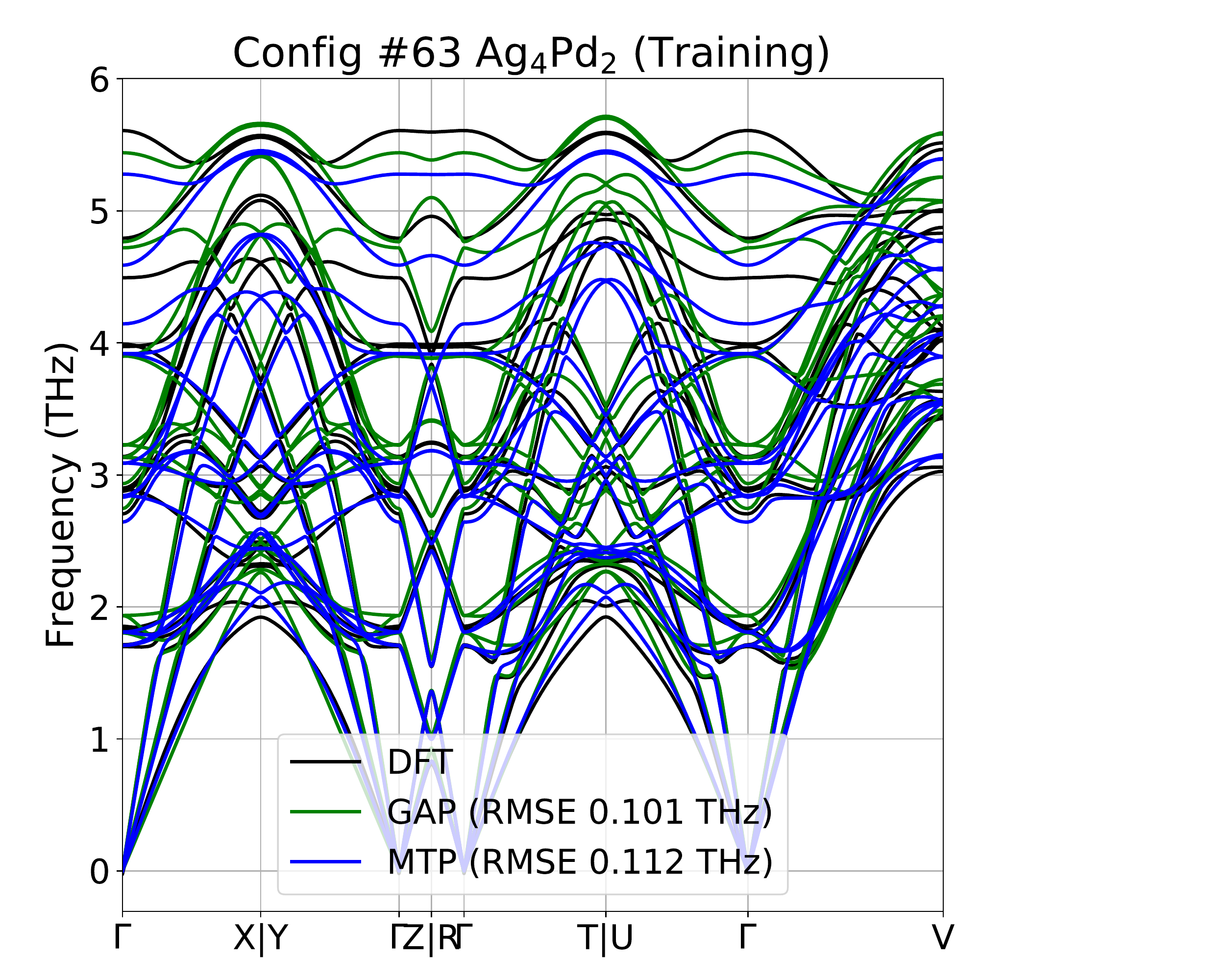}} \\
    \subfloat[]{\includegraphics[width = 3.2in]{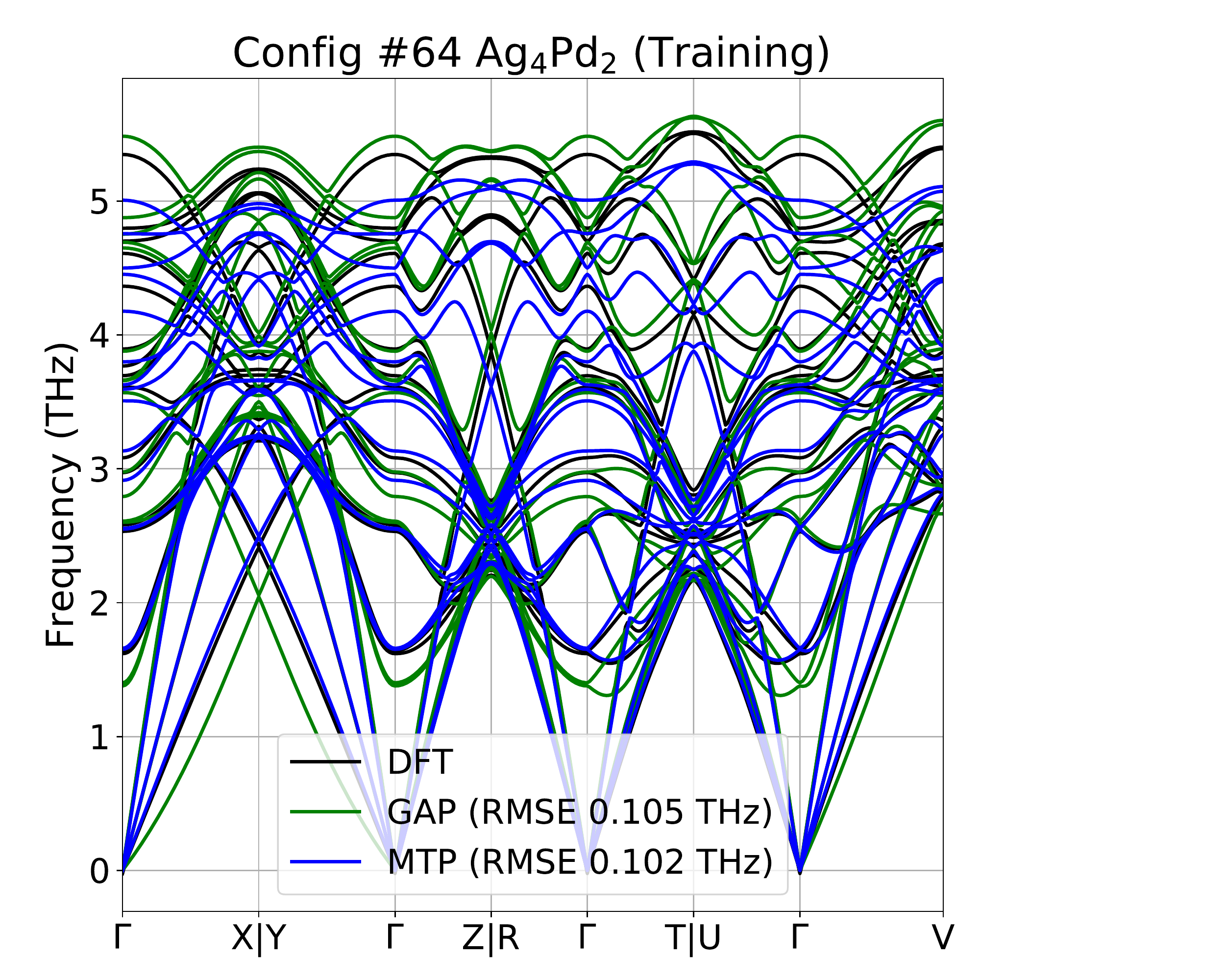}} &
    \subfloat[]{\includegraphics[width = 3.2in]{}}
    \end{tabular}
    \end{minipage}  
    \end{adjustbox}
\end{figure*}
\begin{figure*}
    \begin{adjustbox}{rotate=0}
    \begin{minipage}{\textwidth}  
    \begin{tabular}{ccc}
    \subfloat[]{\includegraphics[width = 3.2in]{figures/phonons-paper-8.pdf}} &
    \subfloat[]{\includegraphics[width = 3.2in]{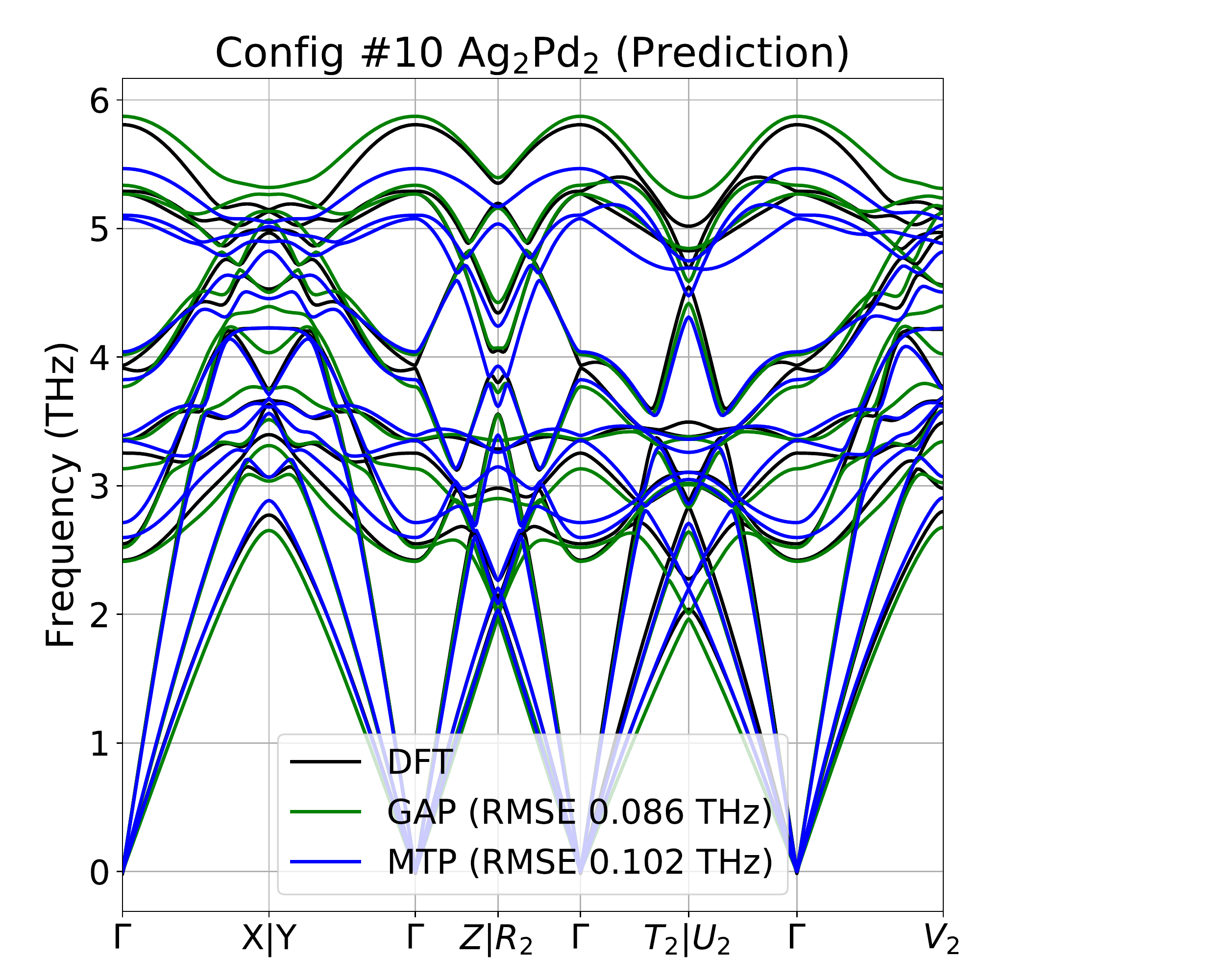}} \\
    \subfloat[]{\includegraphics[width = 3.2in]{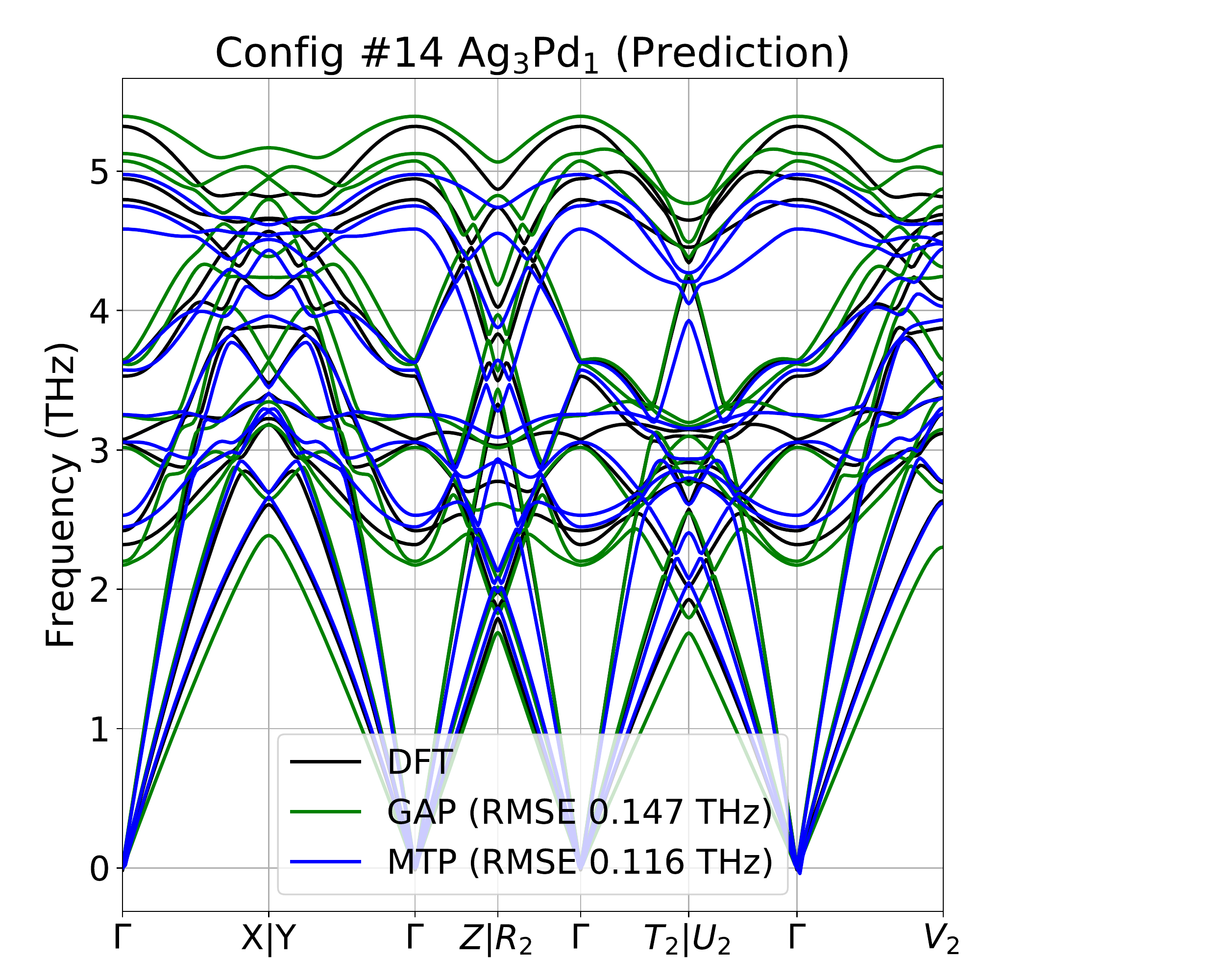}} &
    \subfloat[]{\includegraphics[width = 3.2in]{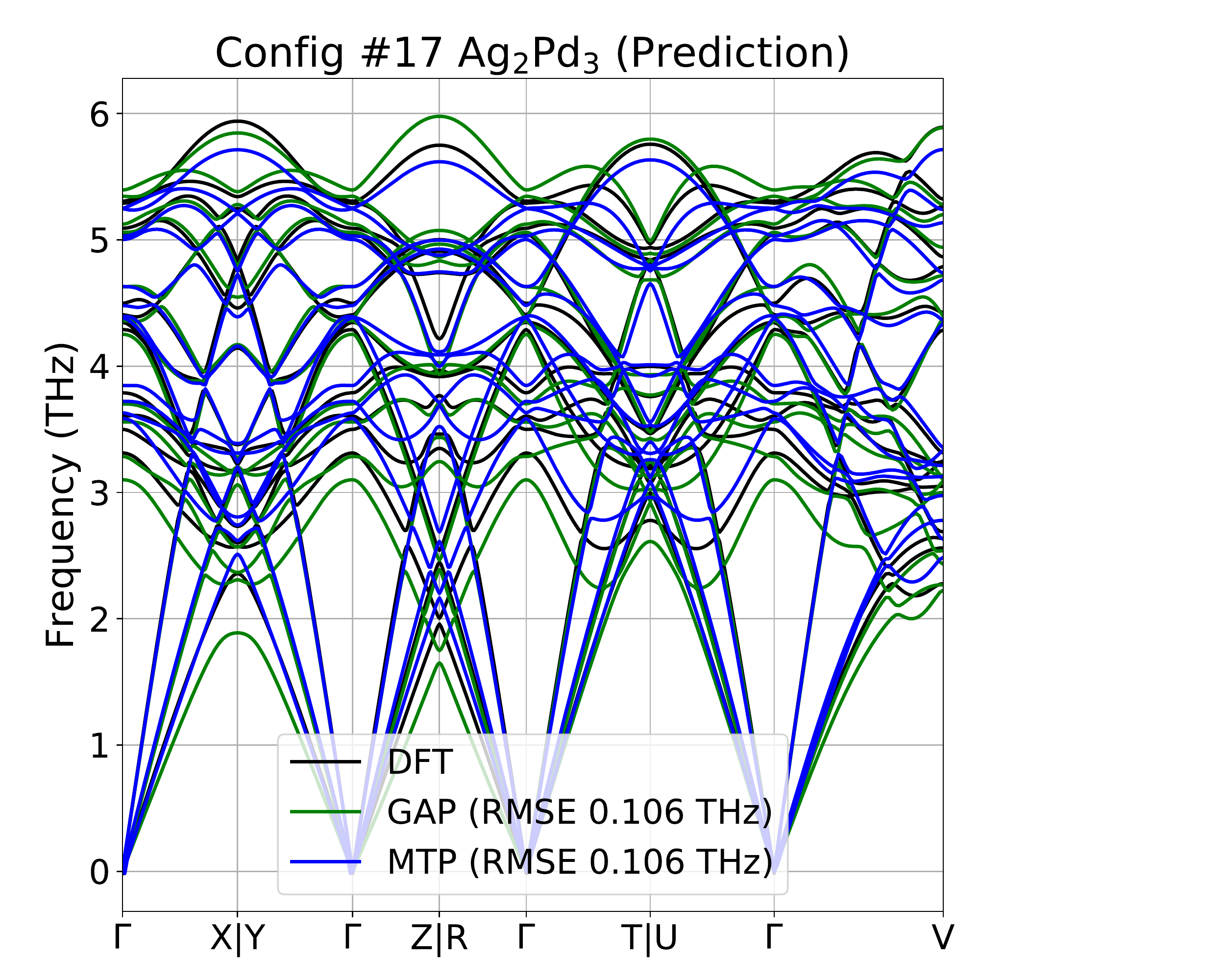}} \\
    \subfloat[]{\includegraphics[width = 3.2in]{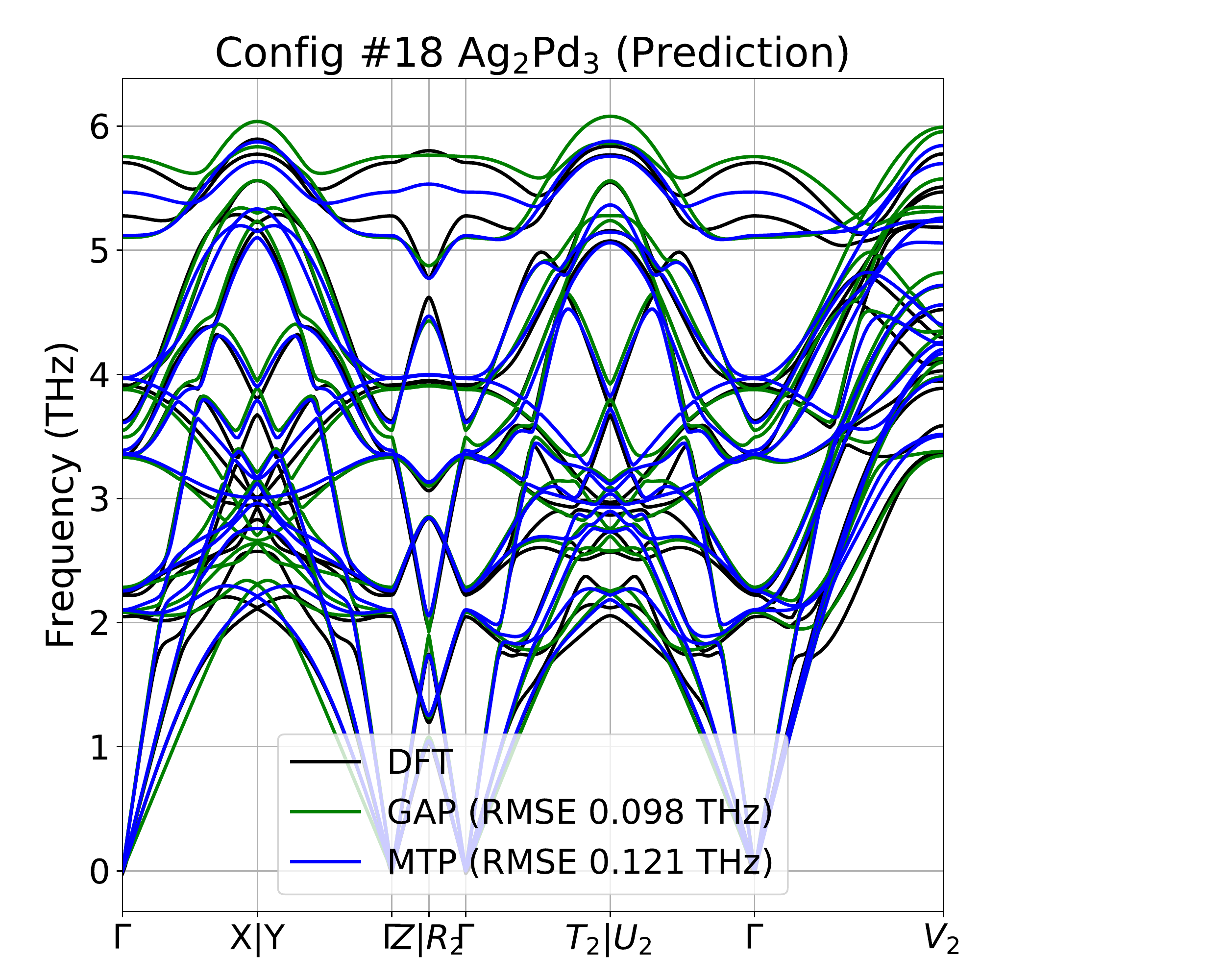}} &
    \subfloat[]{\includegraphics[width = 3.2in]{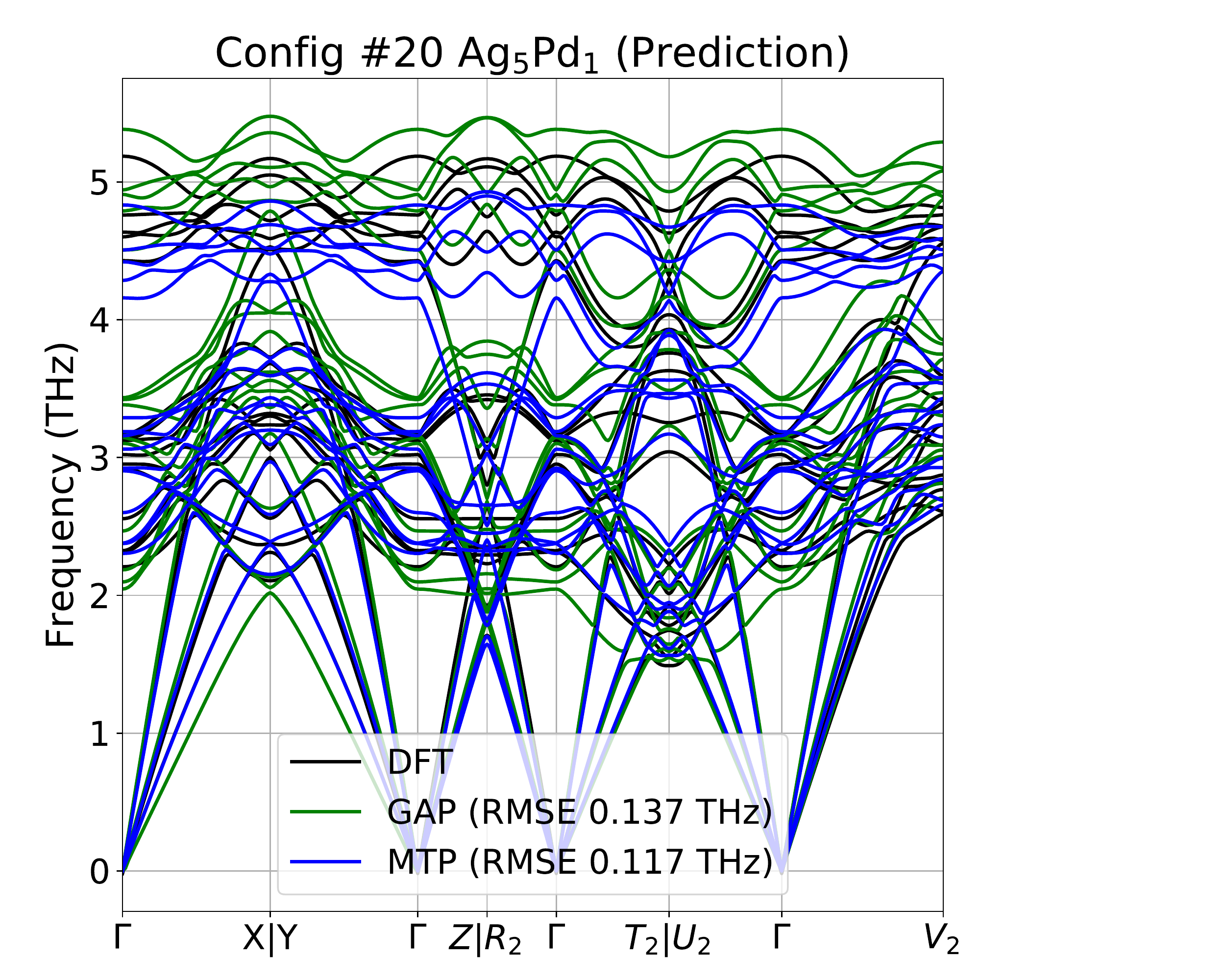}}
    \end{tabular}
    \end{minipage}  
    \end{adjustbox}
\end{figure*}
\begin{figure*}
    \begin{adjustbox}{rotate=0}
    \begin{minipage}{\textwidth}  
    \begin{tabular}{ccc}
    \subfloat[]{\includegraphics[width = 3.2in]{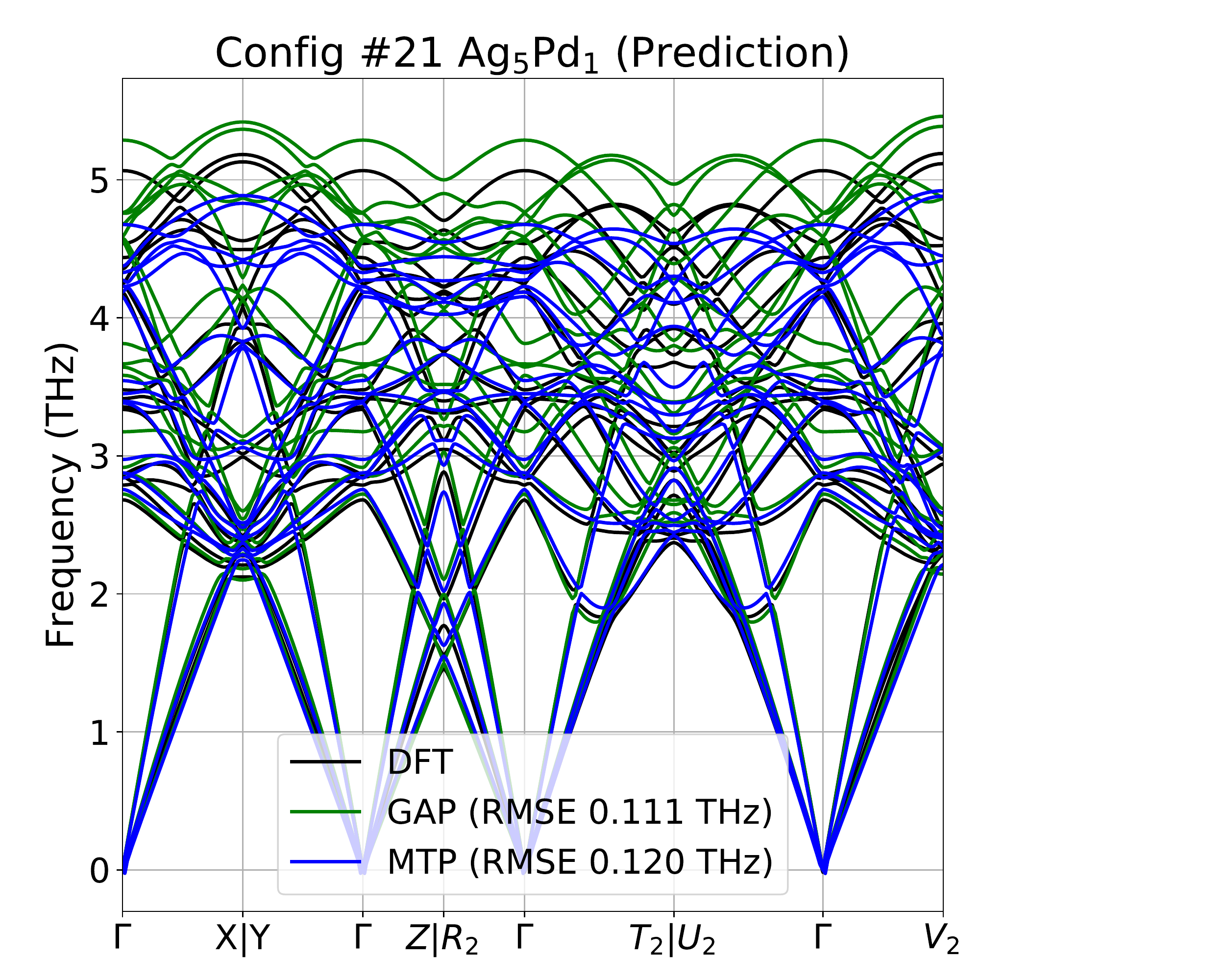}} &
    \subfloat[]{\includegraphics[width = 3.2in]{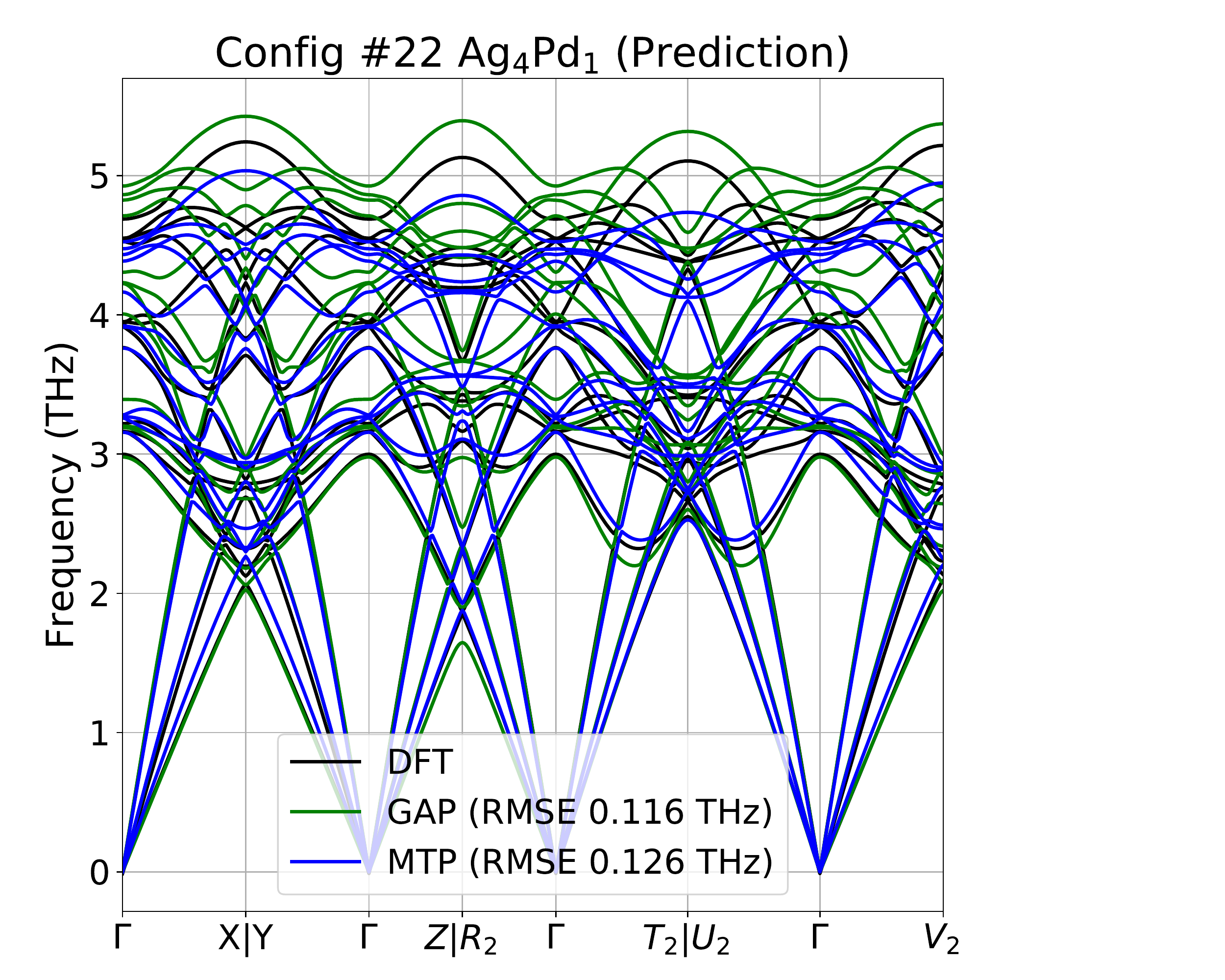}} \\
    \subfloat[]{\includegraphics[width = 3.2in]{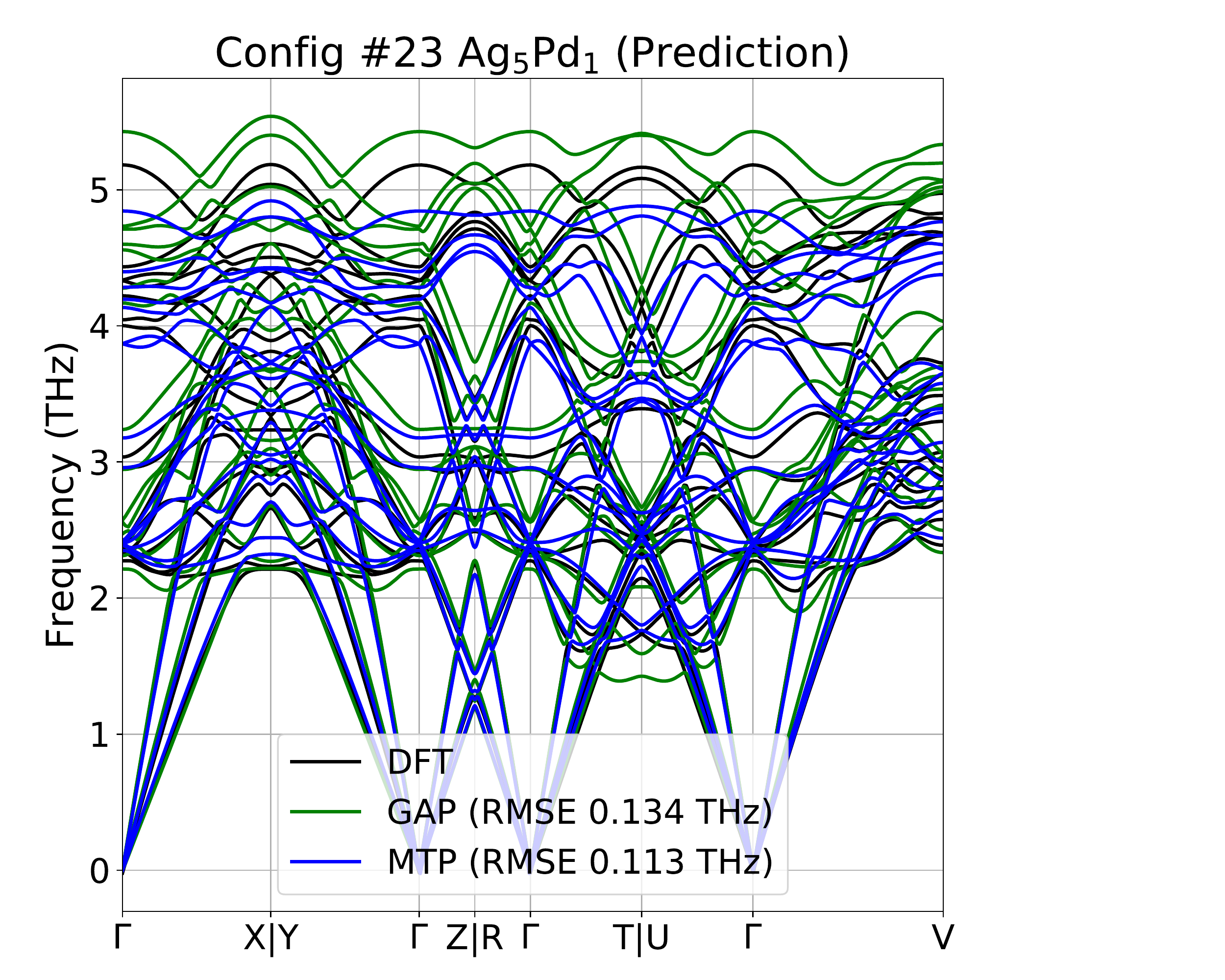}} &
    \subfloat[]{\includegraphics[width = 3.2in]{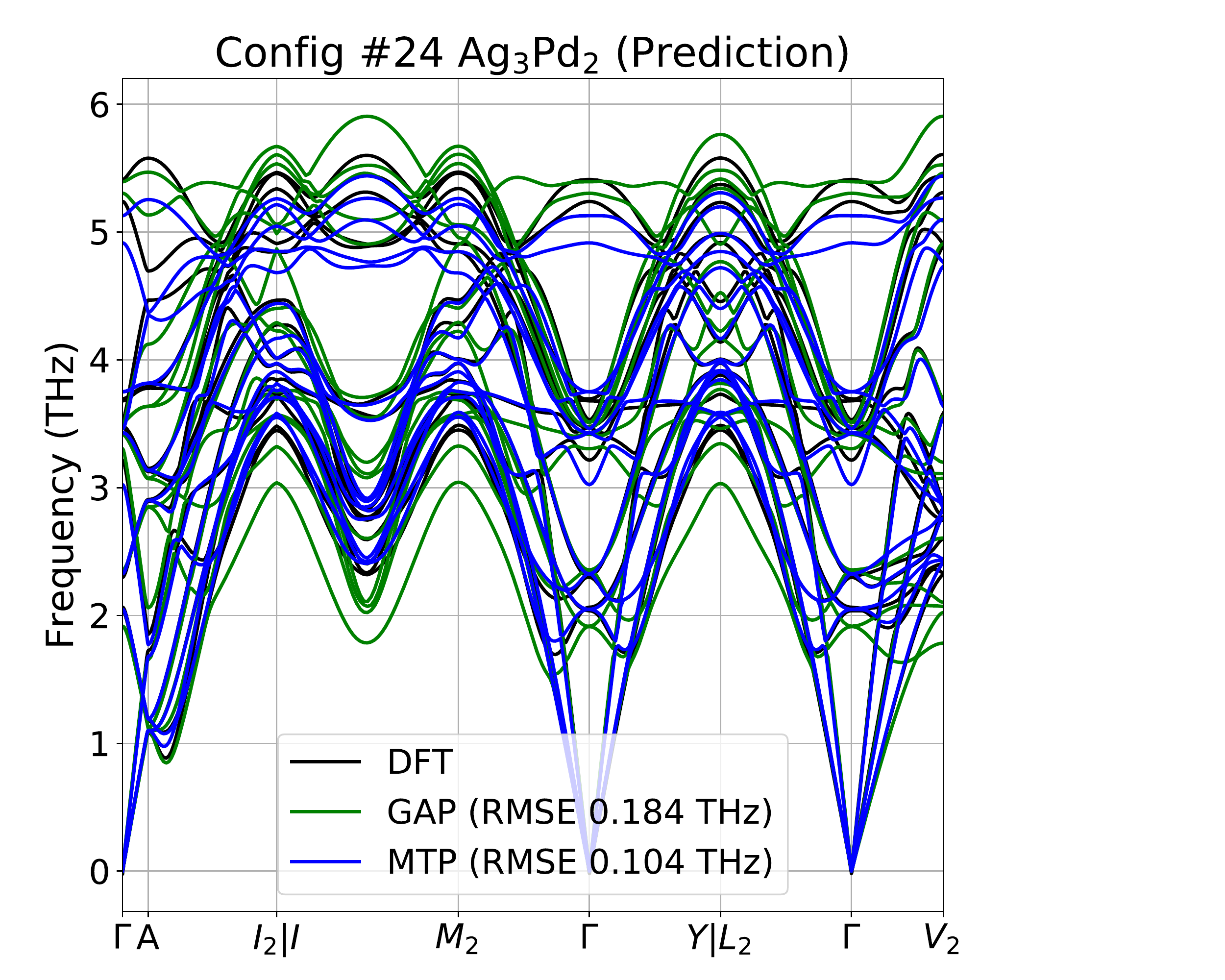}} \\
    \subfloat[]{\includegraphics[width = 3.2in]{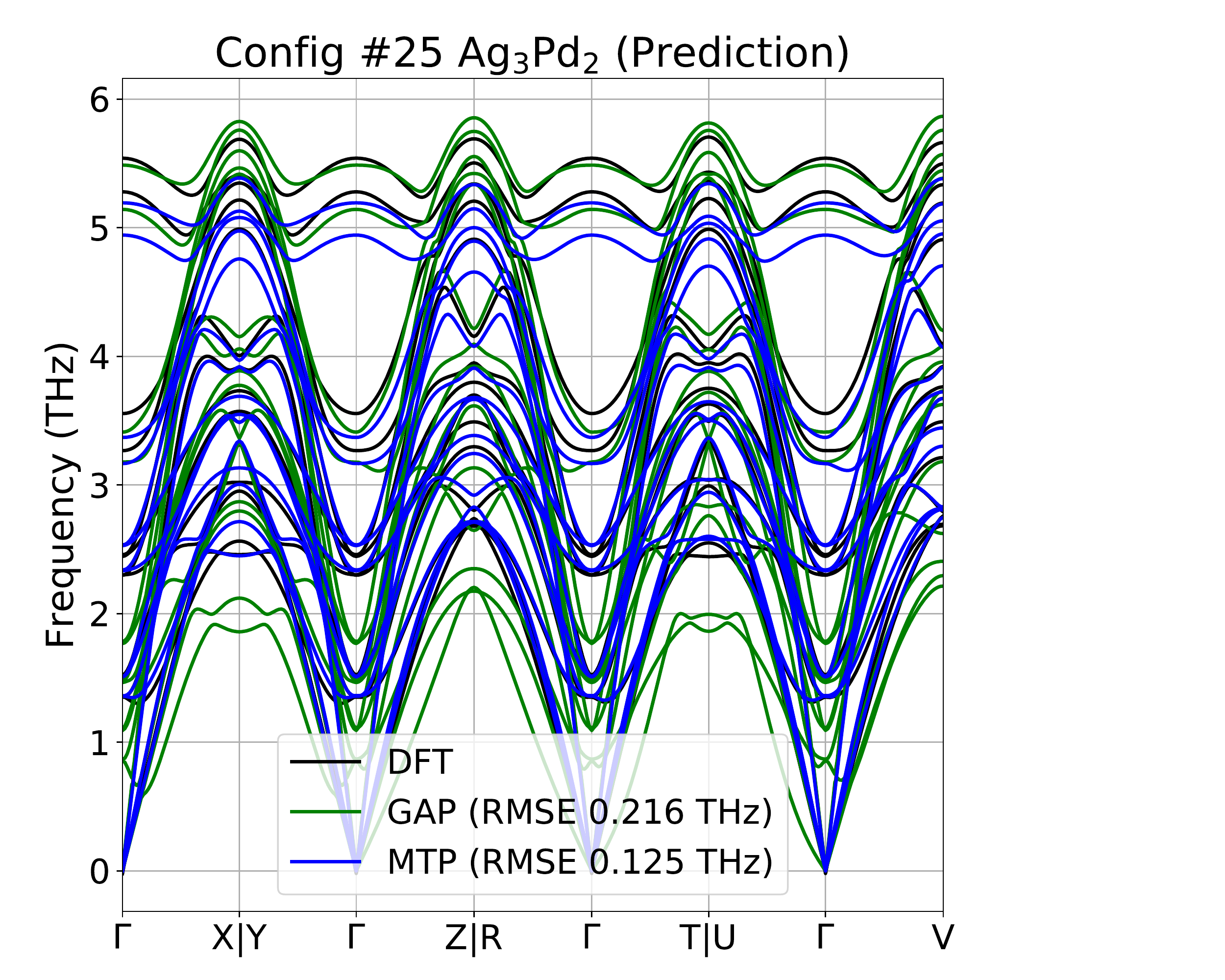}} &
    \subfloat[]{\includegraphics[width = 3.2in]{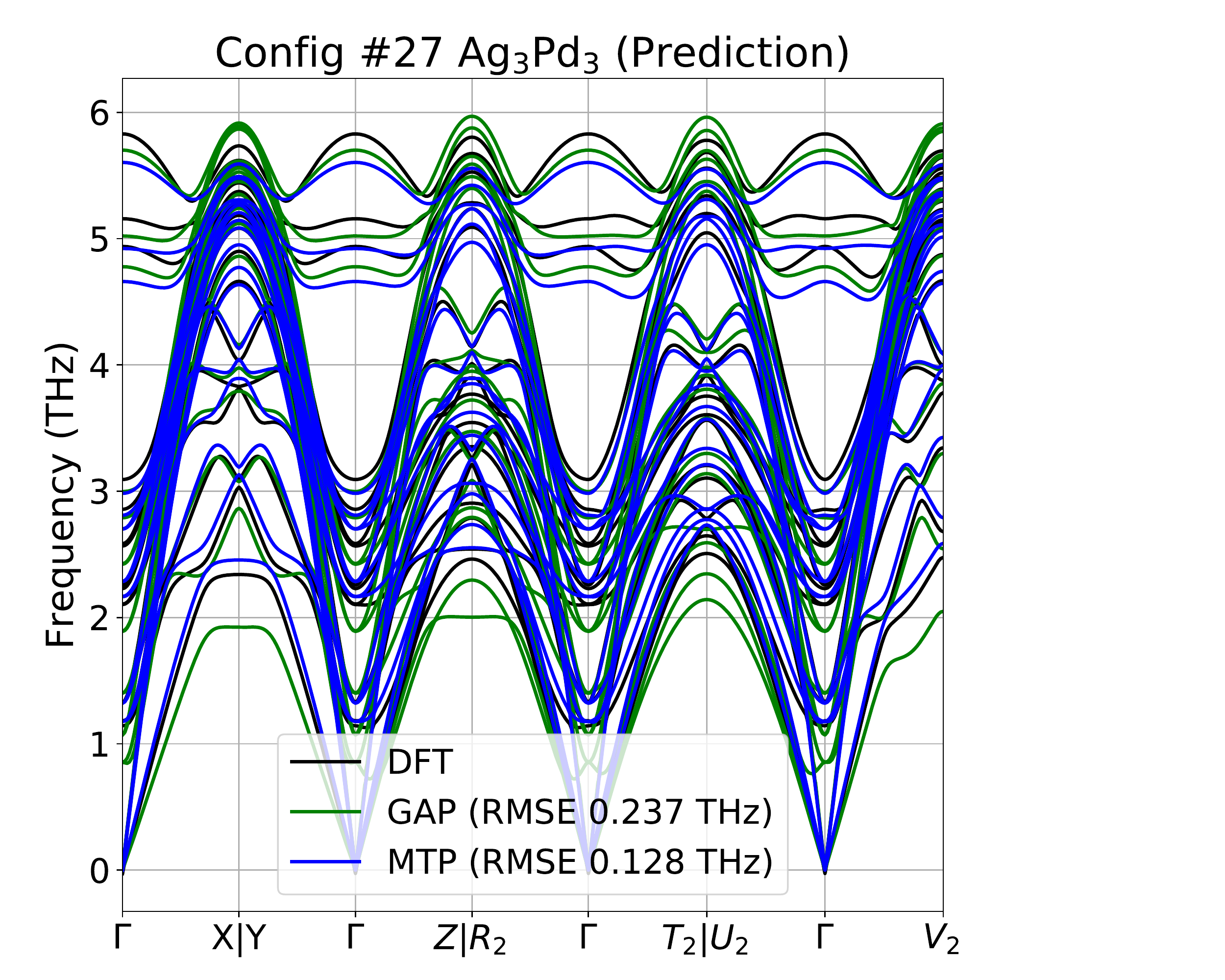}}
    \end{tabular}
    \end{minipage}  
    \end{adjustbox}
\end{figure*}
\begin{figure*}
    \begin{adjustbox}{rotate=0}
    \begin{minipage}{\textwidth}  
    \begin{tabular}{ccc}
    \subfloat[]{\includegraphics[width = 3.2in]{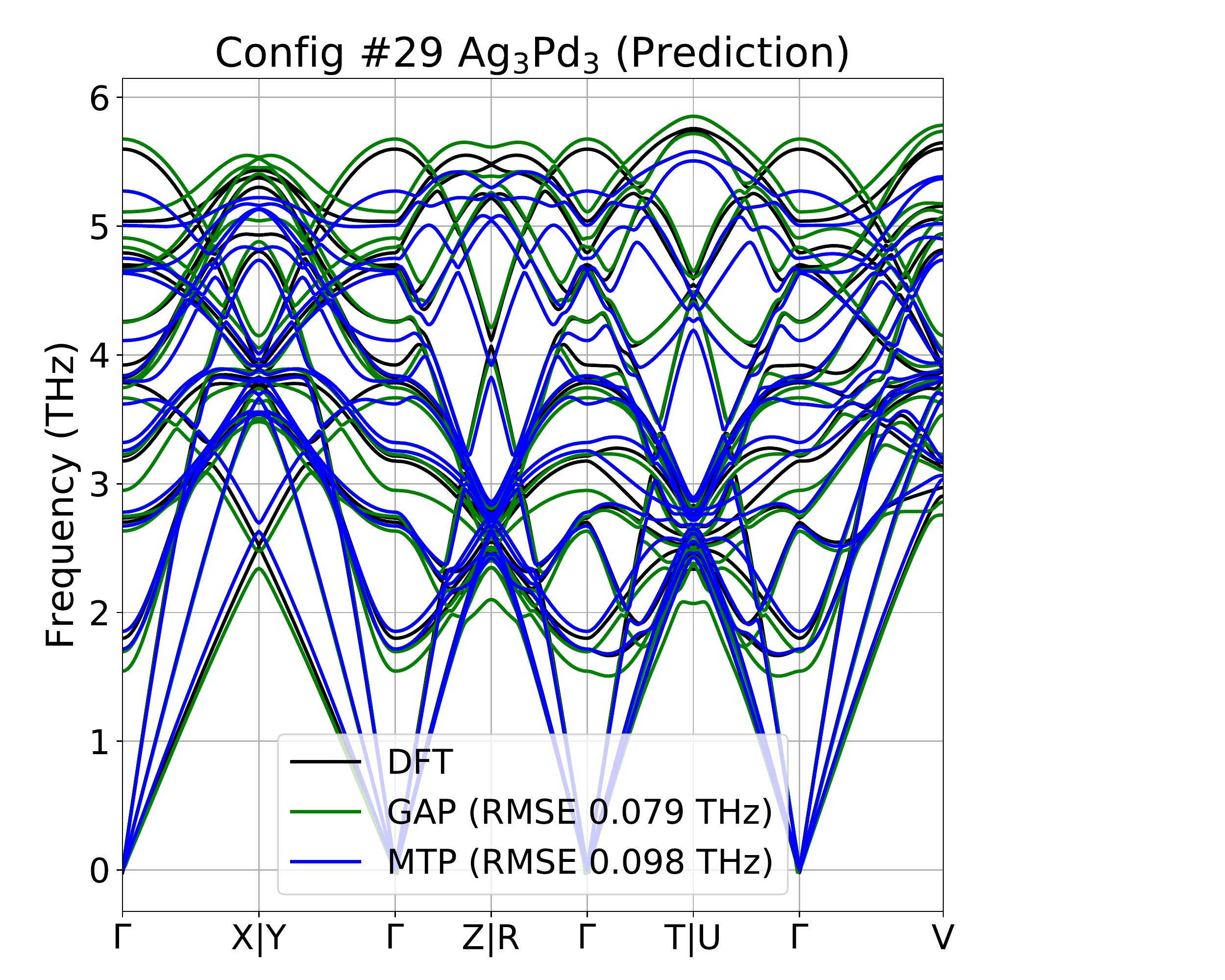}} &
    \subfloat[]{\includegraphics[width = 3.2in]{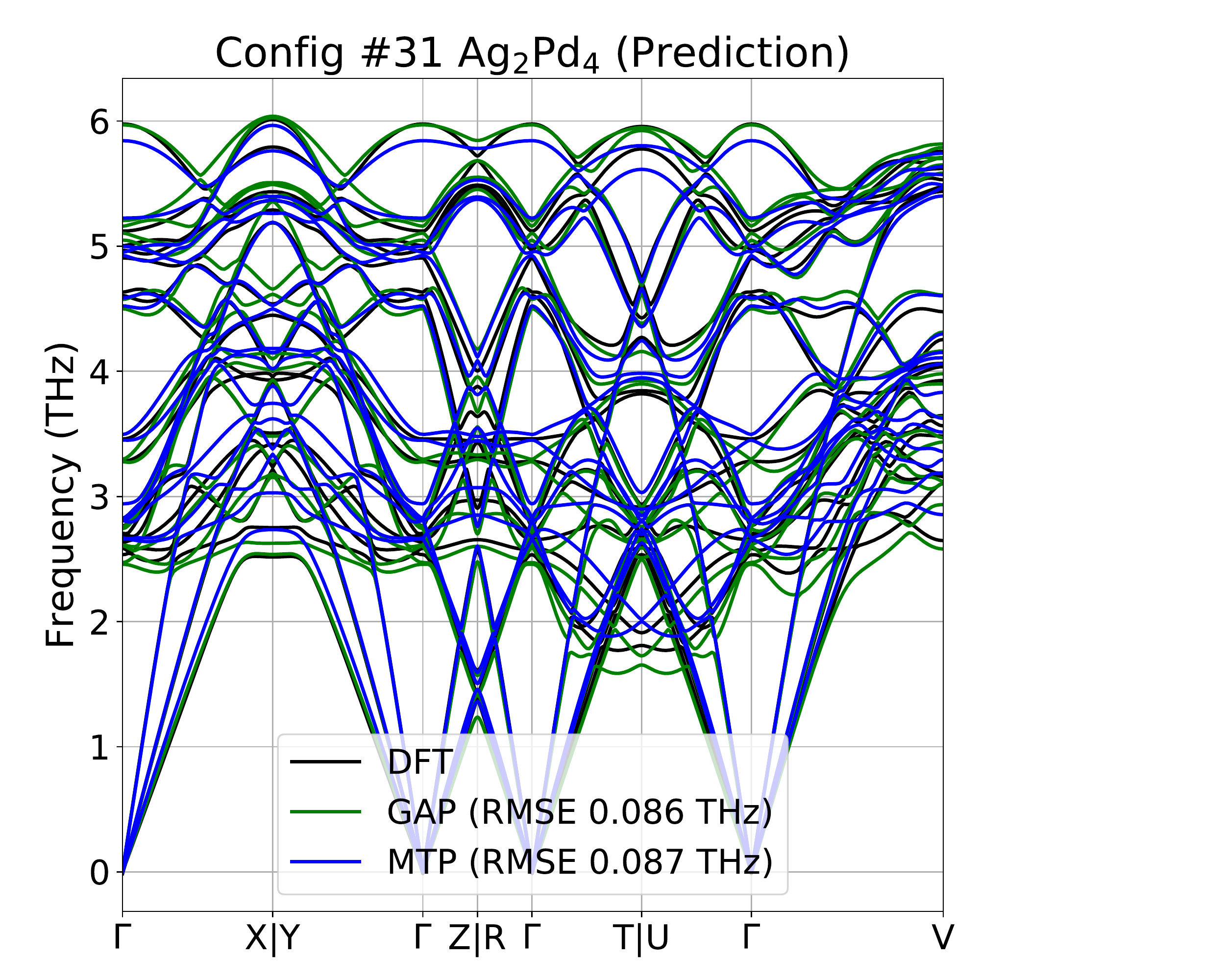}} \\
    \subfloat[]{\includegraphics[width = 3.2in]{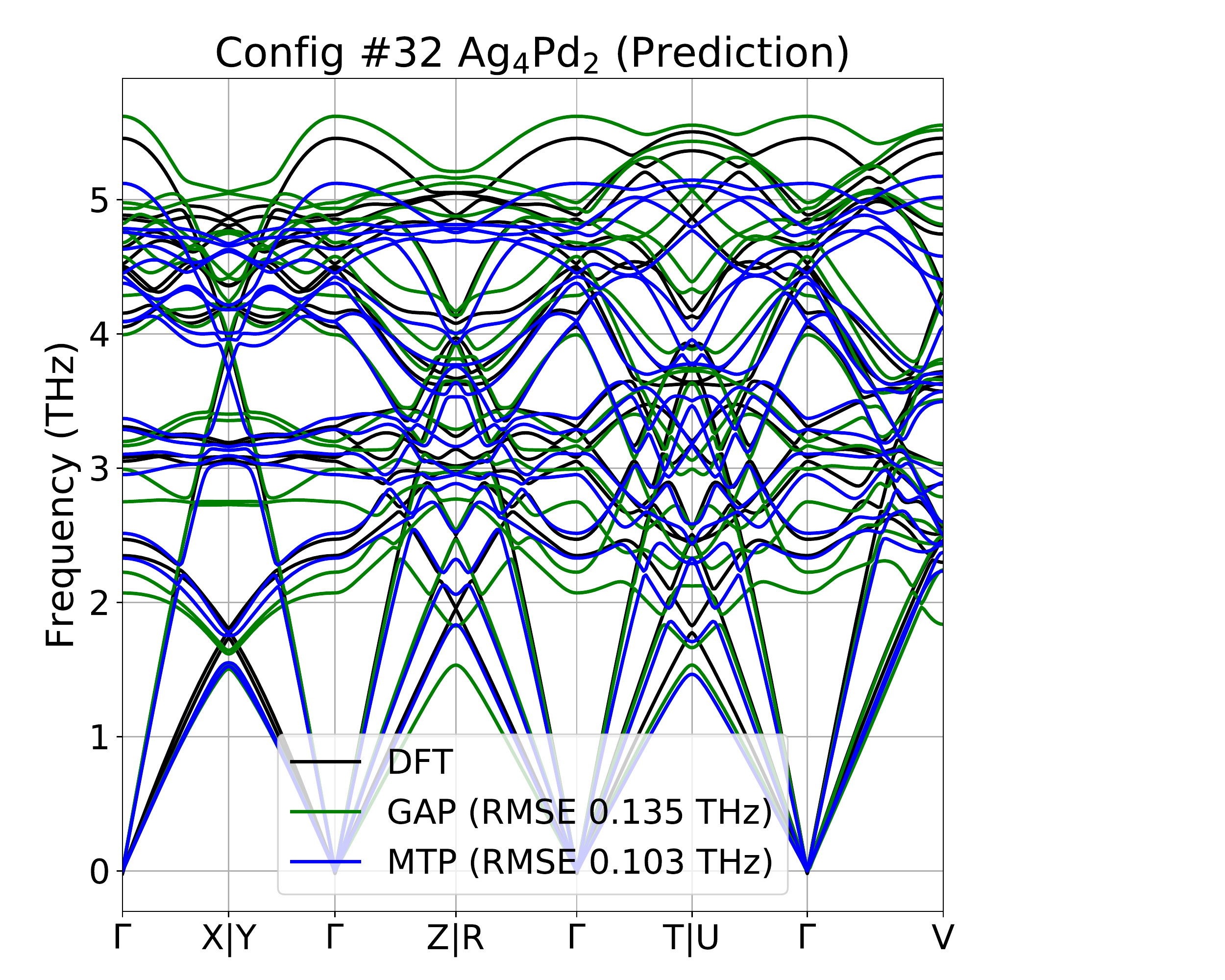}} &
    \subfloat[]{\includegraphics[width = 3.2in]{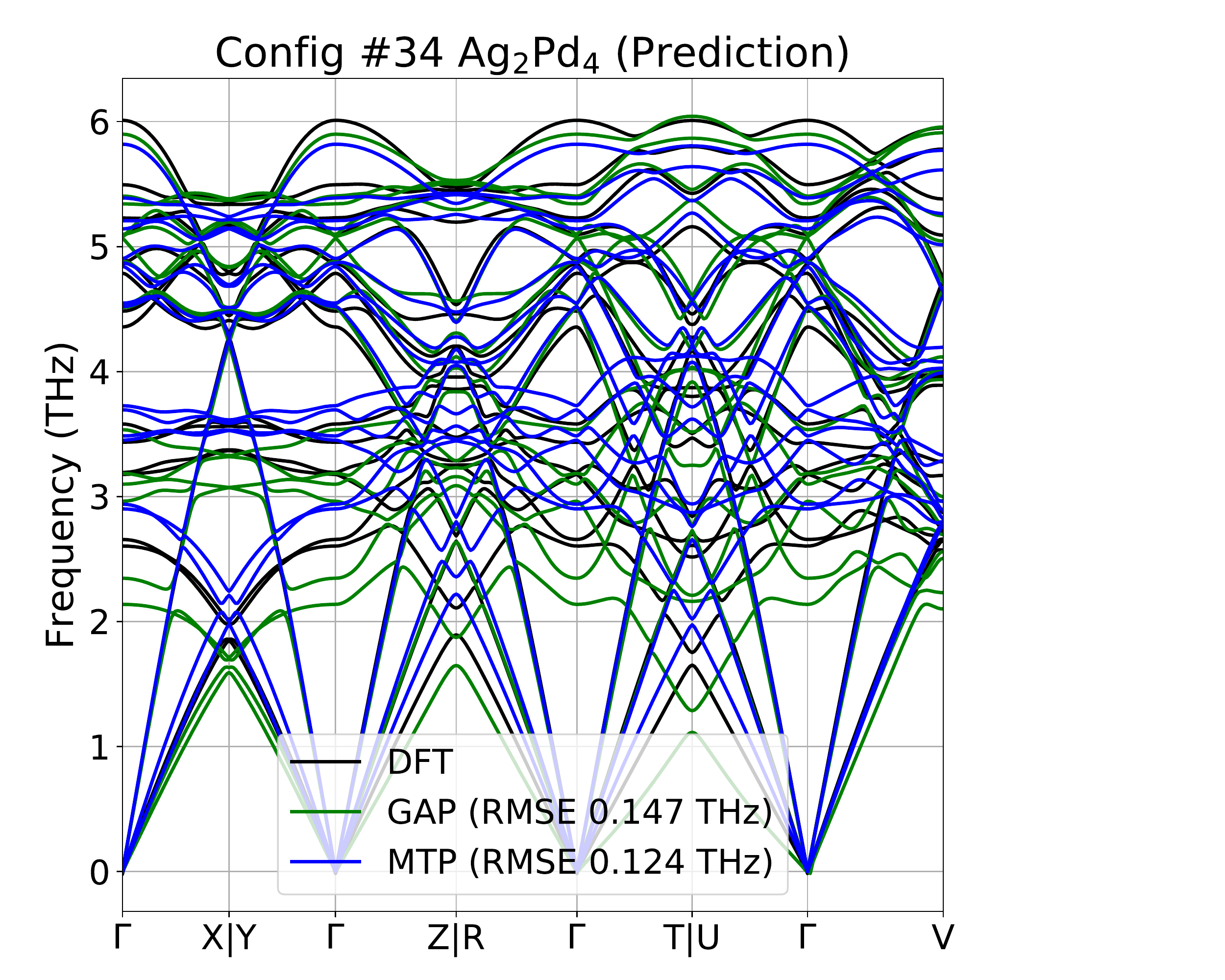}} \\
    \subfloat[]{\includegraphics[width = 3.2in]{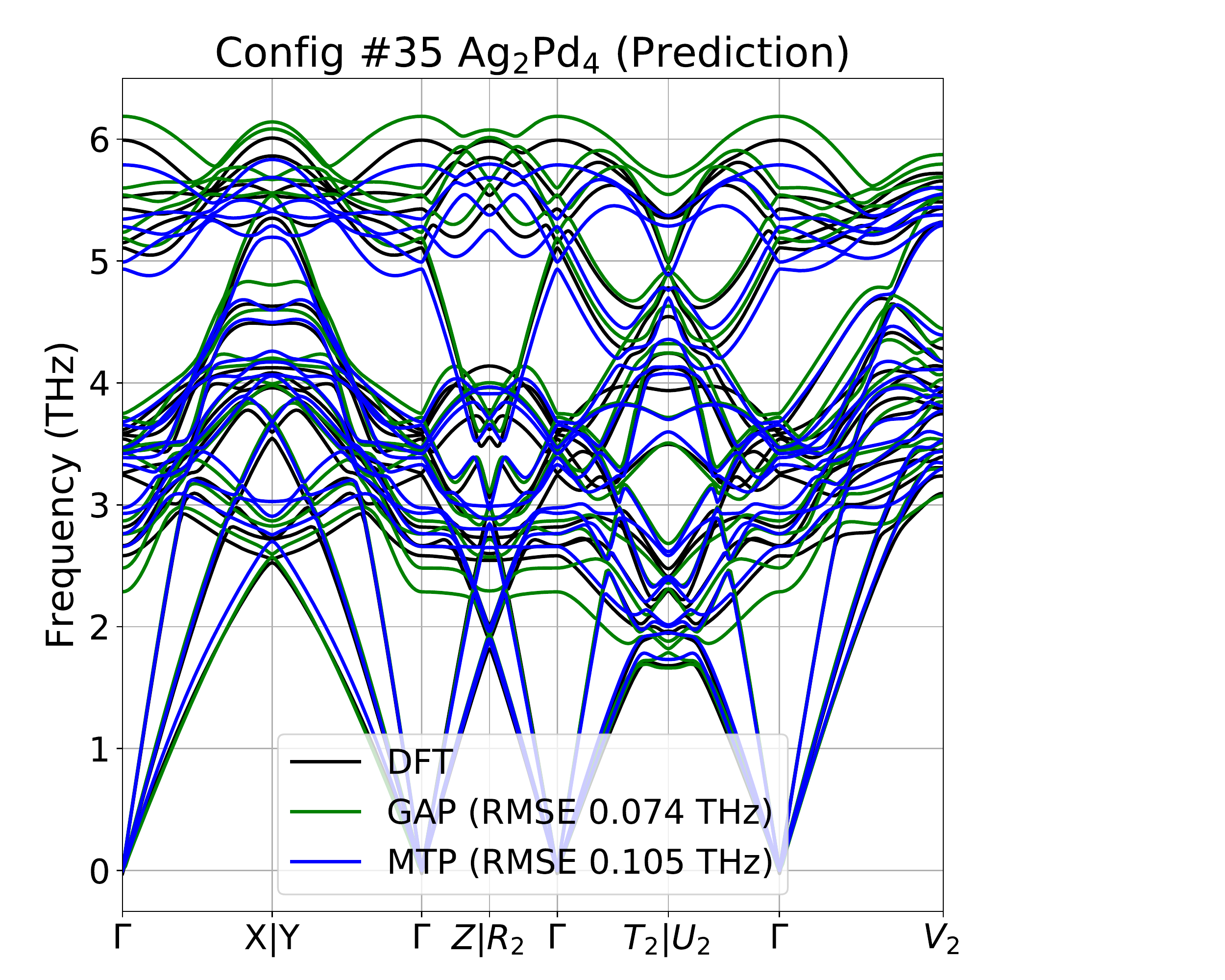}} &
    \subfloat[]{\includegraphics[width = 3.2in]{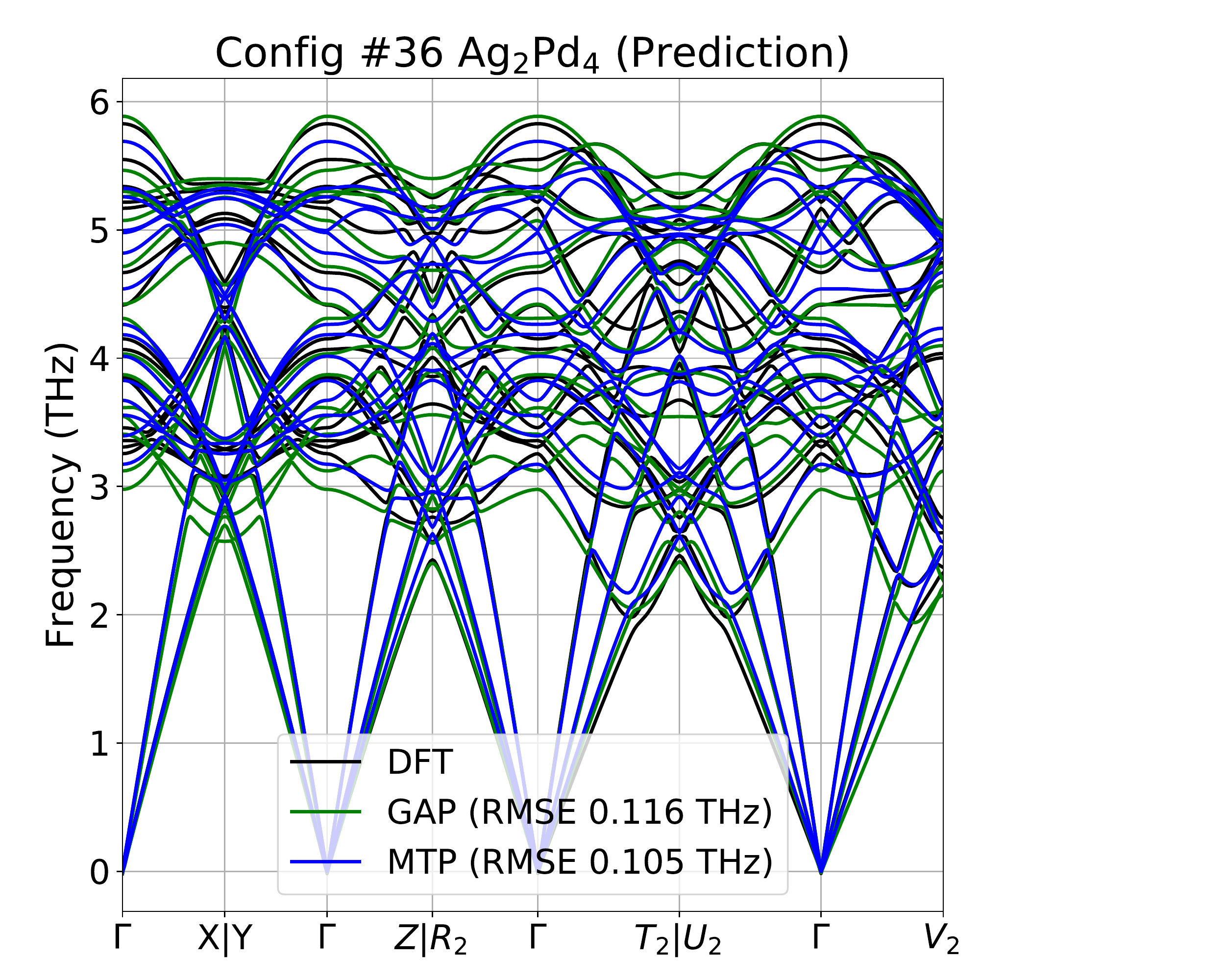}}
    \end{tabular}
    \end{minipage}  
    \end{adjustbox}
\end{figure*}
\begin{figure*}
    \begin{adjustbox}{rotate=0}
    \begin{minipage}{\textwidth}  
    \begin{tabular}{ccc}
    \subfloat[]{\includegraphics[width = 3.2in]{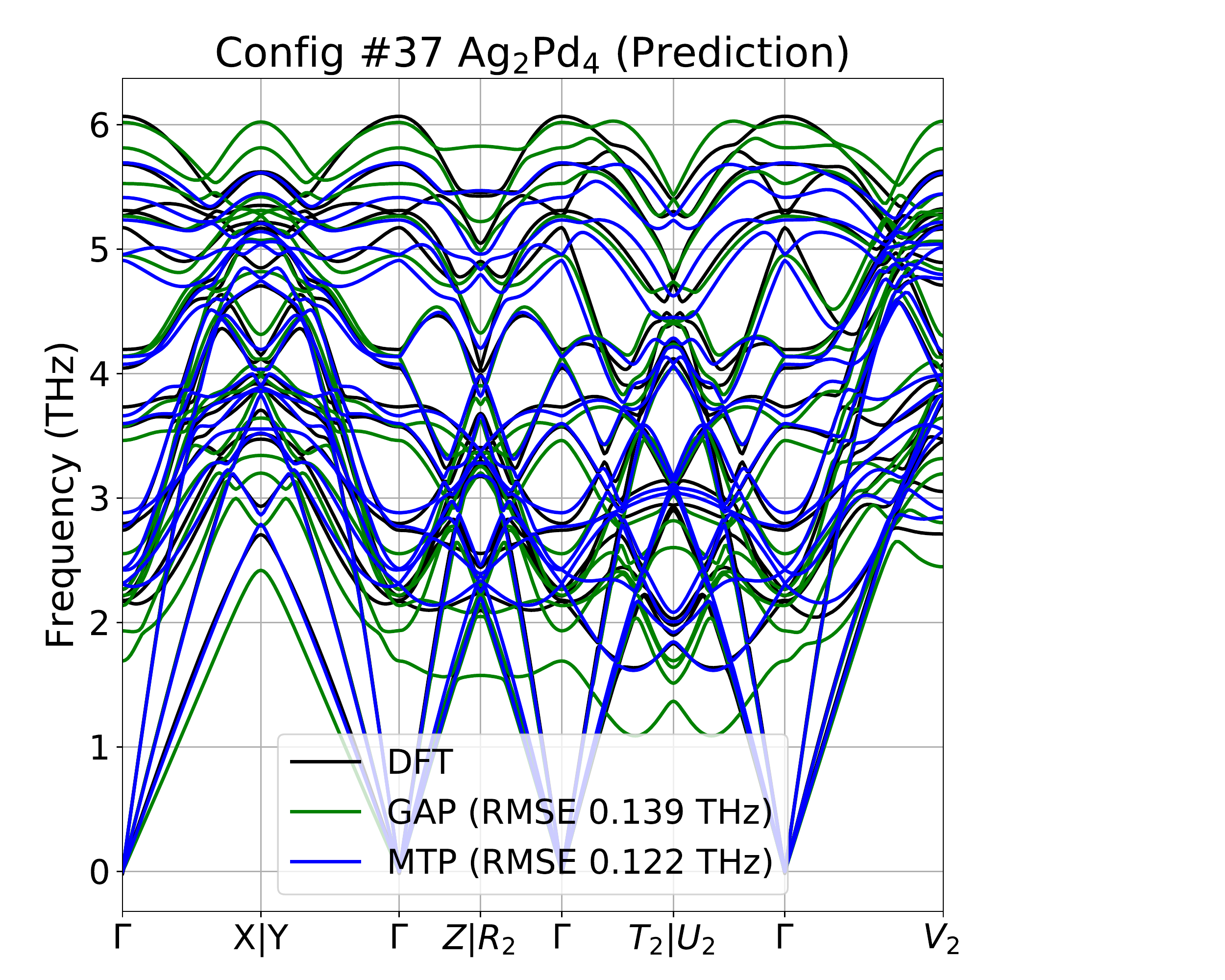}} &
    \subfloat[]{\includegraphics[width = 3.2in]{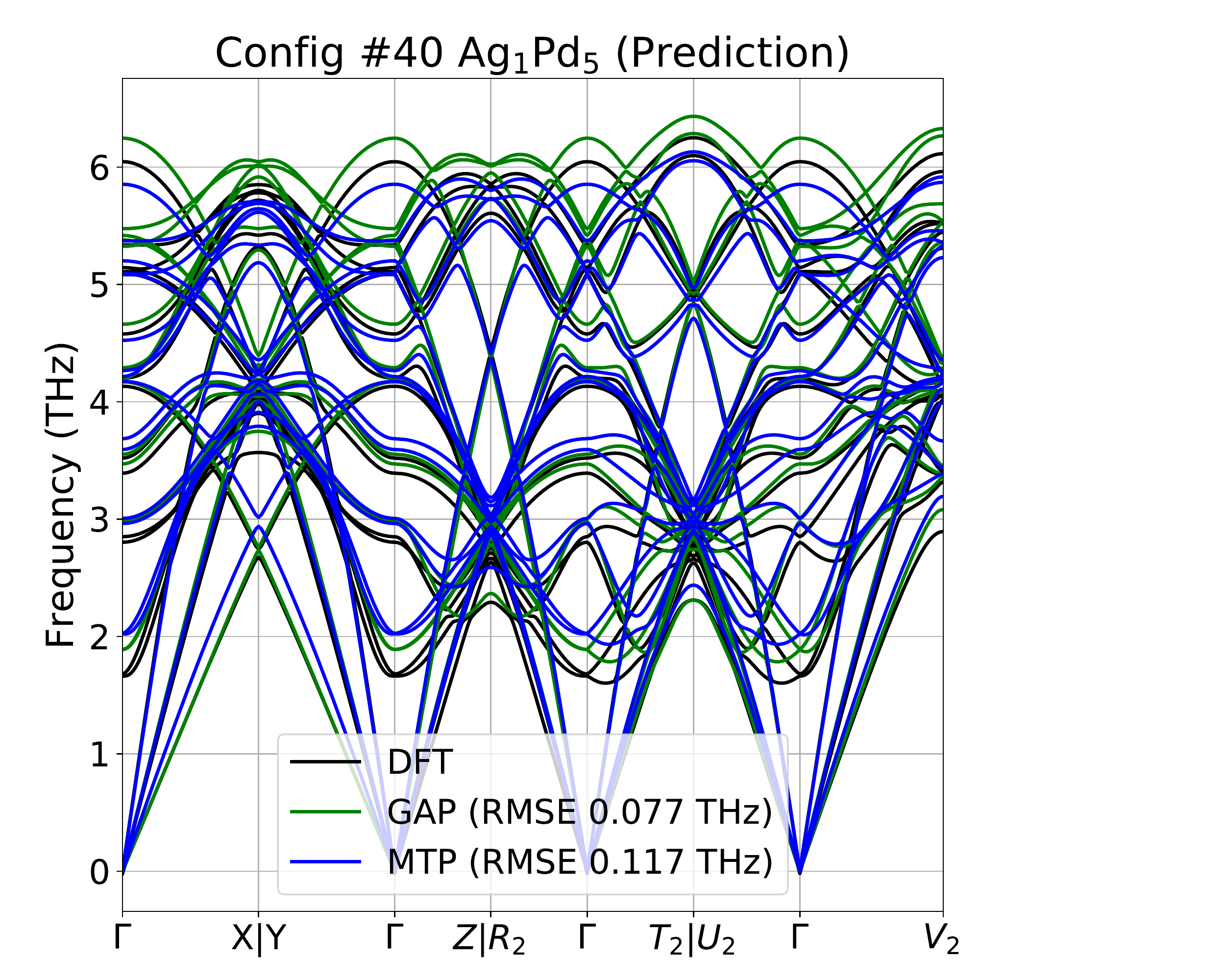}} \\
    \subfloat[]{\includegraphics[width = 3.2in]{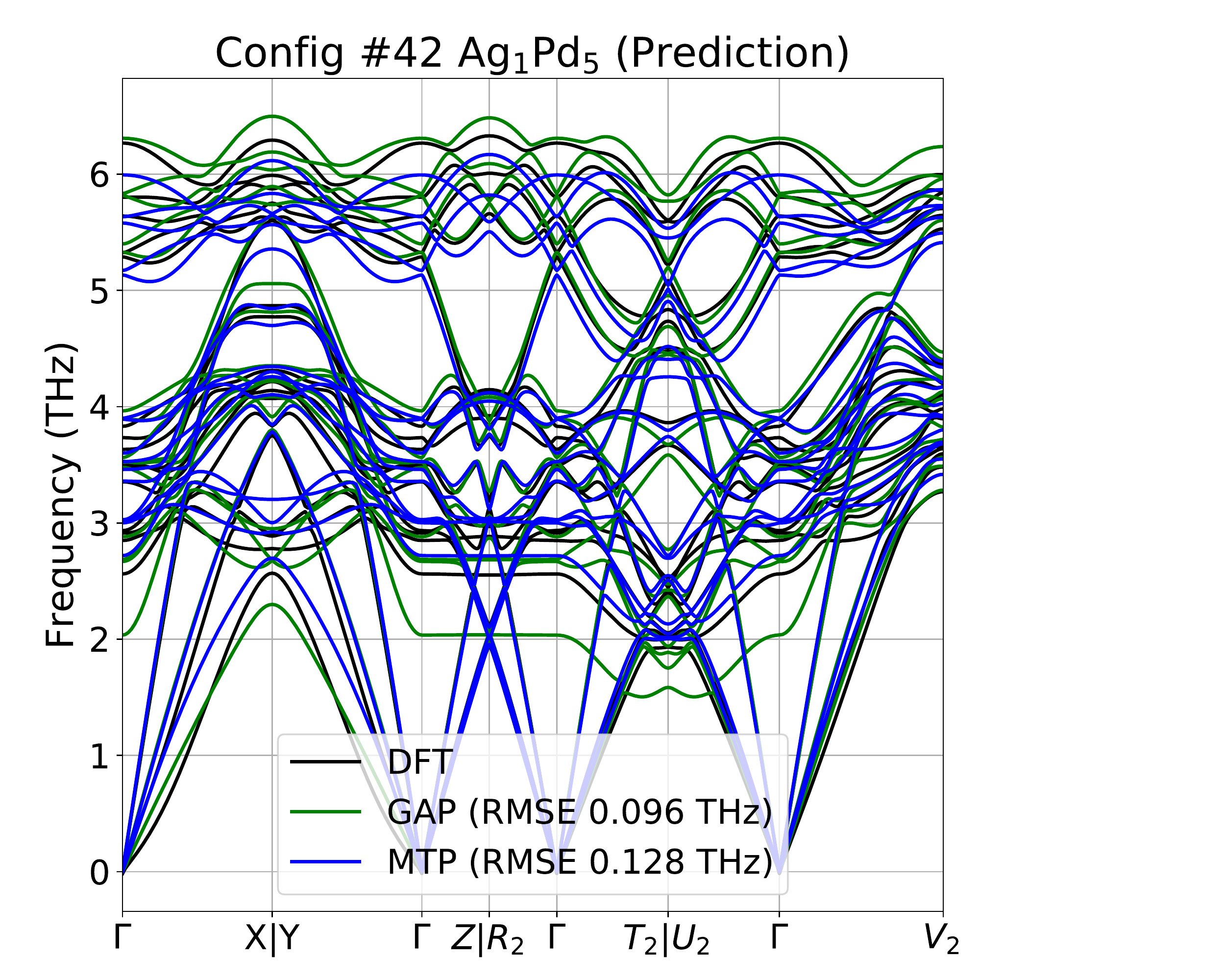}} &
    \subfloat[]{\includegraphics[width = 3.2in]{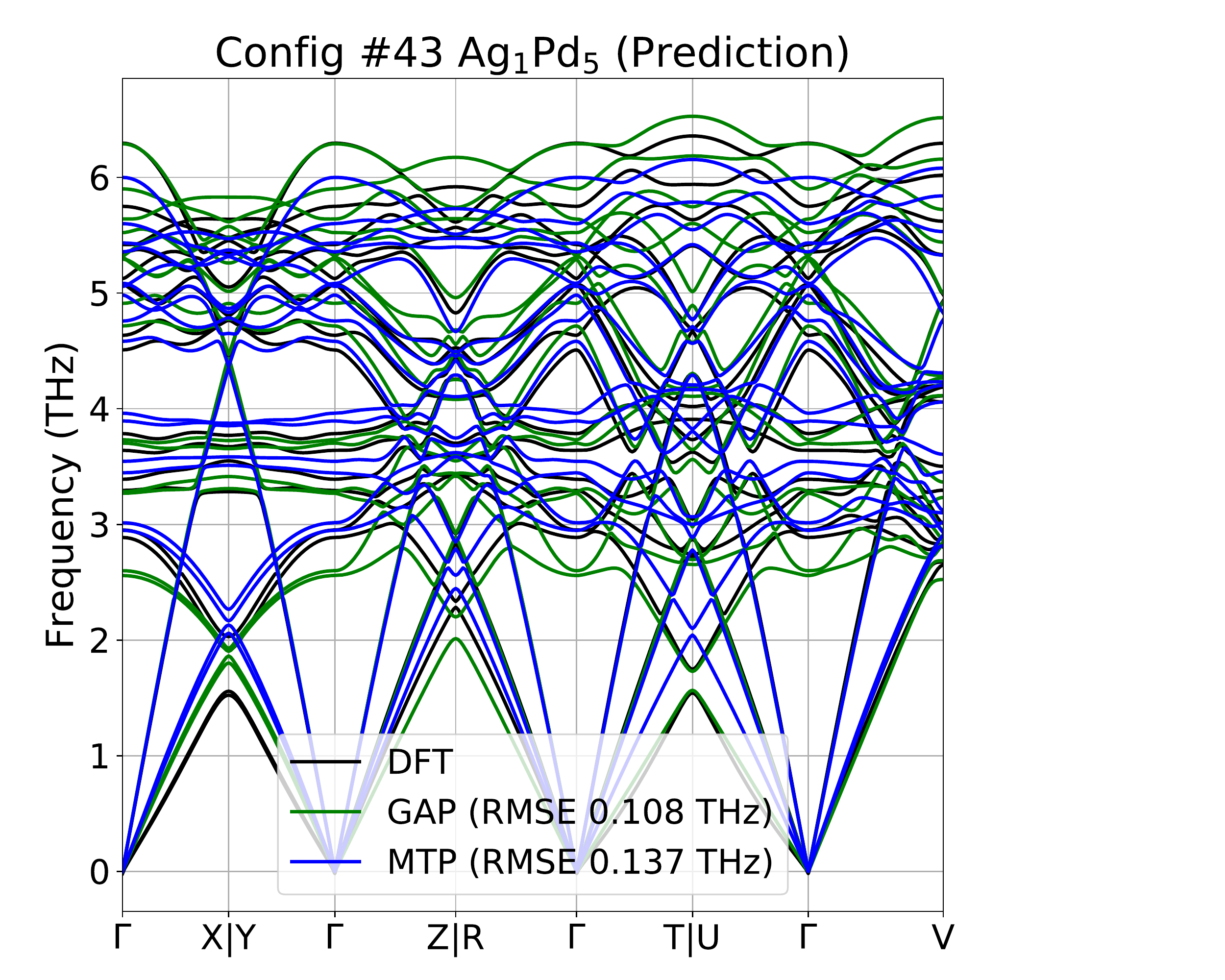}} \\
    \subfloat[]{\includegraphics[width = 3.2in]{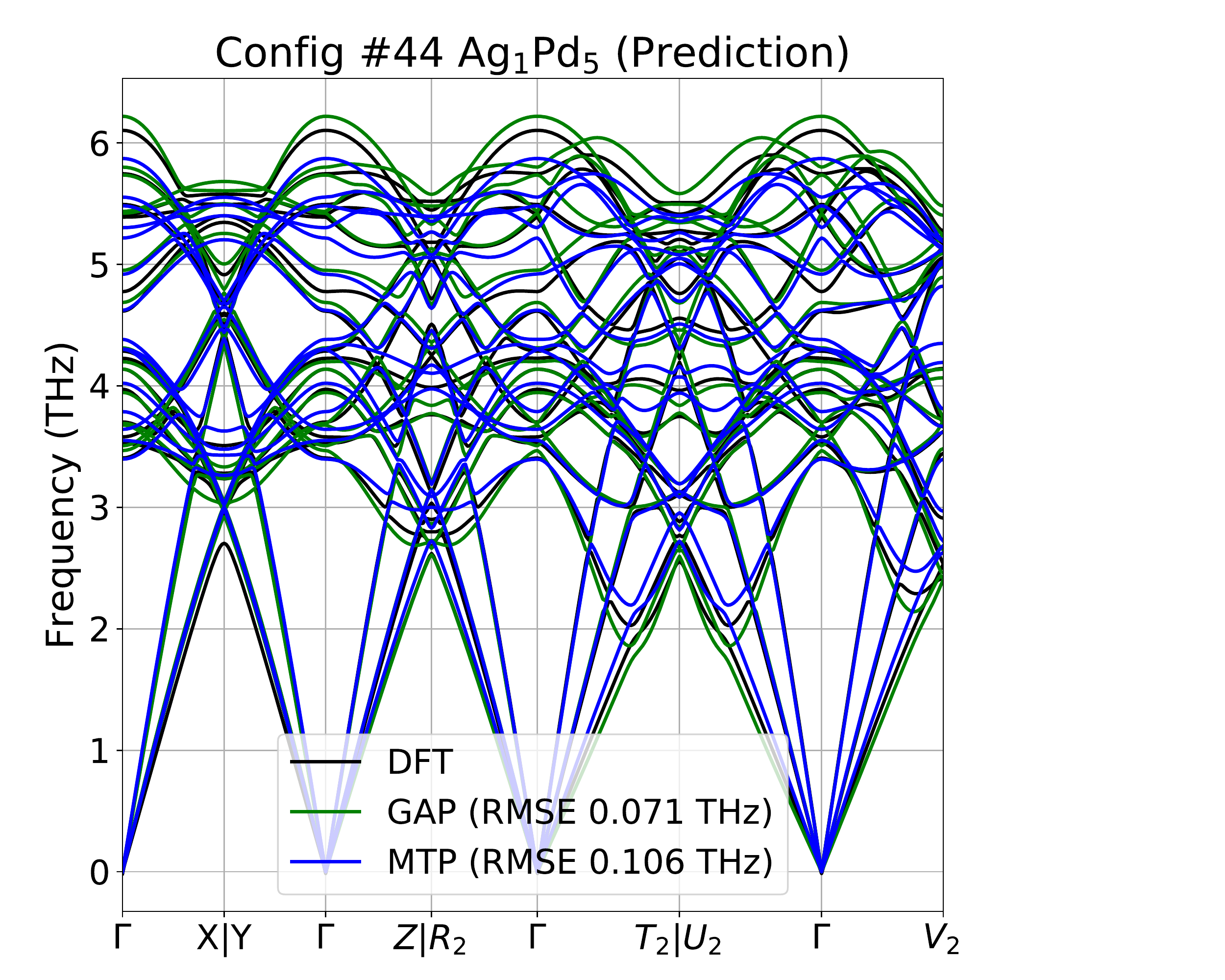}} &
    \subfloat[]{\includegraphics[width = 3.2in]{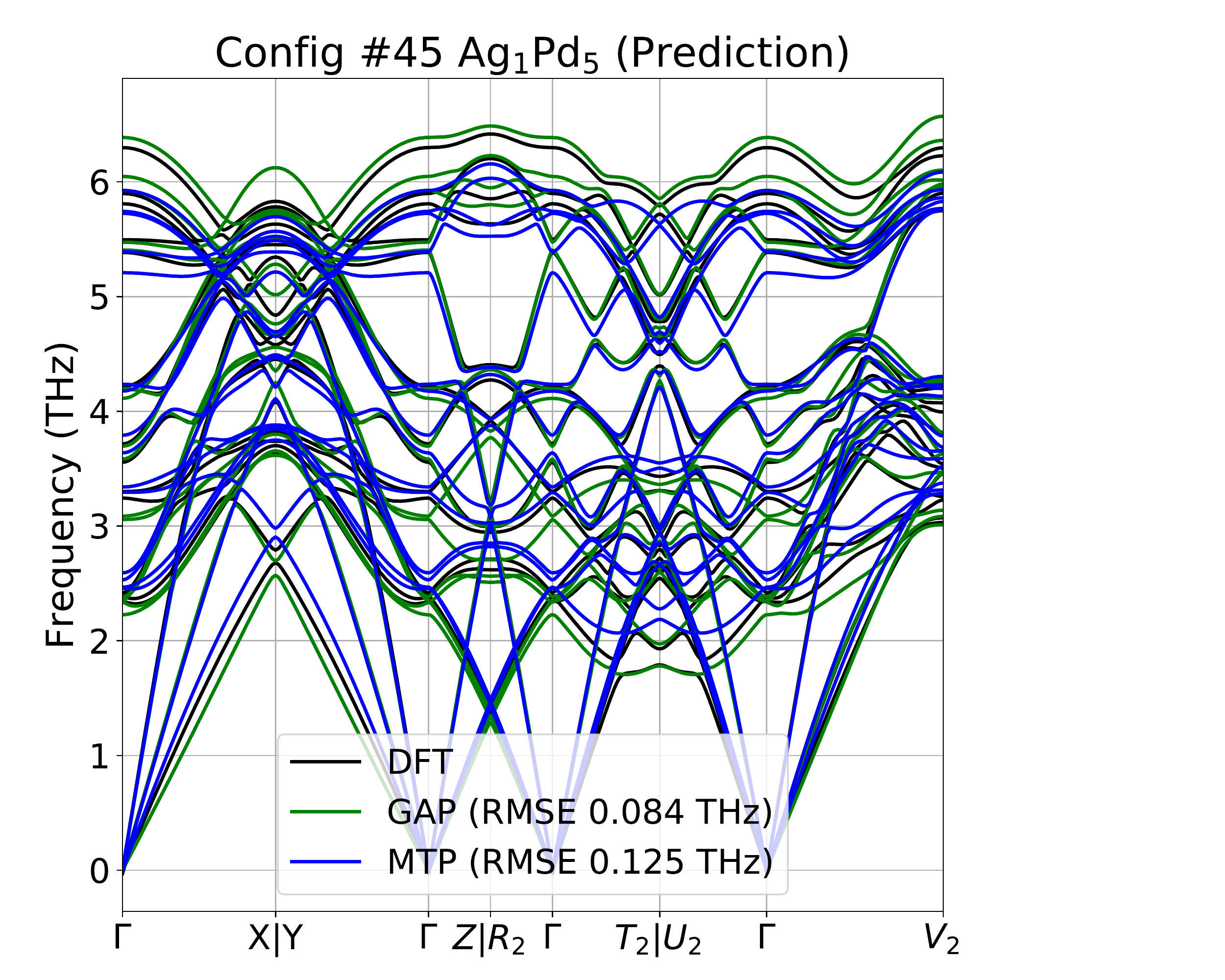}}
    \end{tabular}
    \end{minipage}  
    \end{adjustbox}
\end{figure*}
\begin{figure*}
    \begin{adjustbox}{rotate=0}
    \begin{minipage}{\textwidth}  
    \begin{tabular}{ccc}
    \subfloat[]{\includegraphics[width = 3.2in]{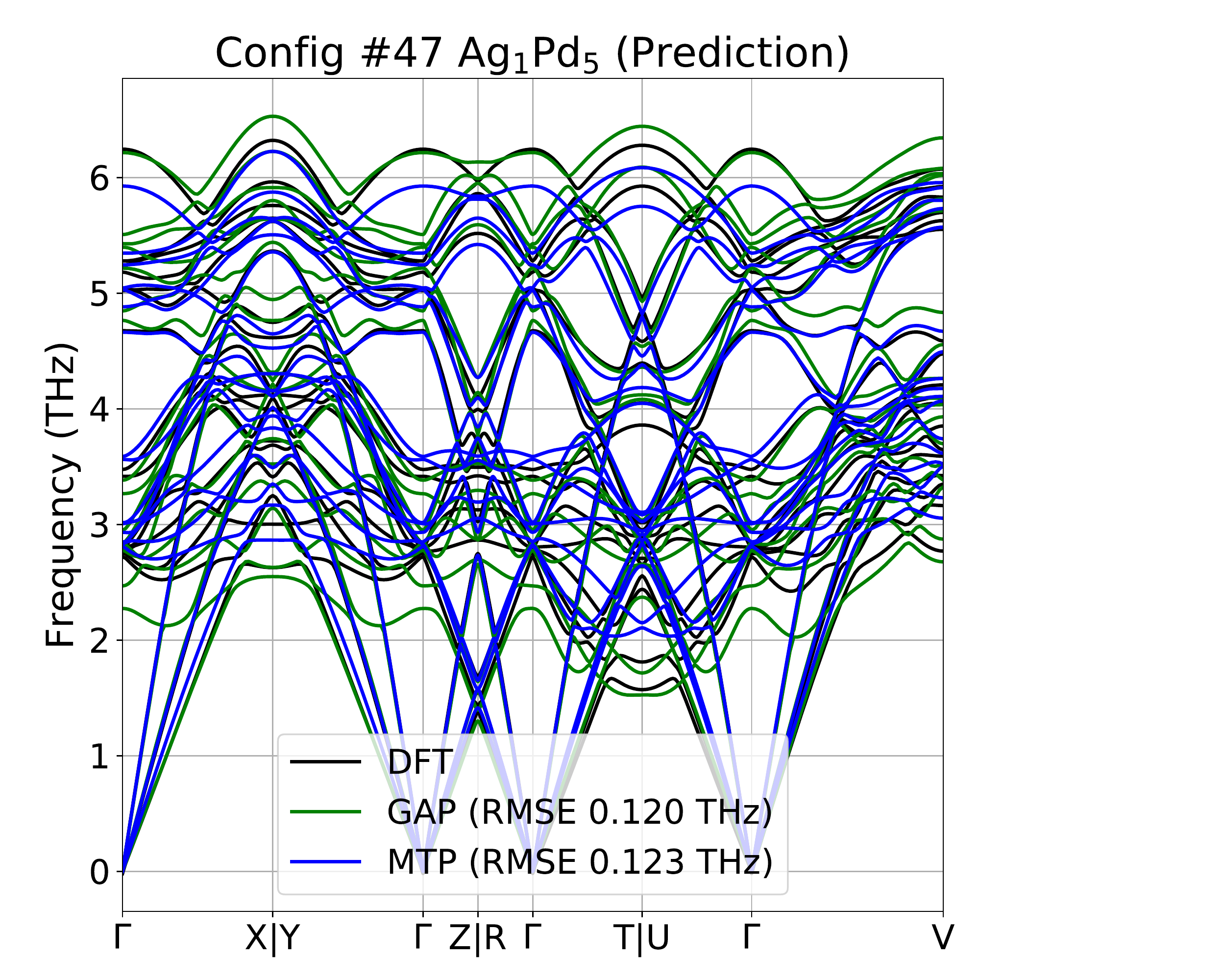}} &
    \subfloat[]{\includegraphics[width = 3.2in]{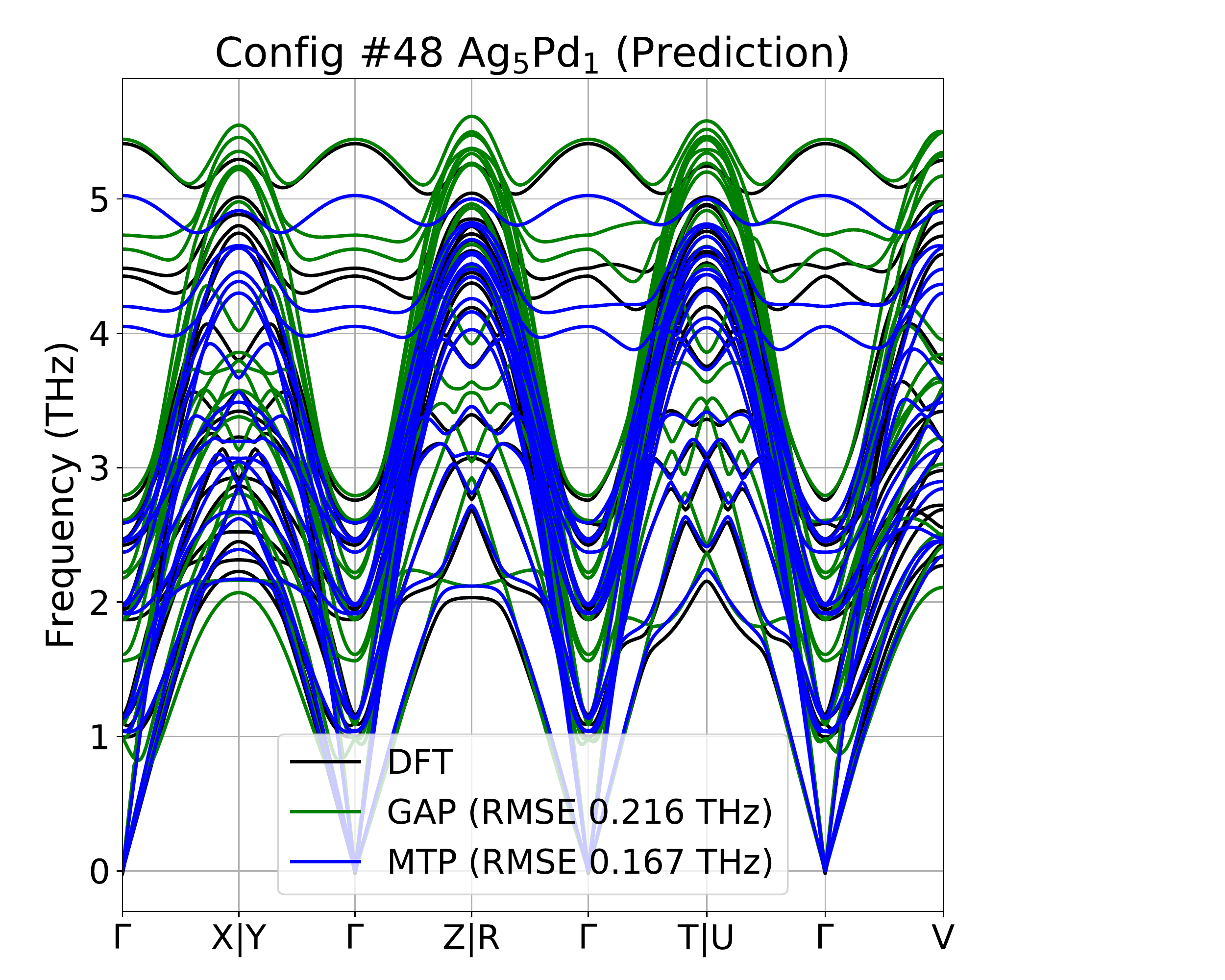}} \\
    \subfloat[]{\includegraphics[width = 3.2in]{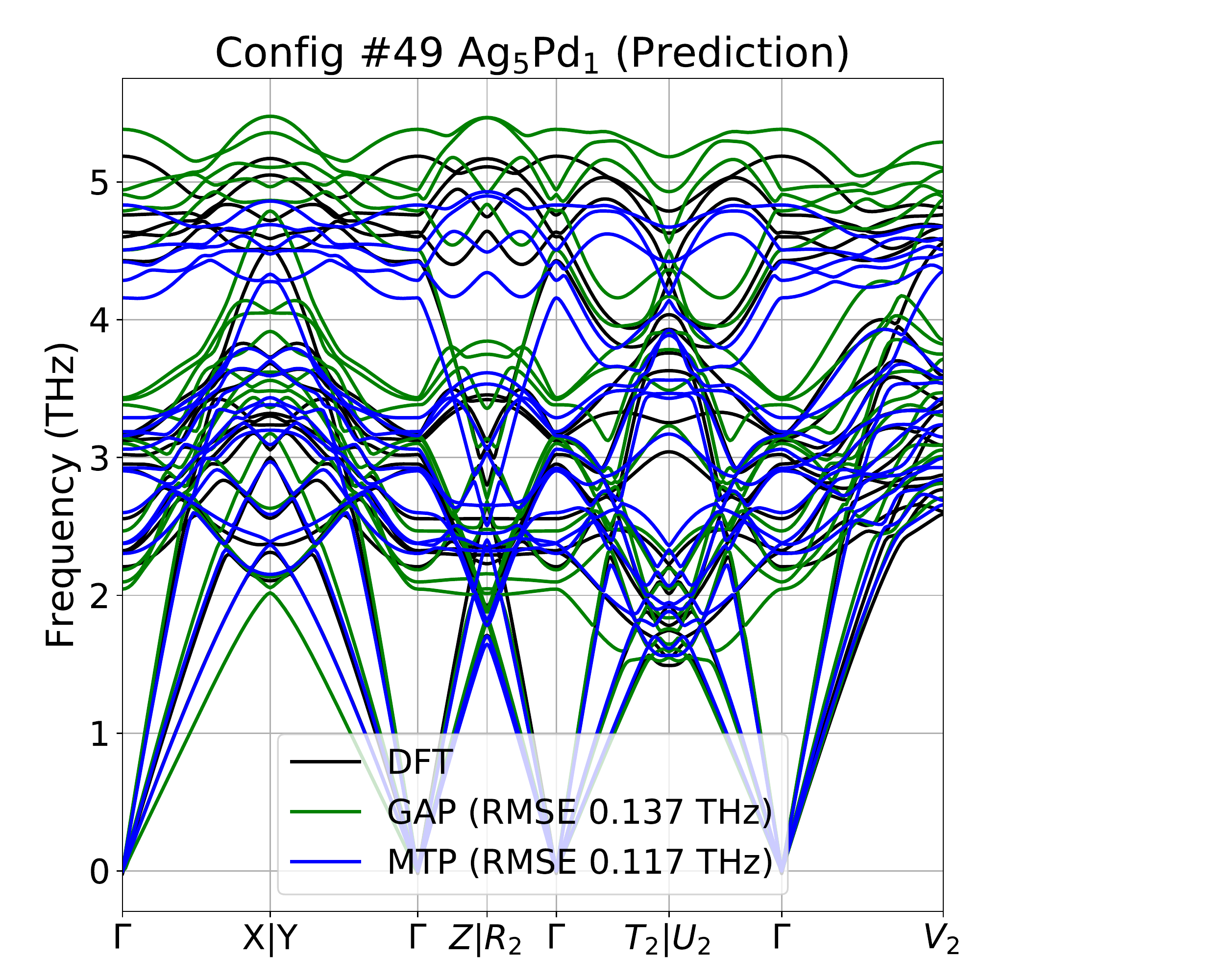}} &
    \subfloat[]{\includegraphics[width = 3.2in]{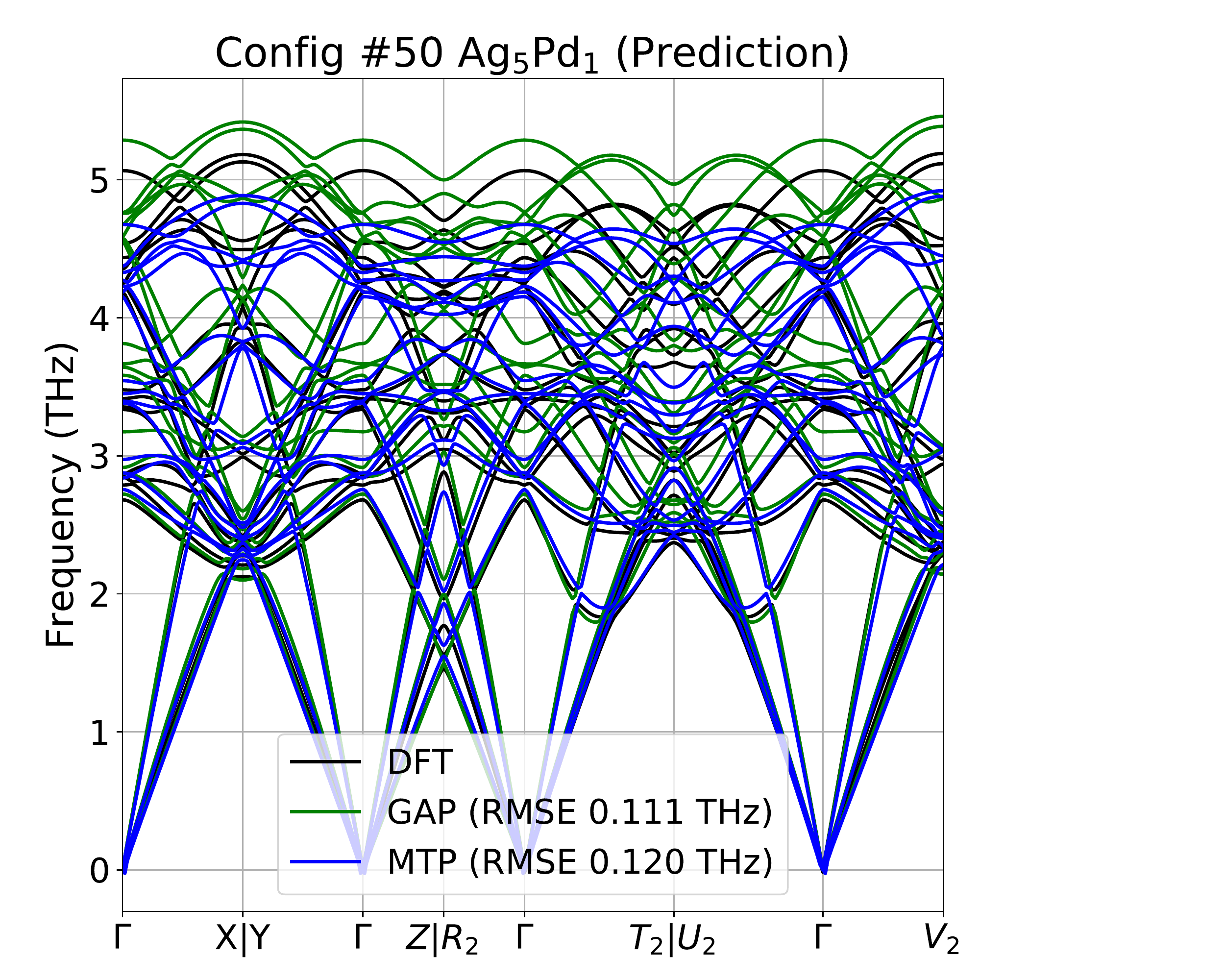}} \\
    \subfloat[]{\includegraphics[width = 3.2in]{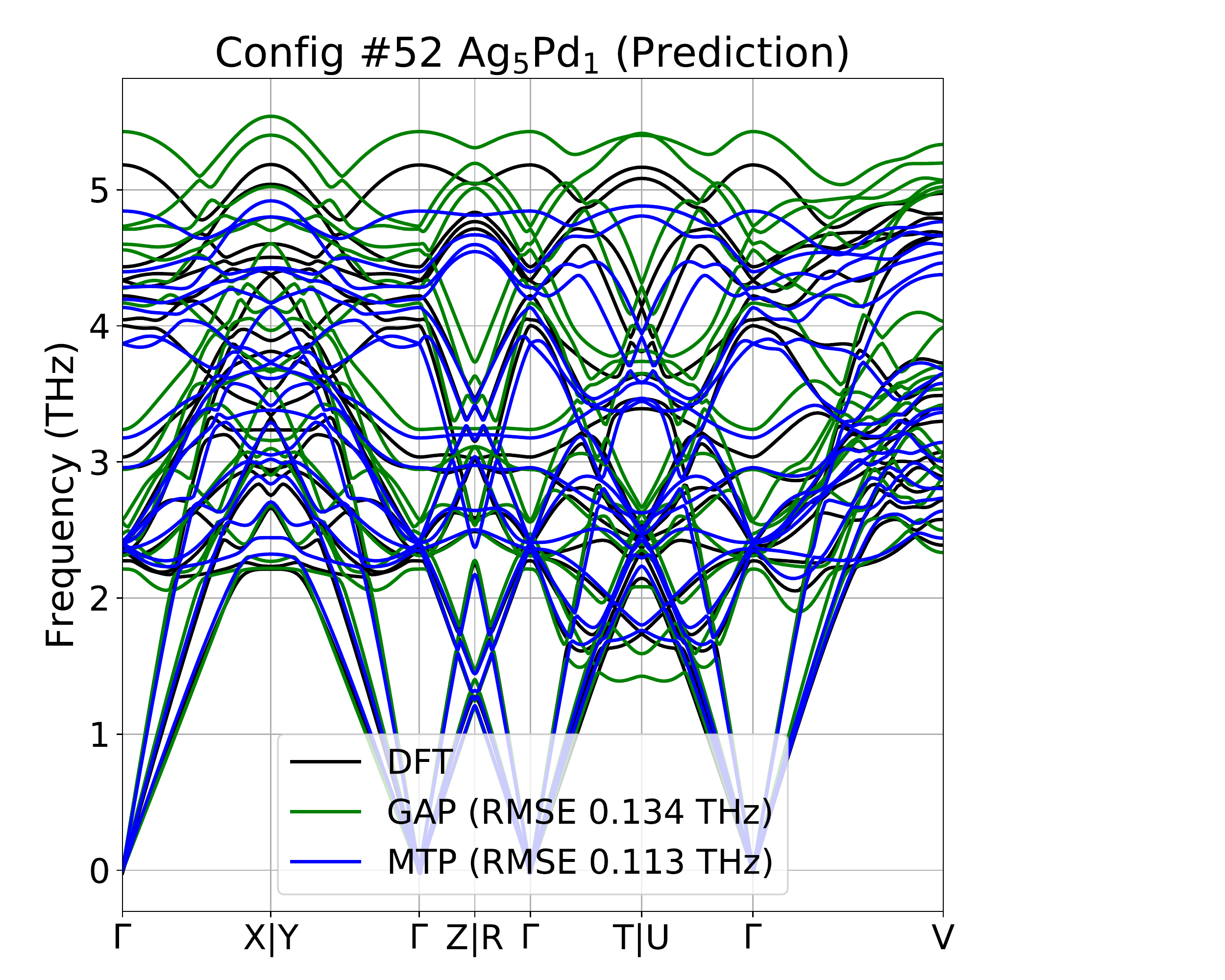}} &
    \subfloat[]{\includegraphics[width = 3.2in]{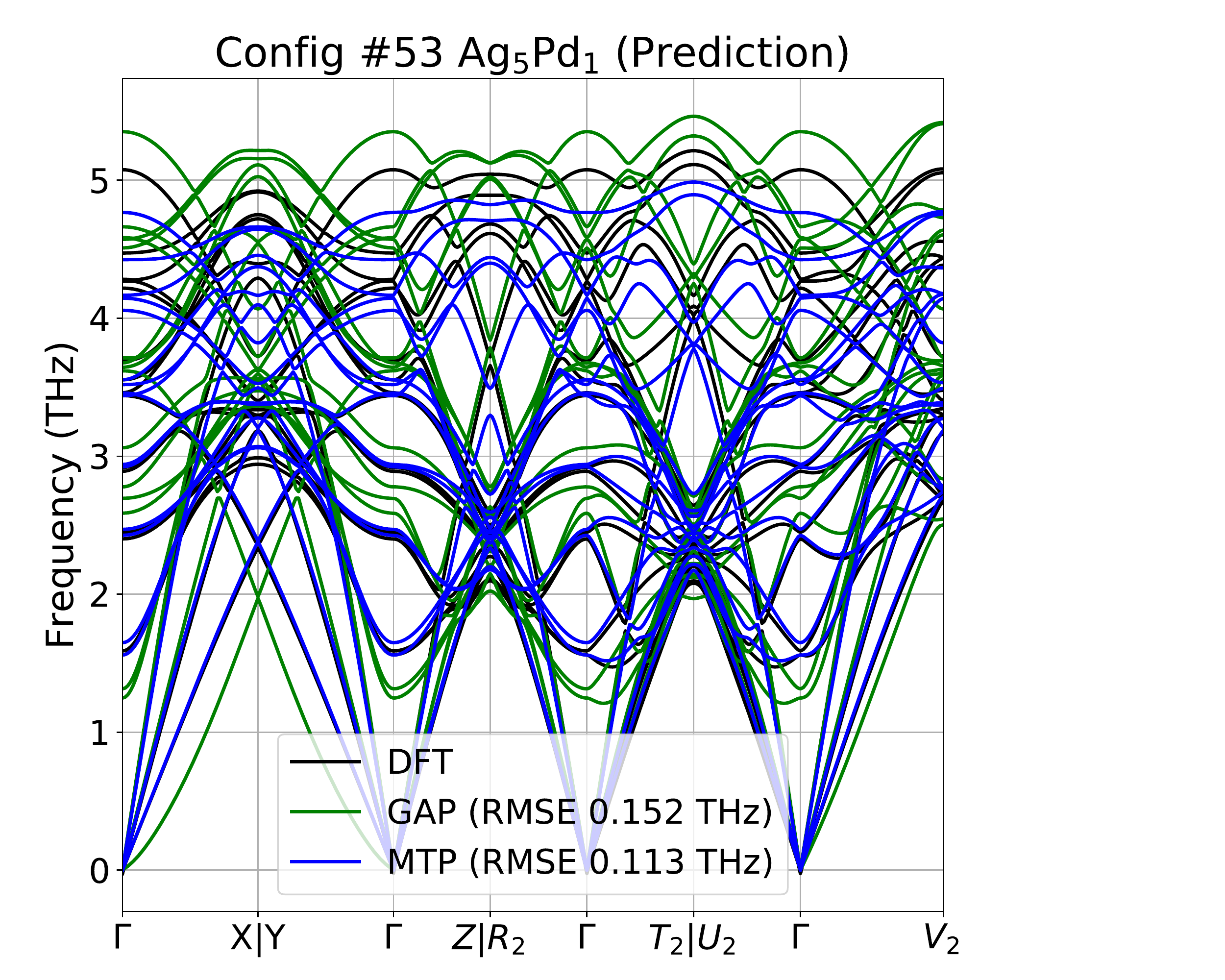}}
    \end{tabular}
    \end{minipage}  
    \end{adjustbox}
\end{figure*}
\begin{figure*}
    \begin{adjustbox}{rotate=0}
    \begin{minipage}{\textwidth}  
    \begin{tabular}{ccc}
    \subfloat[]{\includegraphics[width = 3.2in]{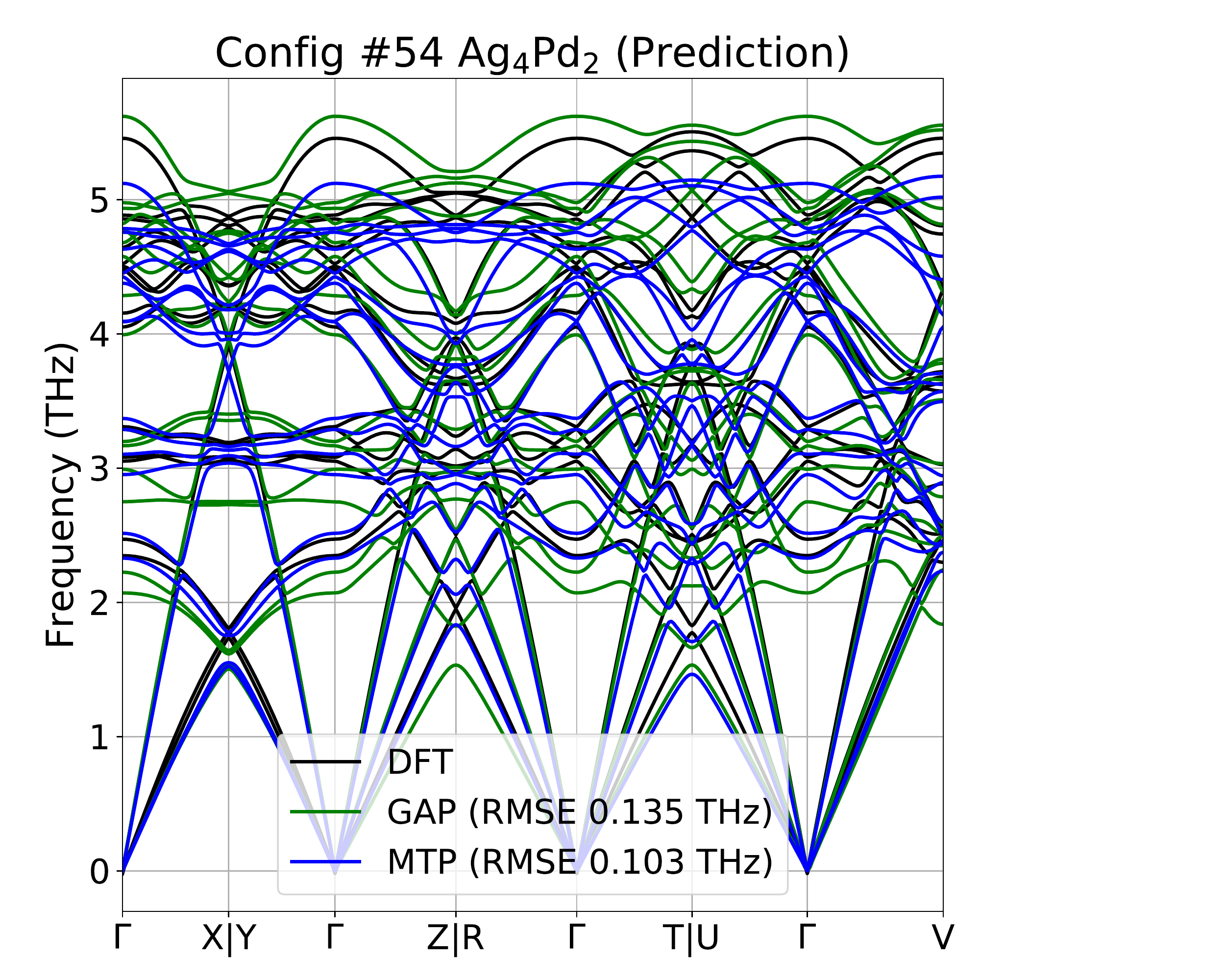}} &
    \subfloat[]{\includegraphics[width = 3.2in]{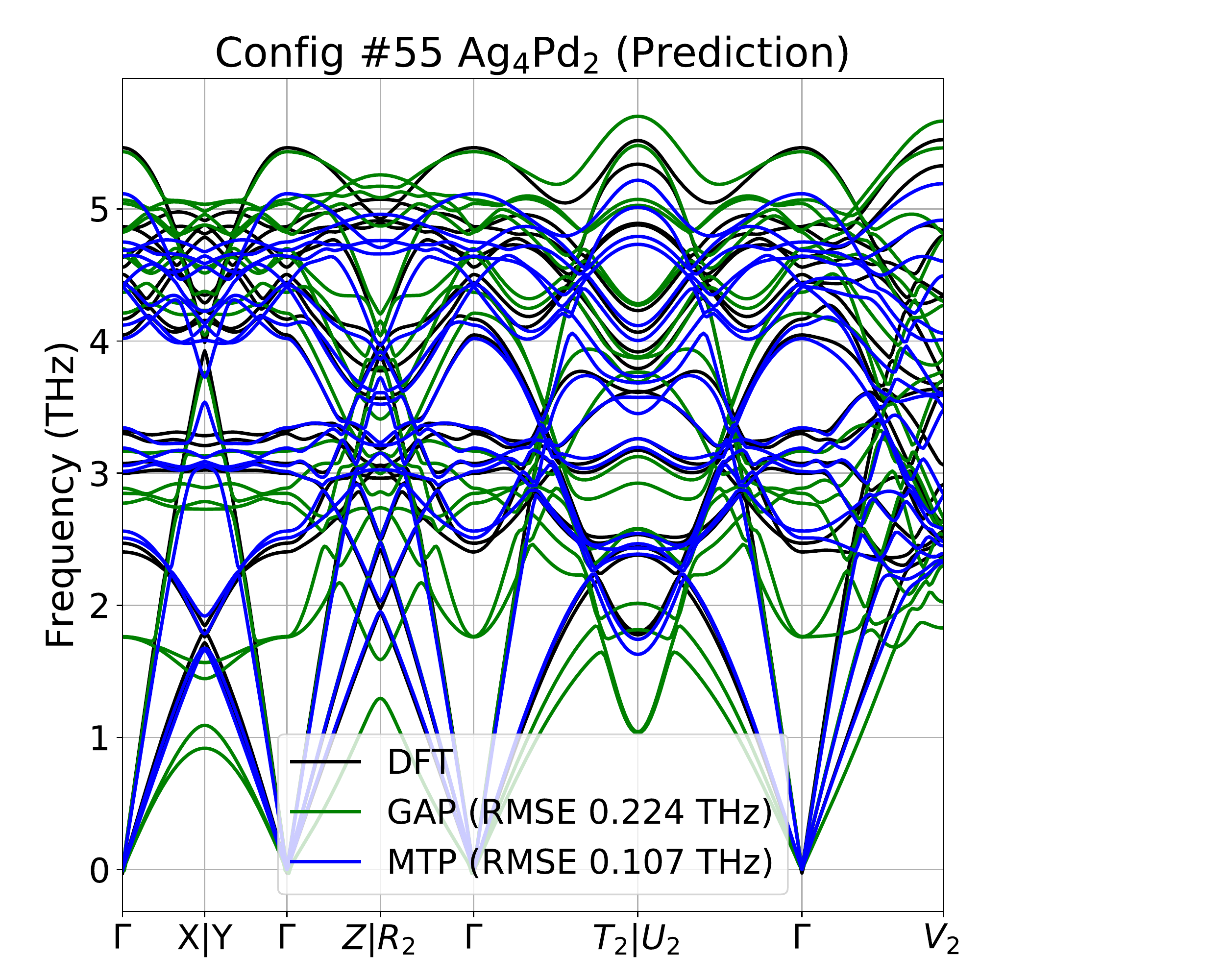}} \\
    \subfloat[]{\includegraphics[width = 3.2in]{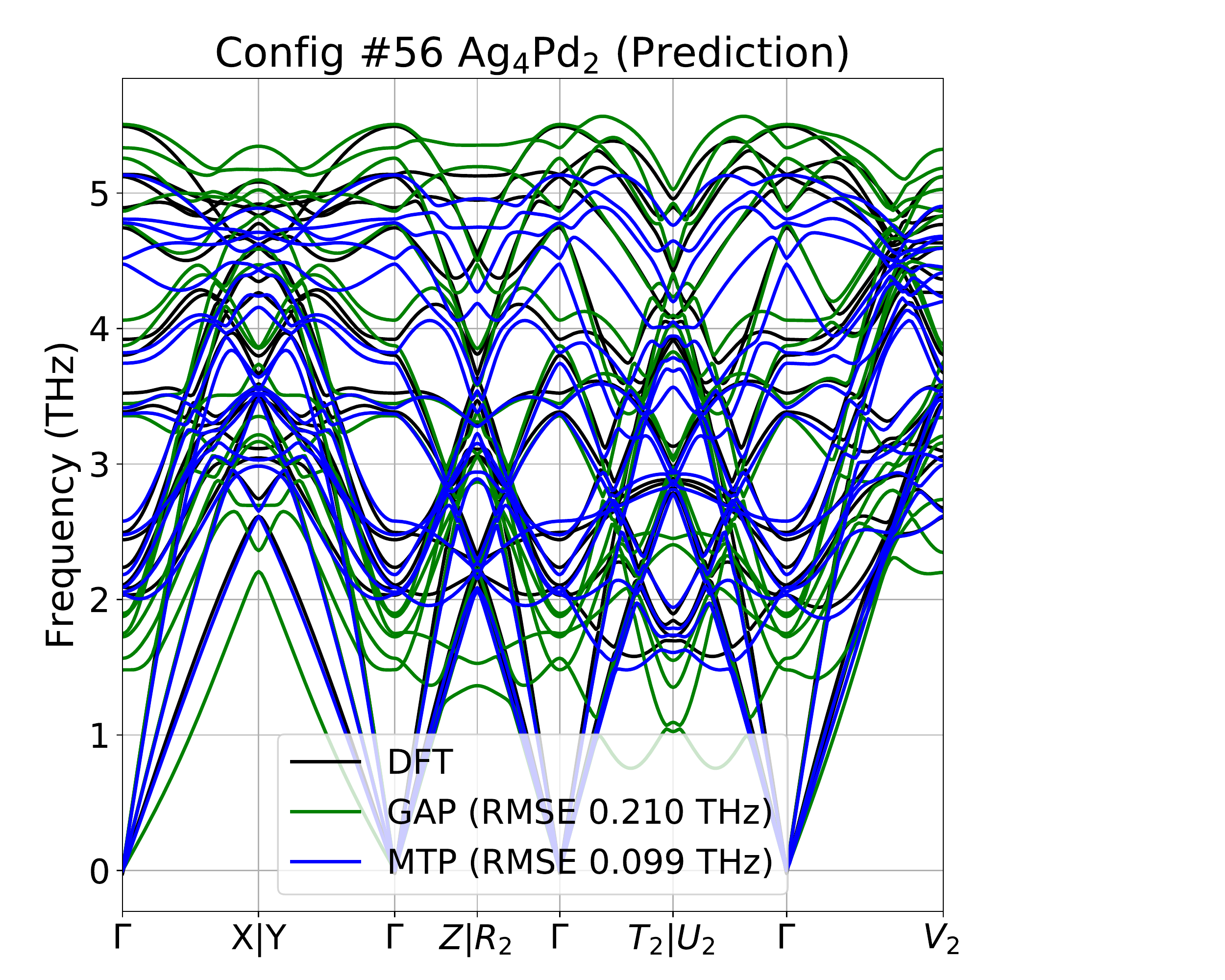}} &
    \subfloat[]{\includegraphics[width = 3.2in]{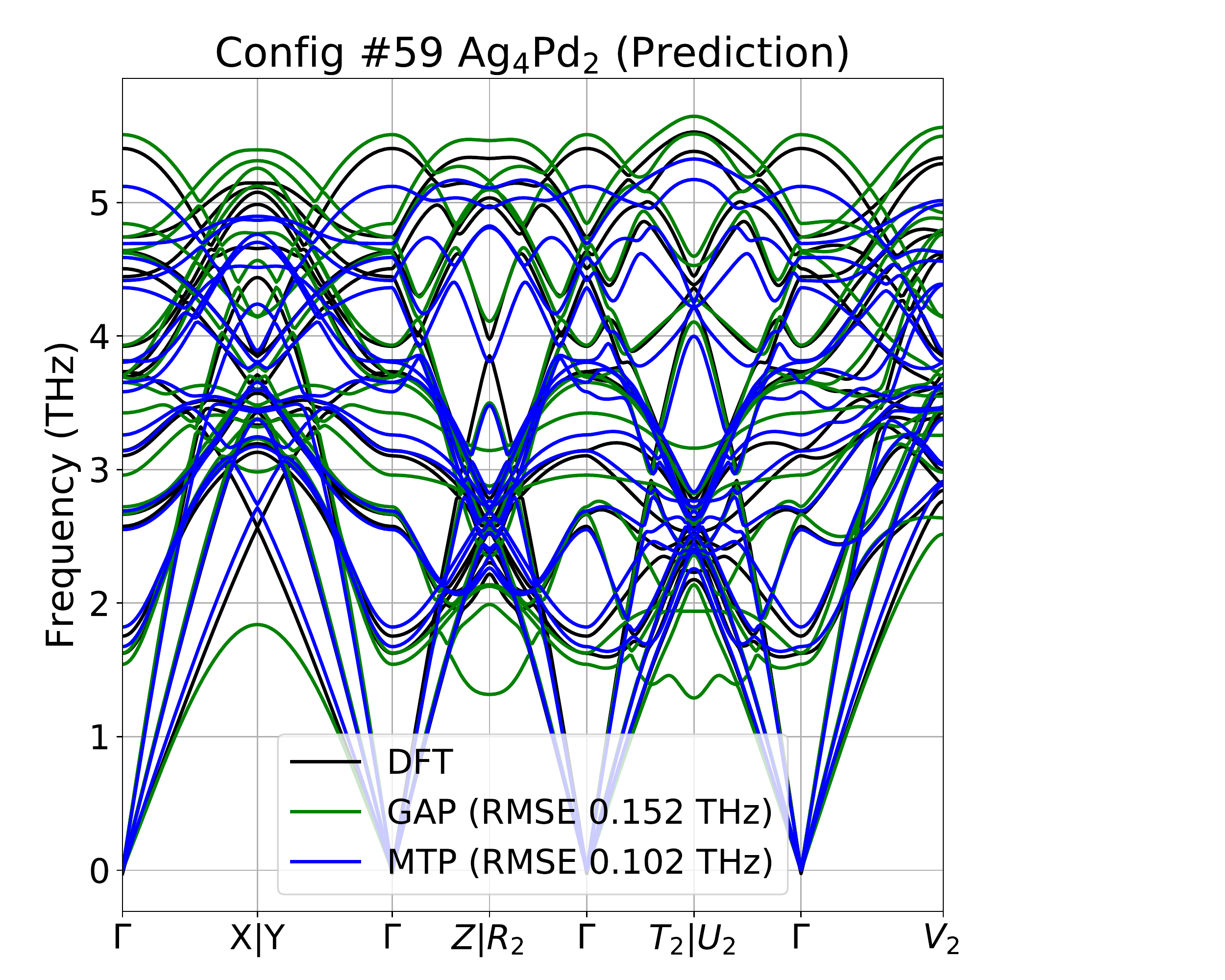}} \\
    \subfloat[]{\includegraphics[width = 3.2in]{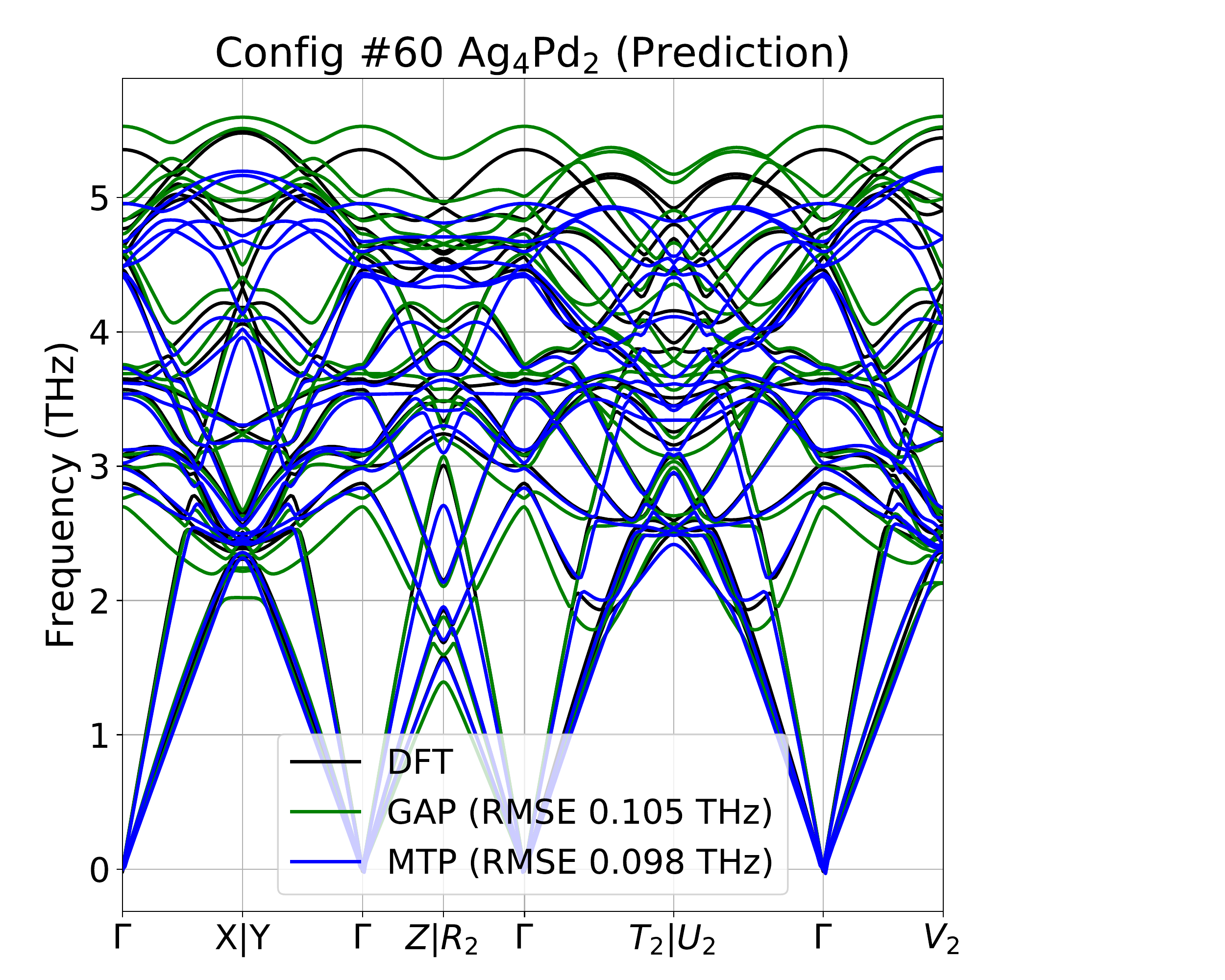}} &
    \subfloat[]{\includegraphics[width = 3.2in]{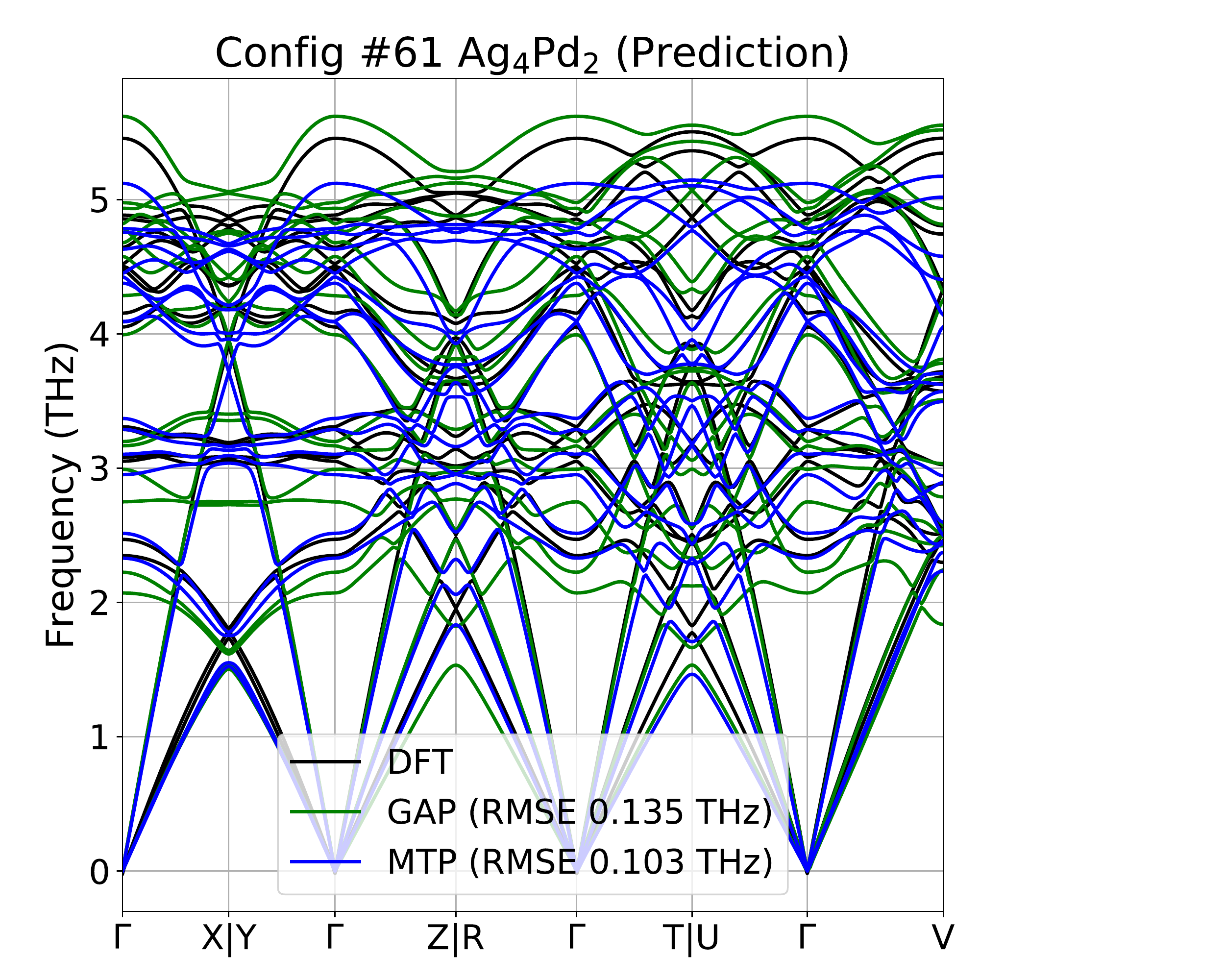}}
    \end{tabular}
    \end{minipage}  
    \end{adjustbox}
\end{figure*}
\bibliography{library}{}
\bibliographystyle{plain}